\documentclass{aa}  

\makeatletter
\renewcommand*\aa@pageof{, page \thepage{} of \pageref*{LastPage}}
\makeatother
\usepackage{lscape}
\usepackage{supertabular}
\usepackage{booktabs}
\usepackage{placeins}
\usepackage{float}
\usepackage{graphicx}
\usepackage{txfonts}
\usepackage{xcolor}
\usepackage{multirow}
\usepackage{hyperref}
\hypersetup{
  colorlinks=true,
    linkcolor=blue,
    citecolor=blue,
    urlcolor=blue}
\usepackage{rotating}

\bibpunct{(}{)}{;}{a}{}{,}

\begin{document}

   \title{Luminaries in the sky: The TESS LEGACY sample of bright stars}
   \subtitle{II. In-depth seismic characterisation of 32 naked-eye stars \\in the PLATO LOP fields}

   \author{E.~Panetier\inst{1} 
           \and
           G.~T.~Hookway\inst{2}
           \and
           E.~Corsaro\inst{3}
           \and
           S.~N.~Breton\inst{3}
           \and
           R.~A.~García\inst{4}
           \and
           B.~Liagre\inst{1}
           \and
           M.~N.~Lund\inst{5}
           \and 
           M.~B.~Nielsen\inst{2}
           \and
           D.~B.~Palakkatharappil\inst{4}
           \and
           L.~Debacker\inst{4,6}
           \and 
           J.~Gosmain\inst{4,7}
           \and
           M.~Chaumard\inst{8}
           \and
           A.~Chontos\inst{9}
           \and
           F.~Grundahl\inst{5}
           \and
           S.~Mathur\inst{10,11}
           \and
           A.~R.~G.~Santos\inst{4}
           }
   \institute{Universit\'e Paris Cit\'e, Universit\'e Paris-Saclay, CEA, CNRS, AIM, 91191, Gif-sur-Yvette, France \\
           \email{eva.panetier@cea.fr}
           \and
           School of Physics and Astronomy, University of Birmingham, Birmingham B15 2TT, UK
           \and
           INAF – Osservatorio Astrofisico di Catania, Via S. Sofia, 78, 95123 Catania, Italy
           \and 
           Universit\'e Paris-Saclay, Universit\'e Paris Cit\'e, CEA, CNRS, AIM, 91191, Gif-sur-Yvette, France
           \and
           Stellar Astrophysics Centre, Department of Physics and Astronomy, Aarhus University, Ny Munkegade 120, DK-8000 Aarhus C, Denmark
           \and
           IMT Atlantique, Technopole Brest-Iroise CS83818, 29238 Brest Cedex 03, France
           \and
           Institut Villebon - \textit{Georges Charpak}, Université Paris-Saclay, 91400 Orsay, France
           \and
           Ecole Centrale-Supelec, Universit\'e Paris-Saclay, 91190 Gif-sur-Yvette, France
           \and
           Department of Physics and Astronomy, Dartmouth College, Hanover, NH USA
           \and
           Instituto de Astrof\'isica de Canarias (IAC), E-38205 La Laguna, Tenerife, Spain
           \and
           Universidad de La Laguna (ULL), Departamento de Astrof\'isica, E-38206 La Laguna, Tenerife, Spain
           }

   \date{Received ; accepted}

  \abstract
   {The Transiting Exoplanet Survey Satellite (TESS) is conducting a nearly full-sky survey, enabling the photometric characterisation of millions of stars. The forthcoming PLAnetary Transits and Oscillations of stars mission (PLATO) will provide long-duration, high-precision photometry of tens of thousands of bright stars to be characterised through asteroseismology. The TESS Luminaries Sample is a catalogue of 196 bright naked-eye (V < 6) main-sequence (MS) and sub-giant (SG) stars exhibiting solar-like oscillations.
   Among them, the subset located within the PLATO long-duration observation phase (LOP) fields constitutes an exceptional set of targets that will be observable by PLATO from the earliest phases of the mission, making them ideal calibrators during commissioning and the first months of science operations. This paper aims to provide an in-depth asteroseismic characterisation of 32 Luminaries stars that fall within the PLATO LOP fields of view.
   Individual mode parameters were extracted using three independent seismic pipelines, one of which was similar to the algorithms used in the official PLATO pipeline. The Peirce criterion and a Z-score were applied to identify the optimal combination of data calibration, observing cadence, and fitting pipeline for each star.
   We analysed 32 MS and SG stars up to TESS Sector 88, with individual mode frequencies derived for the first time for 26 of them. 
   For all stars, we derived large and small separations, the asymptotic phase term, $\varepsilon$, radial mode amplitudes, and mean linewidths per order. Comparisons with previous \textit{Kepler} observations reveal consistent trends in the seismic parameters, confirming the robustness of our analysis based on TESS data. In SGs, mixed-mode identification differs in the three pipelines, revealing extraction inconsistencies requiring longer datasets to improve our mode identifications. 
   These results demonstrate the capability of TESS to deliver high-quality asteroseismic observations for MS and SG stars. The Luminaries stars located in the PLATO LOP fields constitute a unique sample that will play a crucial role in validating, calibrating, and optimising PLATO’s seismic performance from the earliest stages of the mission.
   }
   \keywords{}

   \maketitle
   \nolinenumbers

\section{Introduction}

The PLAnetary Transits and Oscillations of stars \citep[PLATO;][]{rauer_2025} mission, scheduled for launch in January 2027, is designed to detect and characterise exoplanets through high-precision photometry. Its ultimate goal is the detection of terrestrial planets orbiting Sun-like stars, but it will also enable a census of planetary systems around a broader range of solar-like stars, thereby contributing to our understanding of planetary system formation and evolution. A key aspect of its observing strategy is to focus on two dedicated regions of the sky, known as the long-duration observation phase (LOP) fields, each covering an area of $49^\circ \times 49^\circ$ \citep{rauer_2025}. According to the current mission plan, each LOP field will be monitored for approximately two years, although this may be revised following evaluation of initial observations.
The mission is scheduled to begin observations in the southern LOP field, known as LOPS2 \citep{nascimbeni_2025}, and may subsequently continue at LOPN1, the current candidate field proposed by \citet{nascimbeni_2022} in the northern hemisphere.

Given the number of targets expected to be observed \citep{montalto_2021} and seismically characterised \citep{goupil_2024} with the PLATO mission, fully automated pipelines will be essential to identify and characterise stellar oscillations in a consistent way across the entire sample. In this context, it is important that these pipelines are carefully tested and validated, before the launch as well as during the commissioning phase and the early stages of the mission, when the robustness of the data products is first being assessed.
The PLATO mission will therefore strongly benefit from stars whose seismic properties are already well established. Such stars provide useful comparison targets enabling the verification that the results produced by different analysis approaches are consistent with each other, and that the pipelines behave as expected on real data. In this way, they offer a practical way to build confidence in both the instrument performance and the analysis tools.

A particularly valuable set of potential seismic calibrators has been identified within the Transiting Exoplanet Survey Satellite \citep[TESS;][]{ricker_2014} Luminaries sample \citep[TLS;][hereafter Paper~I]{lund_2025}. This sample consists of 196 bright ($V < 6$) main-sequence (MS) and sub-giant (SG) solar-like stars observed by TESS. Among them, 34 stars lie within the planned PLATO LOP fields and therefore represent prime seismic targets for the mission. They are included in the first public release of the all-sky PLATO input catalogue \citep[asPIC 1.1][]{montalto_2021}, and those located within the LOPS2 are also part of the most recent version of the PIC \citep[PIC 2.2.0.1][]{montalto_2026, nascimbeni_2022, nascimbeni_2025, nascimbeni_2026}. In general, these stars are included in the target PIC (tPIC) as part of the core science P2 sample, which comprises dwarf and SG stars with $V<8.5$. These stars are among the brightest TESS seismic targets \citep{lund_2025}, making them well suited for complementary ground-based follow-up observations. Combined with the characterisation of individual oscillation modes, this enables precise determinations of their radii, masses, and ages. 
In the present study, we determine the detailed seismic properties of 32 of these targets, while 171\,Pup and $\chi$\,Dra are analysed in dedicated companion papers (Lund et al., in prep.; Rudrasingam et al., in prep.) because interferometric complementary observations were done providing more constraints for their characterisation. We note that HD 62644, $\chi$\,Dra, $\upsilon$\,Cep, and 99 Her will also be investigated by Appourchaux et al. (in prep.). These stars will play a key role in validating PLATO’s early seismic performances.

The most precise asteroseismic constraints are obtained when individual oscillation mode frequencies are extracted. However, the feasibility and robustness of this extraction depend strongly on the stellar properties and evolutionary stage.
F-like stars present particular challenges, as their oscillation modes exhibit large linewidths due to shorter mode lifetimes \citep{appourchaux_2012a,compton_2019}. The resulting broader and partially overlapping peaks complicate the identification of individual modes and reduce the precision of frequency determinations. Such stars occupy the upper boundary of the MS region of the Hertzsprung–Russell (HR) diagram where solar-like oscillations are observed \citep[$6400\,\mathrm{K}\lesssim T_\mathrm{eff}\,$,][]{appourchaux_2012, breton_2023}.
In contrast, cooler solar-like MS stars \citep[$5000 \lesssim T_\mathrm{eff} \lesssim 6400\,$K,][]{lund_2017} generally display narrower modes that allow for more straightforward frequency extraction. 
As stars evolve towards the SG phase, structural changes in the interior lead to the appearance of pressure (p) and gravity (g) mixed modes \citep{beck_2011,benomar_2013,appourchaux_2020}. These modes introduce additional complexity in the frequency spectrum and require dedicated identification strategies \citep[e.g.][]{handberg_2017,garciasaraviaortizdemontellano_2018,themessl_2018,kallinger_2019,corsaro_2020,nielsen_2021,nielsen_2025}. However, they also provide powerful diagnostics of stellar interiors at this key evolutionary stage, as they probe both the deep core and the outer layers. This offers unique constraints on the internal structure \citep[e.g.][]{deheuvels_2011,bellinger_2021,buchele_2025} and on angular momentum redistribution \citep[e.g.][]{deheuvels_2014,deheuvels_2020,buldgen_2024}, which remains one of the major open questions in stellar evolution. The sample of stars analysed in this work spans a range of effective temperatures and evolutionary stages across the MS and SG branch. This diversity enables an evaluation of frequency extraction and mode identification under conditions representative of PLATO’s future targets.

This paper is organised as follows. In Sect.~\ref{sec:obs}, we describe the sample of stars, including a summary of the data calibration performed in Paper~I and the additional calibration steps required for certain targets. Sect.~\ref{sec:mode_extraction_method} presents the three pipelines used for mode extraction. In Sect.~\ref{sec:results}, we report our peak-bagging results, beginning with the method used to select the adopted dataset for each star, including the cadence, calibration, and analysis pipeline, followed by a discussion of the results. Sect.~\ref{sec:comp_legacy} compares our findings with pre-existing studies. Finally, we summarise our findings in Sect.~\ref{sec:conclusion}.


\section{Observations and data preparation}
\label{sec:obs}

\subsection{Stellar sample description}

\begin{figure}
    \centering
    \includegraphics[width=1\linewidth,trim=0.5cm 0 0.5cm 0,clip]{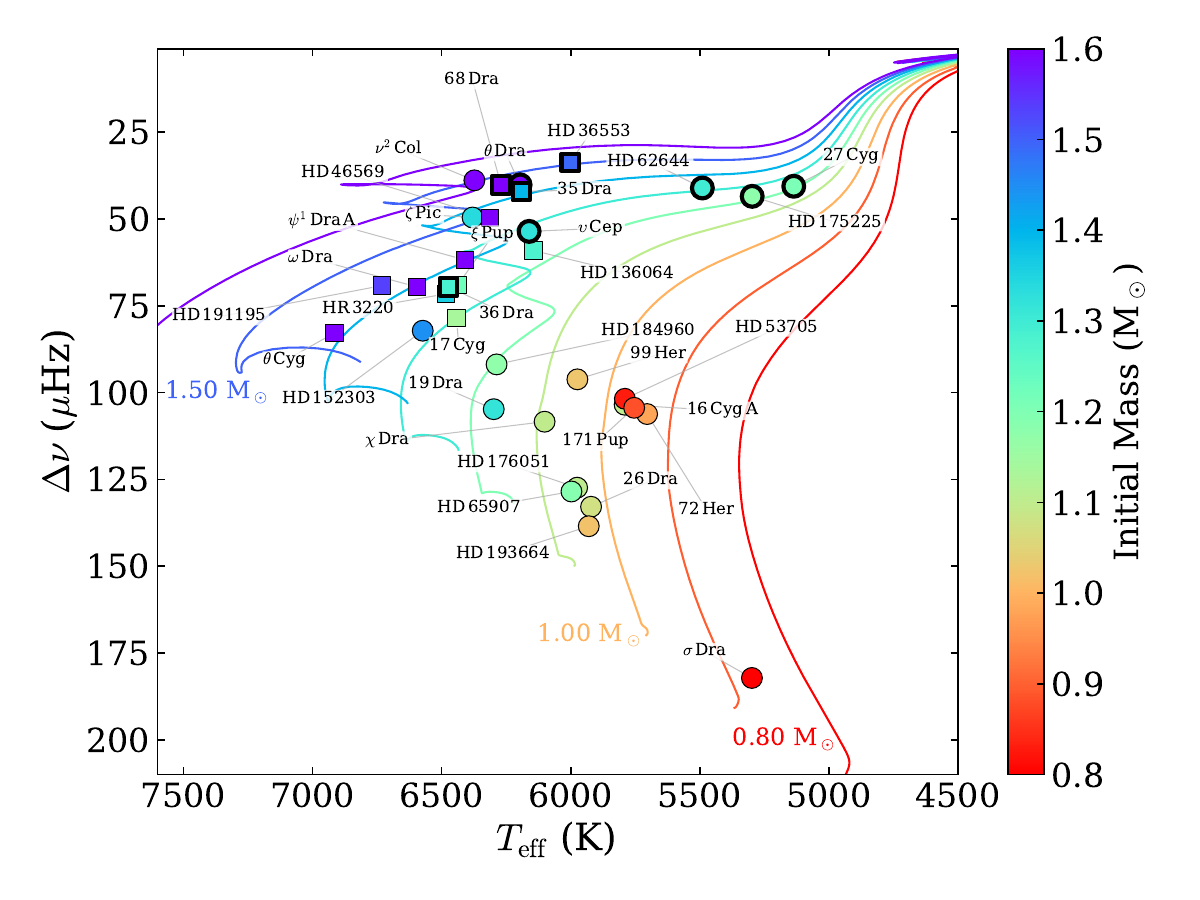}
    \caption{HR diagram illustrating the relation between $\Delta\nu$ and $T_\mathrm{eff}$ for the sample of 34 TLS stars. The evolutionary tracks are taken from the MESA isochrones and stellar tracks (\href{https://waps.cfa.harvard.edu/MIST/}{MIST}) catalogue and assume solar metallicity. The tracks were computed with a mass spacing of 0.1\,M$_\odot$. The data points are colour-coded according to seismic stellar mass. Squares denote stars with F-like, broad oscillation modes, circles correspond to the other stars, and markers with thick edges indicate stars exhibiting mixed modes (see Section~\ref{subsec:starbystar} for details).}
    \label{fig:HR-diagram}
\end{figure}

Our sample comprises 32 MS and SG stars found to exhibit solar-like oscillations as part of Paper~I. Among these, 10 stars are located within the PLATO LOPS2 field, while 24 are situated in the LOPN1 field. The global seismic parameters were derived using the \texttt{pySYD} pipeline \citep{chontos_2022}, as described in Paper~I. 
Here, the frequency of maximum oscillation power, $\nu_\mathrm{max}$, corresponds to the frequency where the oscillation power spectrum reaches its maximum, while the large frequency separation, $\Delta\nu$, is defined as the average spacing between consecutive radial ($\ell=0$) acoustic modes. These global seismic parameters scale with the stellar mass, $M$, radius, $R$, and effective temperature, $T_\mathrm{eff}$, according to the asteroseismic scaling relations \citep{kjeldsen_1995}, such that
\begin{equation}
\frac{\nu_\mathrm{max}}{\nu_{\mathrm{max},\odot}} \approx
\left( \frac{M}{M_\odot} \right)
\left( \frac{R}{R_\odot} \right)^{-2}
\left( \frac{T_{\mathrm{eff}}}{T_{\mathrm{eff},\odot}} \right)^{-1/2}\,,
\label{eq:scaling_nu_max}
\end{equation}
and
\begin{equation}
\frac{\Delta\nu}{\Delta\nu_\odot} \approx
\left( \frac{M}{M_\odot} \right)^{1/2}
\left( \frac{R}{R_\odot} \right)^{-3/2}\,.
\label{eq:scaling_delta_nu}
\end{equation}
The values of $\Delta\nu$ and $\nu_\mathrm{max}$, along with the corresponding fundamental stellar parameters, are listed in Table~\ref{tab:fund_params}. The seismic HR diagram presented in Fig.~\ref{fig:HR-diagram} displays $\Delta\nu$ as a function of the effective temperature, with the colour scale indicating the seismic stellar mass. The masses were computed from Eqs.~\eqref{eq:scaling_nu_max} and \eqref{eq:scaling_delta_nu}, adopting the solar  values $\nu_{\mathrm{max},\odot}=3090\,\mu\mathrm{Hz}$, $\Delta\nu_\odot=135\,\mu\mathrm{Hz}$, and $T_{\mathrm{eff},\odot}=5770\,\mathrm{K}$ \citep{huber_2011}. For this purely illustrative representation in the HR diagram, no correction to $\Delta\nu$ was applied \citep[e.g.][]{white_2011}. The stars span a broad range of temperatures and seismic masses, covering both the MS and the SG branch.

\subsection{Data}

We analysed TESS observations up to and including Sector~88. 
TESS is a NASA mission led by the Massachusetts Institute of Technology (MIT) and launched on April 18, 2018, performing high-precision, time-series photometric observations as part of an all-sky survey \citep{ricker_2014}. 
TESS is designed to achieve a photometric precision of 50 parts per million (ppm) for stars with TESS magnitudes between 9 and 15. Using four identical cameras, it observes the sky sector by sector with a field of view of $24^\circ \times 96^\circ$ per sector. Each sector is monitored for approximately 27 days, with multiple cadence options available.

In this paper, we use both the standard pre-search data conditioning sample aperture photometry (PDCSAP) light curves produced by the TESS science processing operations centre (SPOC) and light curves extracted from target pixel files using custom apertures with the K2P$^2$ pipeline \citep{lund_2015}, following the procedure described in Paper~I. The data include observations obtained at 120-s cadence and, where available from Sector~27 onwards, 20-s cadence. In sectors providing both cadences, the 20-s data were additionally binned to a 120-s cadence to take advantage of their typically lower noise properties \citep{huber_2022}. All original TESS data were downloaded from the Mikulski archive for space telescopes (MAST\footnote{https://archive.stsci.edu}).

For targets requiring improved photometric precision, particularly bright or saturated stars, the custom-aperture light curves provided a significant improvement over the PDCSAP products and, in some cases, enabled the detection of oscillations not apparent in the standard light curves.
All light curves were corrected for long-term and instrumental trends using the \emph{Kepler} asteroseismic science operations centre (KASOC) filter \citep{handberg_2014}. The power spectral density (PSD) was then computed from the filtered time series using a Lomb–Scargle periodogram \citep{lomb_1976,scargle_1982} and normalised according to Parseval’s theorem.

$\theta$ Cyg required special treatment because its PSD exhibits a clear series of harmonics, with a fundamental peak corresponding to a period of about 0.2815 days (or 
6.7560\,hours) close to what was previously reported by \citet{guzik_2016}. Such a harmonic pattern is typically indicative of a non-sinusoidal periodic signal and may be associated with a transit or eclipse event in the light curve, either intrinsic to the target or arising from a contaminating companion within the TESS photometric aperture. Folding the light curve at this frequency reveals a distinct, regularly recurring flux dip. After removing the data points associated with this feature and recomputing the PSD, the harmonics linked to the binarity or transit signature disappear, enabling a reliable seismic analysis.

\section{Mode extraction methods and strategy}
\label{sec:mode_extraction_method}

The properties of the stellar oscillations were inferred through a detailed peak-bagging analysis carried out with three independent frameworks: \texttt{apollinaire} \citep{breton_2022}, FAMED \citep{corsaro_2020}, and \texttt{PBjam} \citep{nielsen_2021, nielsen_2025}. Each pipeline independently models the observed PSD to recover the frequencies, amplitudes, and linewidths of the oscillation modes. Below, we describe how each method was implemented and optimised.

\subsection{\texttt{Apollinaire}}

The \texttt{apollinaire} module \citep{breton_2022} was developed to extract seismic parameters directly in the Fourier domain using an ensemble-sampling Markov chain Monte Carlo (MCMC) approach implemented with \texttt{emcee} \citep{foreman-mackey_2013}. The analysis begins with the removal of the stellar background signal, which is modelled as the sum of one to three Harvey-like profiles \citep{harvey_1985}, a Gaussian envelope centred on $\nu_{\rm max}$ to describe the p-mode power excess, and a constant white-noise component.

Initial estimates of the individual oscillation mode parameters were obtained from the global pattern fit for solar-like MS stars. The detailed procedure is discussed in Appendix~\ref{app:modes_id}. 
For stars exhibiting more complex oscillation spectra, these initial estimates were instead obtained using the \texttt{iechelle}\footnote{Available at \url{https://gitlab.com/dinilbose/iechelle.git}} pipeline, which allows the user to manually select preliminary frequency estimates from échelle diagrams. In this case, the initial mode identification followed the prescription of \citet{white_2012}, with the phase offset, $\varepsilon$, inferred from the estimated mode frequencies, while the large frequency separation, $\Delta\nu$, was adopted from Paper~I together with the effective temperature listed in Table~\ref{tab:fund_params}. The parameter priors were then manually adjusted to maintain distinct frequency windows for neighbouring modes, thereby preventing overlap and improving the stability of the MCMC sampling.

These initial estimates were used as starting points for the simultaneous sampling of the joint posterior distribution. Each oscillation mode, characterised by its angular degree, $\ell$, and radial order, $n$, was modelled as a symmetric Lorentzian profile, with the central frequency, $\nu_\mathrm{n\ell}$, treated as a free parameter. The mode heights were parametrised according to
\begin{align}
    H_{\mathrm{n}, \ell} &= A_\ell\,H_\mathrm{n, 0} \; \text{for} \; \ell=0,1 \, , \\
    H_{\mathrm{n}, \ell} &= A_\ell\,H_\mathrm{n-1, 0} \; \text{for} \; \ell=2,3 \, ,
\end{align}
where $A_0 = 1$ by definition, and the relative amplitudes, $A_\ell$, for $\ell \neq 0$ were fitted. The linewidths, $\Gamma_\mathrm{n\ell}$, were assumed to be common among groups of modes with degrees $\ell=(0,1,2,3)$ and radial orders $(n,n,n+1,n+1)$, respectively. The effects of the observational window function, including gaps in the time series, were explicitly taken into account by convolving the model with the window function prior to the Bayesian inference, ensuring unbiased parameter estimates. This ensures that the extracted mode parameters remain unbiased, as demonstrated by \citet{breton_2022a}.
To ensure uniform priors across all sampled parameters, the MCMC was performed in terms of $\log \Gamma_\mathrm{n,0}$, $\log H_\mathrm{n,0}$, $A_\ell$, and $\nu_{\mathrm{n},\ell}$. Convergence of the sampling and the absence of strong parameter degeneracies were assessed through visual inspection of the marginalised posterior distributions and two-dimensional corner plots.

\subsection{FAMED}

The peak-bagging analysis based on the adoption of the \textsc{Diamonds} code\footnote{\url{https://github.com/EnricoCorsaro/DIAMONDS}} \citep{corsaro_2014} relies on performing a preliminary step that fits the background signal of the stellar PSD. The background signal comprises granulation activity at two different scales, long-trend variation related to instrumental noise, potential magnetic activity and rotational effects, and the white noise. For this purpose, we adopted the \textsc{Background} code extension of \textsc{Diamonds}\footnote{\url{https://github.com/EnricoCorsaro/Background}}. 
Once the background fit is estimated, we adopted the peak-bagging pipeline \textsc{FAMED} \citep{corsaro_2020} for an automated extraction and mode identification of the significant oscillation peaks that can be found in the power excess of the stellar PSD.
\textsc{FAMED} is relying on the so-called multi-modal fitting, which is accomplished by means of the nested sampling Monte Carlo method \citep{Skilling04}, capable of identifying multiple degenerate solutions in the parameter space. Degenerate solutions are built through the adoption of a simple, single-profile, model shaped as a Lorentzian to reproduce the presence of oscillation peaks in the stellar PSD \citep{Corsaro19}. The pipeline operates modularly by progressively improving the level of detail in the analysis, focusing originally on the entire oscillation envelope region and then moving into individual chunks of length defined by the large frequency separation, $\Delta\nu$.
The mode identification method used here is summarised in Sect.~\ref{subsec:data_driven}.

\subsection{\texttt{PBjam}}

\texttt{PBjam}\footnote{\url{https://github.com/PBjam-projects/PBjam}}~\citep{nielsen_2021,nielsen_2025} was also used to perform peak-bagging on this stellar sample, following the methodology described in \citet{Hookway_2025}. The analysis proceeds in two stages: an initial mode-identification step followed by detailed peak-bagging.

During mode identification, a background model consisting of three Harvey-like profiles \citep{harvey_1985} and a white-noise component is fitted to the power spectrum. This model is combined with a sum of Lorentzian profiles describing the radial and quadrupole ($\ell=0,2$) modes \citep{anderson_1990}, whose frequencies are estimated following the procedure described in Appendix~\ref{app:modes_id}.
In this framework, mode identification is formulated as a Bayesian inference problem in which asymptotic parameters and individual mode frequencies are inferred simultaneously from the power spectrum. 
The posterior distributions quantify both the most probable mode identification and the associated uncertainties, allowing ambiguous or noise-dominated peaks to be distinguished from genuine oscillation modes. Posterior sampling is performed with the \texttt{dynesty} nested sampling algorithm \citep{speagle_2020}. The \texttt{PBjam} priors are constructed from asteroseismic parameters of tens of thousands of previously analysed \textit{Kepler}/K2, TESS, and model stars, spanning red giant branch, red clump, SG, and MS evolutionary stages, with $\nu_\mathrm{max}$ values ranging from $20\,\mu$Hz to $4000\,\mu$Hz. These priors therefore encode extensive empirical knowledge of oscillation patterns to guide the identification of mixed modes.
The treatment of dipole ($\ell=1$) modes depends on the evolutionary stage of the star. For stars without visible mixed modes, $\ell=1$ modes are modelled as Lorentzian profiles, similarly to $\ell=0$ and $\ell=2$. For stars exhibiting mixed dipole modes, the asymptotic relation for $\ell=1$ gravity modes is included, as described in Sect.~\ref{subsec:mix_modes_form}.

The parameters inferred during the identification stage are then used as priors for the detailed peak-bagging step, where the model constraints are relaxed. Frequency priors are defined as $\beta$ distributions, i.e. continuous probability distributions bounded on a finite interval and parametrised by two positive shape parameters $(\alpha,\beta)$ that control their symmetry and concentration. These distributions were mapped onto an interval of width $\delta\nu_{02}/2$ centred on the initial estimate. Symmetric priors with $\alpha=\beta=5$ are adopted, favouring the asymptotic frequency, while allowing deviations within the prescribed range, for example due to acoustic glitches \citep{mazumdar_2014}. Normal priors are used for the logarithms of the mode widths and heights, centred on the identification-stage values with a standard deviation of 50\,\%.

Posterior distributions are sampled with the \texttt{emcee} MCMC algorithm. To assess the reliability of the inferred frequencies, the median absolute deviation of the posterior distributions is compared with the standard deviation of the corresponding $\beta$ priors. This posterior-to-prior ratio quantifies the contribution of the data, and modes with ratios below $2/3$ are considered validated and retained for further analysis.


\section{Results and analysis}
\label{sec:results}

To characterise the oscillation spectra, we first applied the \texttt{apollinaire} pipeline to the four different light curves per star: 20-s cadence data calibrated using either the SPOC or K2P$^2$ procedures, and combined 20-s and 120-s cadence data processed with the same two calibrations.
Owing to the low-frequency filtering applied to the light curves, a simplified background description was sufficient, combined with a cut at low frequency adapted for each star. Two Harvey-like components were required for 17 stars, while a single component adequately described the remaining 15, in all cases providing a satisfactory fit to the background over the frequency range of the oscillation power excess. The quality of the background characterisation was assessed from the posterior distributions of the fits and visual inspection of the residuals.
Background fitting converged for at least one light curve for every star in the sample.
Individual oscillation modes were then fitted using the background-corrected power density spectra. Comparing the resulting mode frequencies across light curves enabled us to evaluate the impact of cadence and calibration combination, leading to the a posteriori selection of the adopted light curve for each star based on the internal consistency and quality of the \texttt{apollinaire} results.

\subsection{Selection of the optimal light curve}

The level of agreement between the mode-frequency datasets obtained from the different light curves using \texttt{apollinaire} was quantified. This analysis was restricted to stars with peak-bagging results from at least three out of the four light curves. For each light curve $i$, we evaluated the deviation of its measured mode frequencies with respect to a consensus estimate defined by the remaining datasets. For a given mode $(n,\ell)$, this deviation is defined as
\begin{equation}
    \Delta_{i,\mathrm{n\ell}} = \nu_{i,\mathrm{n\ell}} - \bar{{\nu}}_{j\neq i,\mathrm{n\ell}}\,,
\end{equation}
where $\nu_{i,\mathrm{n\ell}}$ is the frequency measured from light curve $i$, and $\bar{{\nu}}_{j\neq i,\mathrm{n\ell}}$ is the weighted mean of the frequencies obtained from the other light curves, $j\neq i$. 
The significance of the deviation $\Delta_{i,\mathrm{n\ell}}$ is quantified using a Z-score, which represents the number of standard deviations a measured mode frequency is from the mean:
\begin{equation}
    Z_{i,\mathrm{n\ell}} = \frac{\Delta_{i,\mathrm{n\ell}}}{\sigma_{\Delta_{i,\mathrm{n\ell}}}}\,,
\end{equation}
where the uncertainty $\sigma_{\Delta_{i,\mathrm{n\ell}}}$ includes contributions from both light curve $i$ and the weighted mean of the remaining frequency datasets. Denoting by $\sigma_{i,\mathrm{n\ell}}$ and $\sigma_{j,\mathrm{n\ell}}$ the uncertainties associated with the frequency measurements $\nu_{i,\mathrm{n\ell}}$ and $\nu_{j,\mathrm{n\ell}}$, respectively, and assuming independent Gaussian errors, the variance of the deviation can be written as
\begin{equation}
    \sigma_{\Delta_{i,\mathrm{n\ell}}}^2 = \sigma_{i,\mathrm{n\ell}}^2 + \left(\sum_{j\neq i} \frac{1}{\sigma_{j,\mathrm{n\ell}}^2}\right)^{-1}\,.
\end{equation}
This Z-score therefore measures how far the frequency reported from light curve $i$ lies from the weighted mean of all other frequency datasets. Values of $|Z| \lesssim 1$ indicate agreement within the quoted uncertainties, while larger values highlight tensions between frequency measurements for a given mode.
The different light curves for each star do not always cover exactly the same time span, which can introduce small variations in the measured mode frequencies. These differences arise because solar-like oscillations are stochastically excited \citep{goldreich_1977, samadi_2001, belkacem_2008} and intrinsically damped \citep{belkacem_2012}; hence, their amplitudes and phases fluctuate over time. In addition, magnetic activity can shift mode frequencies slightly on timescales comparable to the observation windows \citep[e.g.][]{garcia_2010, regulo_2016,santos_2019}. These intrinsic temporal effects may therefore contribute to the residual deviations captured by the Z-scores.

The optimal light curve was selected by maximising the temporal coverage while minimising systematic deviations with respect to the frequency datasets obtained from the other light curves. Specifically, we aim to maximise the number of observed sectors and minimise the median Z-score, $\mathrm{med}\,|Z_{i,\mathrm{n\ell}}|$, computed over all modes in light curve $i$. The selection procedure is as follows:
\begin{enumerate}
    \item For the frequency dataset of each light curve $i$, compute med\,$|Z_{i,\mathrm{n\ell}}|$ and count the number of modes, $N_i^{\mathrm{min}|Z|}$, for which the dataset minimises $|Z_{i,\mathrm{ n\ell}}|$.
    \item When the number of sectors observed in 120-s cadence exceeds twice that of the 20-s cadence, and fewer than five sectors of 20-s data are available, priority is given to the light curves combining 20-s and 120-s cadences, provided their Z-score statistics are comparable. If $\mathrm{med}\,|Z_{i,\mathrm{n\ell}}|$ instead favours a 20-s-only light curve, the PSD and échelle diagram are inspected to validate this choice, by assessing the S/N and the visibility of the largest possible number of modes.
    \item In all other cases, preference is given to the light curve with the smallest $\mathrm{med}\,|Z_{i,\mathrm{n\ell}}|$ and the largest $N_i^{\mathrm{min}|Z|}$, with particular emphasis on preserving 20-s cadence data when available.
\end{enumerate}
For the four stars ($\theta$\,Dra, $\psi^1$\,Dra\,A, 16\,Cyg\,A, and HD\,53705) for which peak-bagging results were obtained from only two light curves, a similar approach was adopted, combining visual inspection with maximisation of the temporal coverage. For seven additional stars, the mode parameters could be evaluated from only a single light curve, which was therefore adopted as the optimal dataset.
The fitted background parameters for the selected optimal light curve are listed in Table~\ref{tab:back_params} for each star. 
Mode-parameter extraction could not be performed for $\nu^2$\,Col, HD\,65907, and HD\,152303 due to the very low visibility of the oscillation modes, yielding 29 stars with individual mode frequencies. 
Nevertheless, the background could still be constrained, and the corresponding parameters are also reported in Table~\ref{tab:back_params}: for the 120-s cadence data in the case of $\nu^2$\,Col for which no 20-s observations are available, and for the 20-s cadence data with more than ten observed sectors for the two remaining stars, using the K2P$^2$ calibration.
Out of the 32 stars in the sample, the $\nu_\mathrm{max}$ values derived in this step differ by more than 2\,$\sigma$ from the Paper~I values for only six stars ($\theta$~Dra, HD~175225, $\upsilon$~Cep, 36~Dra, 17~Cyg, and $\sigma$~Dra), while the remaining stars are consistent with Paper~I. Similarly, all $\Delta\nu$ values agree with Paper~I, except for $\theta$~Dra and 19~Dra, which exceed the 2\,$\sigma$ difference.

\subsection{Pipeline comparison: Peirce criterion}
\label{subsec:pipeline_comaprison}

Independent frequency estimates for each mode ($n$, $\ell$) were obtained from the optimal light curve of each star by three different fitters and their preferred fitting pipelines (\texttt{apollinaire}, FAMED, and \texttt{PBjam}), yielding $N\leq3$ measurements per mode. Before any comparison, outliers were identified and removed using Peirce’s rejection criterion \citep{peirce_1852,gould_1855}, which is applicable when measurements are drawn from a common parent distribution with random, normally distributed errors. Although the number of measurements per mode is small, the criterion can still be applied because it is designed to identify even a single anomalous observation in a small dataset. In this framework, a data point is rejected if excluding it increases the likelihood of the residual distribution, while accounting for the probability of observing the given number of abnormal measurements.
The formulation introduced by \citet{gould_1855}, which proceeds iteratively, is implemented in the Python package \texttt{pareidolia}\footnote{available at \url{https://gitlab.com/evapanetier/pareidolia.git}}. For each mode, an initial mean frequency and its root-mean-square dispersion were first computed from the three different measurements. A rejection factor was then evaluated under the assumption of one questionable measurement, and values exceeding this threshold were removed. When multiple points were rejected, the rejection factor was updated accordingly, and the procedure was repeated until convergence was reached. This approach yielded, for each mode, a statistically consistent subset of retained measurements. Similar applications of this method in asteroseismology have been shown to be effective for harmonising frequency determinations across independent analyses \citep[e.g.][]{metcalfe_2010,campante_2011,mathur_2011a,appourchaux_2012}.

Following convergence of the rejection process, the remaining frequencies were organised into three categories. The minimal list comprises modes for which all pipelines provide a consistent, non-rejected measurement after outlier rejection (flag 1 in the frequency tables; one example is shown in Appendix~\ref{app:tables_per_star}, with the full set available in machine-readable format). The maximal list includes modes supported by at least two pipelines (flag 2). Modes detected by only a single fitter are flagged separately (flag 3).

For each star, we identified a consistent set of measurements obtained with a single pipeline to be prioritised for subsequent modelling (Bétrisey et al., in prep.). To this end, we evaluated the agreement between individual frequency estimates and the mean frequencies derived from the minimal list. For each fitter, $k$, we computed the normalised root-mean-square deviation of its frequencies following the prescription of \citet{appourchaux_2012}:
\begin{equation}
    \sigma_{\mathrm{normdev,k}} = \sqrt{ \frac{1}{N_\mathrm{k}} \sum_{n,\ell} \frac{\left|\nu^{k}_\mathrm{n,\ell} - \langle \nu_\mathrm{n,\ell} \rangle\right|^2}{(\sigma^{k}_\mathrm{n,\ell})^2}}\,,
\end{equation}
where \(\nu^{k}_\mathrm{n,\ell}\) and \(\sigma^{k}_\mathrm{n,\ell}\) denote the measured frequency value and its uncertainty, and \(N_\mathrm{k}\) is the number of modes available in the dataset reported by fitter $k$.
The pipeline yielding the smallest normalised deviation with respect to the minimal list was then adopted as the reference to be used for subsequent modelling.
When mode parameters were obtained by only two pipelines for a given star, the minimal list was formed from the modes fitted by both. In this case, the dataset containing the largest number of fitted modes was adopted as reference and the corresponding fitting pipeline is marked with an asterisk in Table~\ref{tab:dnu_tls}.

\subsection{Sample analysis}
\label{subsec:starbystar}

\begin{table*}
\centering
\caption{Number of fitted modes per pipeline for the optimal light curve of the 29 stars, together with the number of modes retained in the minimal list.}
\label{tab:dnu_tls}
\begin{tabular}{l|cccc|ccc}
\toprule
\multirow{2}{*}{Name} & \multicolumn{4}{c|}{Number of fitted modes} & \multirow{2}{*}{Reference} & \multirow{2}{*}{Note} & $\Delta\nu$ \\
 & \texttt{apollinaire} & FAMED & \texttt{PBjam} & Minimal list &  &  & ($\mu\mathrm{Hz}$) \\
\midrule
HD$\,$36553 & 32 & 25 & 15 & 15 & \texttt{apollinaire} & F-like, mixed & $33.54 \pm 0.06$ \\
$\theta\,$Dra & 34 & 35 & 15 & $\neq$ mode id. & FAMED* & Solar-like, mixed & $39.38 \pm 0.10$ \\
HD$\,$62644 & 37 & 23 & 24 & 22 & \texttt{apollinaire} & Solar-like, mixed & $41.12 \pm 0.04$ \\
68$\,$Dra & 38 & 35 & 10 & 8 & \texttt{apollinaire} & F-like, mixed & $39.66 \pm 0.11$ \\
35$\,$Dra & 39 & 36 & 38 & 31 & \texttt{apollinaire} & F-like, mixed & $41.87 \pm 0.06$ \\
27$\,$Cyg & 27 & 26 & 17 & 13 & \texttt{apollinaire} & Solar-like, mixed & $39.85 \pm 0.05$ \\
HD$\,$175225 & 40 & 35 & 26 & 24 & \texttt{apollinaire} & Solar-like, mixed & $43.51 \pm 0.03$ \\
$\zeta\,$Pic & 45 & 38 & 17 & 16 & \texttt{apollinaire} & Solar-like, mixed & $49.35 \pm 0.13$ \\
HD$\,$46569 & 35 & 27 & 15 & 13 & \texttt{apollinaire} & F-like & $49.92 \pm 0.11$ \\
$\upsilon\,$Cep & 29 & 52 & 17 & 15 & \texttt{apollinaire} & Solar-like, mixed & $53.79 \pm 0.13$ \\
HD$\,$136064 & 30 & 32 & 16 & 16 & \texttt{apollinaire} & F-like & $59.38 \pm 0.09$ \\
$\psi^1\,$Dra$\,$A & 30 & 35 & 28 & 20 & \texttt{apollinaire} & F-like & $61.87 \pm 0.21$ \\
HD$\,$191195 & 47 & -- & -- & -- & \texttt{apollinaire}* & F-like & $71.24 \pm 0.21$ \\
HD$\,$50223 & 47 & 22 & 11 & 10 & \texttt{apollinaire} & F-like & $69.04 \pm 0.15$ \\ 
36$\,$Dra & 39 & 50 & 17 & 17 & \texttt{PBjam} & F-like & $69.99 \pm 0.23$ \\
HR$\,$3220 & 24 & 38 & 17 & $\neq$ mode id. & FAMED* & F-like & $71.17 \pm 0.02$ \\
17$\,$Cyg & 39 & 29 & 13 & 13 & \texttt{apollinaire} & F-like & $78.99 \pm 0.14$ \\
$\theta\,$Cyg & 29 & 35 & 32 & 16 & \texttt{apollinaire} & F-like & $83.80 \pm 0.11$ \\ 
HD$\,$184960 & 31 & 20 & 17 & 13 & \texttt{apollinaire} & Solar-like, low S/N & $91.73 \pm 0.12$ \\
$\omega\,$Dra & 24 & -- & 21 & 7 & \texttt{apollinaire}* & F-like, low S/N & $89.17 \pm 0.41$ \\
99$\,$Her & 26 & 19 & 26 & 17 & \texttt{apollinaire} & Solar-like & $96.04 \pm 0.16$ \\
HD$\,$53705 & 27 & 10 & 16 & 10 & \texttt{apollinaire} & Solar-like & $101.55 \pm 0.13$ \\
16$\,$Cyg$\,$A & 24 & 16 & 16 & 10 & \texttt{apollinaire} & Solar-like & $103.27 \pm 0.21$ \\
72$\,$Her & 47 & 11 & 15 & 9 & \texttt{apollinaire} & Solar-like & $106.58 \pm 0.28$ \\
19$\,$Dra & 24 & 20 & 23 & 15 & \texttt{PBjam} & Solar-like, low S/N & $115.40 \pm 0.10$ \\ 
HD$\,$176051 & 30 & 8 & 19 & 7 & \texttt{apollinaire} & Solar-like, low S/N & $126.55 \pm 0.20$ \\ 
HD$\,$193664 & 32 & 3 & 17 & 3 & \texttt{PBjam} & Solar-like, low S/N & $137.18 \pm 0.11$ \\
26$\,$Dra & 14 & -- & 6 & 6 & \texttt{apollinaire}* & Solar-like, low S/N & $131.26 \pm 0.13$ \\
$\sigma\,$Dra & 24 & 21 & 22 & 16 & \texttt{apollinaire} & Solar-like, low S/N & $182.07 \pm 0.10$ \\ 
\bottomrule
\end{tabular}
\tablefoot{When available, the minimal list comprises modes fitted by all three pipelines and classified as non-outliers according to the Peirce criterion. If only two pipelines successfully analysed a star, it includes the modes common to those two. For two stars, no consistent mode identification was achieved across pipelines, and a minimal list was therefore not defined.
The pipeline used to fit the selected reference dataset, chosen following the procedure described in Sect.~\ref{subsec:pipeline_comaprison}, is also indicated. If only one or two pipelines yielded results, the dataset with the largest number of extracted modes was adopted as reference and marked with an asterisk. We additionally report the large frequency separations, $\Delta\nu$, measured from the reference dataset defined in Sect.~\ref{subsec:pipeline_comaprison}. The stars are ordered by $\nu_\mathrm{max}$ from Table~\ref{tab:back_params}.}
\end{table*}

This section presents the seismic parameters extracted from the optimal light curve of the 29 stars, enabling a detailed characterisation of their oscillation properties and evolutionary states. 
The number of fitted modes per star and per pipeline is listed in Table~\ref{tab:dnu_tls}. The $\Delta\nu$ reported in the same table was derived from the $\ell=0$ mode frequencies of the optimal light curve through a weighted linear fit as a function of radial order. The weights are defined as the inverse square of the frequency uncertainties, with symmetrised errors taken as the maximum of the upper and lower uncertainties. The fit was restricted to modes within $\pm 3\,\Delta\nu$ of $\nu_{\rm max}$ (from Table~\ref{tab:back_params}), and the uncertainty on $\Delta\nu$ was estimated from the covariance matrix.
For two stars ($\theta$ Dra and HR 3220), the three pipelines did not converge on a consistent mode identification. This is partly due to their location in a temperature regime where the asymptotic phase term, $\varepsilon$, is highly sensitive to small changes in effective temperature \citep[e.g.][]{white_2012}. This is discussed further in the next section. For these stars, the échelle diagrams in Appendix~\ref{app:results-per-star} show the detected modes with their respective identifications, and the reference dataset was taken to be the one with the largest number of fitted modes, corresponding in both cases to the results obtained with FAMED.

From a visual inspection of the échelle diagrams for the full sample, 13 stars were classified as F-like (Table~\ref{tab:dnu_tls}). These stars exhibit broad, overlapping ridges caused by the combination of large linewidths and small inter-degree frequency separations, and they occupy the upper region of the HR diagram (Fig.~\ref{fig:HR-diagram}), corresponding to higher effective temperatures and lower large frequency separations. This behaviour reflects the shorter mode lifetimes expected in hotter stars with shallower convective envelopes, and is consistent with the observational classification proposed in the literature \citep[e.g.][]{appourchaux_2012}. This F-like versus solar-like distinction is therefore used here as a descriptive, observational classification rather than a strict physical boundary.
In contrast, nine stars show evidence of mixed modes detected by at least one pipeline and are highlighted in Fig.~\ref{fig:HR-diagram}. These mixed modes appear as deviations from the regular ridge structure in the échelle diagrams and arise from the coupling of p- and g-mode cavities as stars evolve off the MS \citep[e.g.][]{benomar_2013}. Among these stars, three are classified as F-like and five as solar-like, spanning a range of evolutionary stages. Some exhibit a strongly perturbed $\ell=1$ ridge (HD~62644, 27~Cyg, and HD~175225), others retain a recognisable ridge while showing several mixed modes outside the ridge ($\theta$~Dra, 68~Dra, and $\upsilon$~Cep), and the remaining stars likely display only a single $\ell=1$ mode outside the ridge (HD~36553, 35~Dra, and 36~Dra).

The identification of mixed modes differs significantly between pipelines, reflecting both methodological choices described in Appendix~\ref{app:modes_id}, and the intrinsic complexity of mixed-mode spectra. Using \texttt{PBjam}, a dedicated mixed-mode fitting was performed for seven stars, while FAMED also detected out-of-the-ridge modes in seven stars, which do not entirely overlap with the \texttt{PBjam} sample. 
Since \texttt{apollinaire} does not automatically detect mixed modes, unlike the other two pipelines, we applied the procedure described in Sect.~\ref{subsec:mix_modes_form} to model mixed-mode frequencies and compare them with the observations, allowing genuine modes to be distinguished from noise. The key results regarding mixed-mode detection are listed below:
\begin{itemize}
    \item Both \texttt{PBjam} and FAMED identify mixed modes in $\theta$~Dra, HD~62644, 68~Dra, 27~Cyg, and HD~175225. HD~36553 and $\zeta$~Pic are detected only with \texttt{PBjam}, while 35~Dra and $\upsilon$~Cep are identified solely by FAMED.
    \item The procedure used for \texttt{apollinaire} results successfully identified $\ell=1$ mixed-mode patterns in HD~62644 and HD~175225. For the remaining seven stars, the method did not converge, either because an insufficient number of mixed modes was available to constrain the fit or, in the case of 27~Cyg and $\zeta$ Pic, despite the presence of numerous detected modes. In these latter cases, the lack of convergence is primarily attributed to structural glitches and deviations from regular mixed-mode patterns, which break the assumptions of the asymptotic description and prevent a stable, well-constrained fit.
    \item For several targets, discrepancies between pipelines mainly concern the number of detected mixed modes. In $\theta$~Dra and $\upsilon$~Cep, FAMED identified five and 12 $\ell=1$ modes, respectively, located outside the main ridges and not recovered by the other pipelines. By contrast, \texttt{apollinaire} identified additional mixed modes not recovered by FAMED or \texttt{PBjam}, reporting seven such modes in HD~62644 that are all consistent with the mixed-mode identification procedure, and a further seven $\ell=1$ modes in $\zeta$~Pic that do not show the same consistency with this procedure.
    \item In other cases, discrepancies reflect marginal detections rather than fundamentally different interpretations. For 68~Dra, mode detection varies across pipelines, with several modes detected by only one method. Similarly, for 27~Cyg and HD~175225, three (resp. two) mixed modes were identified by \texttt{PBjam}, whereas six (resp. four) additional modes were detected by the other two pipelines but not recovered by \texttt{PBjam}.
    \item For 35~Dra, a single mode located outside the $\ell=2$ ridge was classified as $\ell=1$ by FAMED and as $\ell=2$ by \texttt{apollinaire}. Given its systematic offset from the $\ell=2$ ridge and its similarity to mixed modes observed in comparable stars \citep[e.g.][]{appourchaux_2020}, this mode is most likely of mixed character. Residual power in the échelle diagram further suggests that additional mixed modes may remain undetected.
\end{itemize}
The diversity in mixed-mode morphology reflects differences in core structure and evolutionary state across the sample. Thus, stars exhibiting a larger number of detectable mixed modes tend to be more evolved.

The oscillation power envelope of at least eight stars in our sample (HD~36553, $\theta$~Dra, $\zeta$ Pic, $\upsilon$~Cep, HD~136064, HD~50223, 17~Cyg, and $\theta$~Cyg) departs from the classical Gaussian-like shape, instead exhibiting a double-humped or plateau-like structure, as seen in their respective PSDs in Appendix~\ref{app:results-per-star}. Similar broad, non-Gaussian envelopes have long been observed in the bright F5 SG Procyon, which has become a benchmark case for such atypical solar-like oscillation profiles \citep[e.g.][]{bedding_2005}. Comparable features have also been reported in several stars observed by CoRoT and \textit{Kepler} \citep[e.g.][]{mathur_2010, garcia_2019}, and were specifically noted by \citet{guzik_2016} for $\theta$~Cyg. Stars displaying this type of envelope tend to be hotter and, in some cases, slightly evolved SGs.

Two stars in our sample were also observed by \textit{Kepler}: $\theta$ Cyg \citep{guzik_2016} and 16 Cyg A \citep{metcalfe_2012, davies_2015, lund_2017}.  The \textit{Kepler} and TESS observations were not combined in the present analysis because of their substantially different S/Ns and the temporal gap of several years between the datasets. Such differences may introduce systematic variations in the mode amplitudes and background properties, making an independent analysis more appropriate. We therefore chose instead to compare the independently extracted frequency sets. The literature frequencies are shown alongside our measurements in the échelle diagrams of the stars (Figs.~\ref{fig:27014182_echelle} and \ref{fig:27533341_echelle}). For 16 Cyg A, all three studies report very similar values, so only the most recent set \citep{lund_2017} is displayed. Twenty-four modes are common between our measurements and the literature for this latter star, all agreeing within 2\,$\sigma$. In the case of $\theta$ Cyg, 26 modes were extracted from both \textit{Kepler} and TESS observations. Of these, 12 modes from our analysis are consistent within 2\,$\sigma$ of the \citet{guzik_2016} values, while nine modes differ by more than 3\,$\sigma$. These discrepancies are primarily associated with modes at the edges of the p-mode envelope, where the oscillation amplitudes are lower, and the S/N is reduced. In addition, as an F-type star, $\theta$ Cyg exhibits intrinsically broad oscillation modes, which overlap in the power spectrum and are more challenging to fit accurately. The combination of low amplitude and broad mode width likely accounts for the observed deviations from the \citet{guzik_2016} measurements. Similarly, $\sigma$ Dra was also analysed by \citet{hon_2024} using both TESS photometry (sectors 41–60) and radial velocities from the Keck Planet Finder (KPF). Their extracted frequencies are shown alongside ours in Fig.~\ref{fig:259237827_echelle}.
Comparing the two TESS-based frequency sets, we identify two additional low-frequency modes, while they reported one additional mode at a higher frequency. We also detect quadrupolar modes that were not reported in their analysis. Aside from these differences, all commonly extracted frequencies agree within less than 1~$\sigma$.

\begin{figure}
    \centering
    \includegraphics[width=1\linewidth]{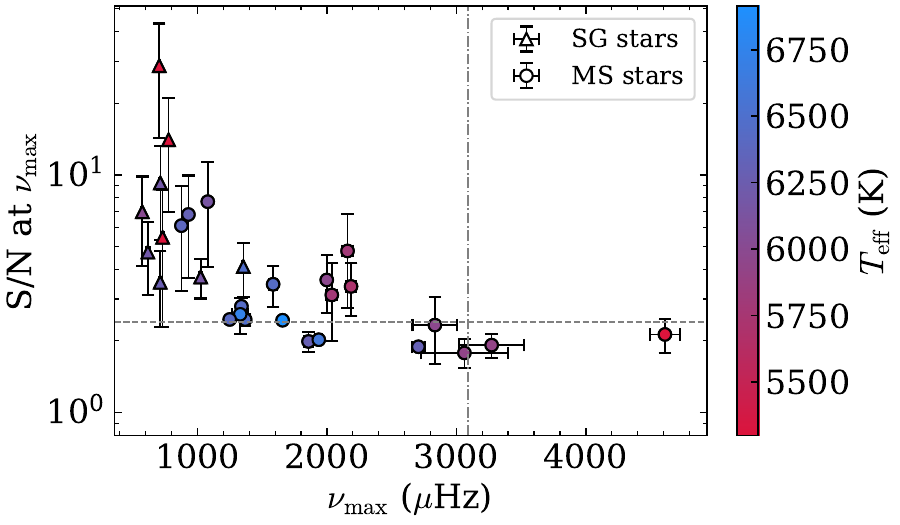}
    \caption{S/N of the radial mode ($\ell=0$) closest to $\nu_\mathrm{max}$, plotted as a function of $\nu_\mathrm{max}$ for each star. Markers are coloured according to the effective temperature of the stars listed in Table~\ref{tab:fund_params}, and indicated by the colour bar. SG stars are shown as triangles, while MS stars are circles. The low S/N threshold of 2.4 is indicated by a dashed horizontal line, and the solar value $\nu_{\mathrm{max},\odot}=3090\,\mu$Hz is marked by a dash-dotted vertical line.}
    \label{fig:SNR_at_numax}
\end{figure}

The S/N of the radial mode closest to $\nu_\mathrm{max}$ was computed for each star using the \texttt{apollinaire} results and is shown as a function of $\nu_\mathrm{max}$ in Fig.~\ref{fig:SNR_at_numax}. Uncertainties were estimated by propagating the errors in both the mode height (mean of lower and upper bounds) and the background parameters. As expected, stars with lower $\nu_\mathrm{max}$ tend to exhibit a higher S/N, even though their apparent magnitudes are relatively similar ($V \in [3.55,5.99]$, Paper~I), confirming that the S/N trend is primarily driven by $\nu_\mathrm{max}$ rather than brightness differences. Seven stars have S/N values below 2.4 (Table~\ref{tab:dnu_tls}), corresponding to cases where the mode ridges become increasingly difficult to discern in the échelle diagrams. For these stars, we added an additional panel in the échelle diagrams (Appendix~\ref{app:results-per-star}) showing the collapsed échelle \citep[e.g.][]{bedding_2005}, which highlights the presence of mode ridges despite the low S/N. 
Stars in the solar $\nu_\mathrm{max}$ regime systematically exhibit low S/N in TESS observations. This limited S/N restricts our ability to constrain mode widths and therefore damping properties, which are key diagnostics of near-surface convection. These targets are therefore prime candidates for PLATO, whose higher photometric precision and longer time series will enable a more robust seismic characterisation. This aligns with PLATO’s core objective of precisely determining stellar masses, radii, and ages for solar-like stars, thereby improving stellar model calibration and the characterisation of their planetary systems.
For some of these stars (i.e. 19 Dra, HD 176051, 26 Dra, and $\sigma$ Dra), the mode widths are not well constrained, as the base of the mode peaks is buried within the noise. Nevertheless, the extracted frequencies remain consistent with the peaks seen in the summed échelle power. Notably, despite its low S/N and high $\nu_\mathrm{max}$, close to the Nyquist frequency of the instrument at this cadence, mode frequencies were successfully extracted for $\sigma$~Dra. This star therefore represents one of the coolest MS stars observed with TESS for which individual mode parameters have been robustly extracted and modelled \citep{hon_2024}.

\section{Discussion and comparisons with other studies}
\label{sec:comp_legacy}

To place these results in a broader context, we further compare the seismic properties of our sample with stars drawn from several literature catalogues: (1) the \textit{Kepler} LEGACY catalogue \citep{lund_2017}, which includes 66 MS stars observed by \textit{Kepler} with exceptionally high-S/N data, providing some of the most precise seismic constraints currently available for stellar modelling \citep{silvaaguirre_2017}; (2) the sample studied by \citet{white_2012} which provided $\varepsilon$ and $\Delta\nu$ for 46 \textit{Kepler} MS oscillators; (3) 36 \textit{Kepler} SG stars from \citet{li_2020}; and (4) the 147 MS and SG stars from the KEYSTONE sample \citep{lund_2024a,Hookway_2025}, observed during the K2 campaigns. These catalogues were selected for their relevance to our sample, as they include both MS and SG stars and span a similar range of effective temperatures. They combine observations from different space missions, primarily \textit{Kepler} and K2, complementing our TESS-based sample, and cover a range of seismic characterisation levels, from detailed individual mode parameters to global measurements. Taken together, they represent the vast majority of currently available MS and SG stars with data of sufficient quality to reliably determine the seismic parameters discussed in this section.
In what follows, we restrict the analysis to the reference dataset selected for each star (as defined in Sect.~\ref{subsec:pipeline_comaprison}).

\subsection{Mode amplitudes and linewidths}
\begin{figure}
    \centering
    \includegraphics[width=1\linewidth]{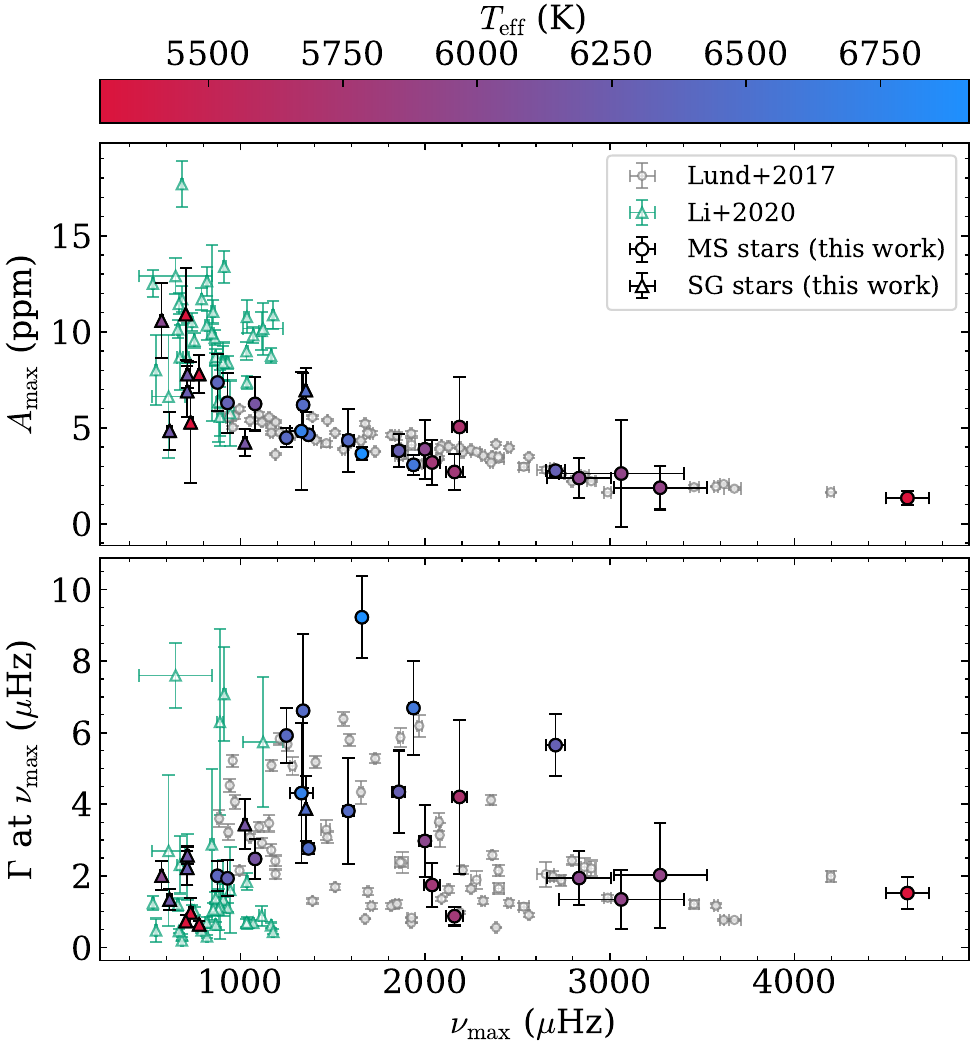}
    \caption{Amplitude (top) and linewidth (bottom) of the radial mode closest to $\nu_\mathrm{max}$, plotted as a function of $\nu_\mathrm{max}$ for each star. Markers follow the same legend as in Fig.~\ref{fig:SNR_at_numax}.
    For comparison, MS stars from the \textit{Kepler} LEGACY sample \citep{lund_2017} and SG stars from \citet{li_2020} are shown as grey circles and green triangles, respectively, in the background of the two panels.}
    \label{fig:l0_at_numax}
\end{figure}

The amplitude and linewidth of the radial mode closest to $\nu_\mathrm{max}$ are shown as a function of $\nu_\mathrm{max}$ (values from Table~\ref{tab:back_params}) in Fig.~\ref{fig:l0_at_numax} for our sample. For comparison, we include MS stars from the \textit{Kepler} LEGACY sample and SG stars from \citet{li_2020}. Consistent with these previous studies, mode amplitudes decrease with increasing $\nu_\mathrm{max}$, and for MS stars also decrease with decreasing effective temperature. Stars exhibiting mixed modes occupy the upper left region of the amplitude panel and are associated with relatively small linewidths in the bottom panel, consistent with their more evolved nature and longer mode lifetimes.
The linewidths display an overall bell-shaped dependence on $\nu_\mathrm{max}$: they are small at low $\nu_\mathrm{max}$, increase and become more dispersed at intermediate $\nu_\mathrm{max}$, and decrease again at higher $\nu_\mathrm{max}$. This behaviour reflects the known dependence of mode damping on stellar structure and effective temperature \citep{belkacem_2012,houdek_2019}, and is recovered here despite differences in data quality and analysis conditions with respect to the comparative \textit{Kepler} studies.
As discussed in the previous section, most stars with low S/N have poorly constrained linewidths, which likely explains the unusually large widths measured for the two stars near 1650 and 2700\,$\mu$Hz ($\theta$~Cyg and 19~Dra). Excluding these cases, all stars in our sample are consistent with the distribution defined by the literature samples in terms of both $A_\mathrm{max}$ and $\Gamma$ at $\nu_\mathrm{max}$, with differences primarily reflecting the larger and more heterogeneous uncertainty budget in our dataset. These findings extend the empirical relations to a complementary sample of bright stars observed under different conditions, and improving their statistical coverage in regions of parameter space that were previously sparsely populated.

\subsection{Average seismic parameters}

\subsubsection{Fitting procedure}
\label{subsec:fitting_procedure}

For each star, we computed the average small frequency separations, $\delta\nu$, and $\varepsilon$, following the procedure described in \citet{lund_2017}. This was done by fitting the observed mode frequencies with
\begin{multline}
    \nu_{n,\ell} \simeq \left(n+\frac{\ell}{2}+\varepsilon\right)\Delta\nu_0-\delta\nu_{0,\ell} \\
    -\frac{\rm{d}\delta\nu_{0,\ell}}{\rm{d}n}\,(n-n_{\nu_{\rm{max},\ell}}) + \frac{\rm{d}\Delta\nu/\rm{d}n}{2}\,(n-n_{\nu_{\rm{max},\ell}})^2\,,
    \label{eq:fit_separation}
\end{multline}
where $\Delta\nu_0$ is the large frequency separation evaluated at $\nu_\mathrm{max}$, and $n_{\nu_{\rm{max}}}$ is the (generally non-integer) radial order corresponding to $\nu_\mathrm{max}$. The small separations, $\delta\nu_{0,1}$ and $\delta\nu_{0,2}$, were optimised independently.
In practice, we implemented this functional form through a parametric model in which $\Delta\nu_0$, $\varepsilon$, $\mathrm{d}\Delta\nu/\mathrm{d}n$, and $n_{\nu_{\rm{max}}}$ were treated as global parameters, while $\delta\nu_{0,\ell}$ and $\mathrm{d}\delta\nu_{0,\ell}/\mathrm{d}n$ were fitted separately for $\ell=1$ and $\ell=2$. Radial modes therefore constrain only the global parameters and the curvature term, while the dipole and quadrupole modes additionally constrain their respective small separations and gradients.
The fit was performed by minimising the residuals between observed and modelled frequencies, weighted by the individual frequency uncertainties. We first obtained a maximum-likelihood solution using a least-squares minimisation \citep[with \texttt{lmfit},][]{newville_2025}, and subsequently explored the joint posterior distribution of the parameters using an MCMC approach. A Gaussian prior on $\nu_\mathrm{max}$ was included by requiring the modelled radial-mode frequency at $n_{\nu_{\rm{max}}}$ to reproduce the observed $\nu_\mathrm{max}$ (from Table~\ref{tab:back_params}) within its uncertainty. Final parameter estimates and uncertainties were derived from the marginalised posterior distributions and are listed in Table~\ref{tab:freq_sep}.

For the three stars showing numerous mixed modes (HD 62644, 27 Cyg, and HD 175225), for which a simple asymptotic description is inadequate, only $\delta\nu_{0,2}$ was estimated together with the global parameters. We emphasise that in other stars where mixed modes were detected, $\delta\nu_{0,1}$ and its variation in order $n$ were still measured because a clear ridge associated with the underlying p-mode pattern was identified in their échelle diagram. In those cases, the presence of mixed modes did not prevent the extraction of reliable frequency separations.

\setlength{\tabcolsep}{3pt}
\begin{table*}
\centering
\caption{Values from the fit of Eq.~\eqref{eq:fit_separation} to the mode frequencies.}
\label{tab:freq_sep}
\begin{tabular}{lccccccc}
\toprule
\multirow{2}{*}{Name} & $\Delta\nu_0$ & \multirow{2}{*}{$\varepsilon$} & d\,$\Delta\nu$/d$n$ & $\delta\nu_{0,1}$ & d\,$\delta\nu_{0,1}$/d$n$ & $\delta\nu_{0,2}$ & d\,$\delta\nu_{0,2}$/d$n$ \\
 & ($\mu$Hz) & & ($\mu$Hz) & ($\mu$Hz) & ($\mu$Hz) & ($\mu$Hz) & ($\mu$Hz) \\
\midrule
HD$\,$36553 & $33.74$\footnotesize{ $\pm 0.06$ } & $1.15$\footnotesize{ $\pm 0.03$ } & $0.230$\footnotesize{ $\pm 0.041$ } & $2.471$\footnotesize{ $\pm 0.105$ } & $-0.184$\footnotesize{ $\pm 0.042$ } & $2.481$\footnotesize{ $\pm 0.237$ } & $-0.237$\footnotesize{ $\pm 0.163$ } \\
$\theta\,$Dra & $40.26$\footnotesize{ $\pm 0.06$ } & $1.43$\footnotesize{ $\pm 0.03$ } & $0.028$\footnotesize{ $\pm 0.014$ } & $-2.033$\footnotesize{ $\pm 0.032$ } & $0.427$\footnotesize{ $\pm 0.009$ } & $2.334$\footnotesize{ $\pm 0.059$ } & $0.082$\footnotesize{ $\pm 0.019$ } \\
HD$\,$62644 & $41.11$\footnotesize{ $\pm 0.03$ } & $1.41$\footnotesize{ $\pm 0.01$ } & $0.102$\footnotesize{ $\pm 0.038$ } & -- & -- & $3.841$\footnotesize{ $\pm 0.069$ } & $0.136$\footnotesize{ $\pm 0.048$ } \\
68$\,$Dra & $40.17$\footnotesize{ $\pm 0.05$ } & $0.93$\footnotesize{ $\pm 0.02$ } & $0.153$\footnotesize{ $\pm 0.025$ } & $2.273$\footnotesize{ $\pm 0.151$ } & $0.208$\footnotesize{ $\pm 0.055$ } & $4.237$\footnotesize{ $\pm 0.260$ } & $-0.249$\footnotesize{ $\pm 0.083$ } \\
35$\,$Dra & $42.14$\footnotesize{ $\pm 0.05$ } & $1.15$\footnotesize{ $\pm 0.02$ } & $0.156$\footnotesize{ $\pm 0.027$ } & $3.435$\footnotesize{ $\pm 0.061$ } & $0.143$\footnotesize{ $\pm 0.016$ } & $3.244$\footnotesize{ $\pm 0.167$ } & $-0.273$\footnotesize{ $\pm 0.054$ } \\
27$\,$Cyg & $39.91$\footnotesize{ $\pm 0.08$ } & $1.32$\footnotesize{ $\pm 0.04$ } & $0.194$\footnotesize{ $\pm 0.047$ } & -- & -- & $3.414$\footnotesize{ $\pm 0.184$ } & $-0.146$\footnotesize{ $\pm 0.081$ } \\
HD$\,$175225 & $43.60$\footnotesize{ $\pm 0.04$ } & $1.33$\footnotesize{ $\pm 0.01$ } & $0.242$\footnotesize{ $\pm 0.016$ } & -- & -- & $3.685$\footnotesize{ $\pm 0.046$ } & $-0.198$\footnotesize{ $\pm 0.017$ } \\
$\zeta\,$Pic & $49.63$\footnotesize{ $\pm 0.07$ } & $1.04$\footnotesize{ $\pm 0.02$ } & $-0.035$\footnotesize{ $\pm 0.029$ } & $3.699$\footnotesize{ $\pm 0.119$ } & $0.444$\footnotesize{ $\pm 0.048$ } & $5.100$\footnotesize{ $\pm 0.227$ } & $0.535$\footnotesize{ $\pm 0.096$ } \\
HD$\,$46569 & $50.05$\footnotesize{ $\pm 0.10$ } & $1.00$\footnotesize{ $\pm 0.04$ } & $0.304$\footnotesize{ $\pm 0.037$ } & $3.051$\footnotesize{ $\pm 0.137$ } & $-0.196$\footnotesize{ $\pm 0.048$ } & $2.354$\footnotesize{ $\pm 0.339$ } & $-0.297$\footnotesize{ $\pm 0.112$ } \\
$\upsilon\,$Cep & $53.86$\footnotesize{ $\pm 0.16$ } & $0.99$\footnotesize{ $\pm 0.06$ } & $0.419$\footnotesize{ $\pm 0.049$ } & $3.046$\footnotesize{ $\pm 0.130$ } & $-0.123$\footnotesize{ $\pm 0.040$ } & $4.448$\footnotesize{ $\pm 0.826$ } & $-0.264$\footnotesize{ $\pm 0.156$ } \\
HD$\,$136064 & $59.61$\footnotesize{ $\pm 0.12$ } & $1.11$\footnotesize{ $\pm 0.04$ } & $0.442$\footnotesize{ $\pm 0.059$ } & $3.459$\footnotesize{ $\pm 0.134$ } & $0.160$\footnotesize{ $\pm 0.059$ } & $4.447$\footnotesize{ $\pm 0.199$ } & $-0.285$\footnotesize{ $\pm 0.118$ } \\
$\psi^1\,$Dra$\,$A & $61.81$\footnotesize{ $\pm 0.10$ } & $0.90$\footnotesize{ $\pm 0.03$ } & $0.117$\footnotesize{ $\pm 0.028$ } & $1.048$\footnotesize{ $\pm 0.264$ } & $-0.058$\footnotesize{ $\pm 0.063$ } & -- & -- \\
HD$\,$191195 & $70.77$\footnotesize{ $\pm 0.34$ } & $1.00$\footnotesize{ $\pm 0.09$ } & $0.372$\footnotesize{ $\pm 0.030$ } & $1.032$\footnotesize{ $\pm 0.259$ } & $-0.363$\footnotesize{ $\pm 0.060$ } & $5.048$\footnotesize{ $\pm 0.624$ } & $-0.378$\footnotesize{ $\pm 0.157$ } \\
$\xi\,$Pup & $69.14$\footnotesize{ $\pm 0.11$ } & $0.91$\footnotesize{ $\pm 0.03$ } & $0.341$\footnotesize{ $\pm 0.023$ } & $1.964$\footnotesize{ $\pm 0.181$ } & $-0.178$\footnotesize{ $\pm 0.035$ } & $3.459$\footnotesize{ $\pm 0.372$ } & $-0.389$\footnotesize{ $\pm 0.063$ } \\
36$\,$Dra & $70.21$\footnotesize{ $\pm 0.18$ } & $0.94$\footnotesize{ $\pm 0.05$ } & $0.063$\footnotesize{ $\pm 0.119$ } & $4.148$\footnotesize{ $\pm 0.177$ } & $0.398$\footnotesize{ $\pm 0.084$ } & $5.953$\footnotesize{ $\pm 0.545$ } & $-0.416$\footnotesize{ $\pm 0.355$ } \\
HR$\,$3220 & $71.49$\footnotesize{ $\pm 0.08$ } & $1.10$\footnotesize{ $\pm 0.02$ } & $0.312$\footnotesize{ $\pm 0.002$ } & $5.573$\footnotesize{ $\pm 0.013$ } & $-0.126$\footnotesize{ $\pm 0.003$ } & $4.909$\footnotesize{ $\pm 0.038$ } & $-0.350$\footnotesize{ $\pm 0.009$ } \\
17$\,$Cyg & $78.59$\footnotesize{ $\pm 0.09$ } & $1.30$\footnotesize{ $\pm 0.03$ } & $0.143$\footnotesize{ $\pm 0.035$ } & $3.206$\footnotesize{ $\pm 0.241$ } & $-0.582$\footnotesize{ $\pm 0.080$ } & $5.708$\footnotesize{ $\pm 0.258$ } & $0.027$\footnotesize{ $\pm 0.095$ } \\
$\theta\,$Cyg & $83.98$\footnotesize{ $\pm 0.15$ } & $0.77$\footnotesize{ $\pm 0.04$ } & $0.556$\footnotesize{ $\pm 0.051$ } & $0.887$\footnotesize{ $\pm 0.158$ } & $-0.255$\footnotesize{ $\pm 0.025$ } & $5.931$\footnotesize{ $\pm 0.536$ } & $-0.640$\footnotesize{ $\pm 0.071$ } \\
HD$\,$184960 & $91.83$\footnotesize{ $\pm 0.21$ } & $1.22$\footnotesize{ $\pm 0.05$ } & $0.426$\footnotesize{ $\pm 0.105$ } & $3.190$\footnotesize{ $\pm 0.251$ } & $-0.340$\footnotesize{ $\pm 0.097$ } & $9.774$\footnotesize{ $\pm 0.751$ } & $-0.415$\footnotesize{ $\pm 0.164$ } \\
$\omega\,$Dra & $88.62$\footnotesize{ $\pm 0.12$ } & $0.94$\footnotesize{ $\pm 0.03$ } & $-0.115$\footnotesize{ $\pm 0.108$ } & $2.828$\footnotesize{ $\pm 0.261$ } & $-0.359$\footnotesize{ $\pm 0.085$ } & $8.216$\footnotesize{ $\pm 0.990$ } & $0.370$\footnotesize{ $\pm 0.369$ } \\
99$\,$Her & $96.12$\footnotesize{ $\pm 0.16$ } & $1.33$\footnotesize{ $\pm 0.03$ } & $0.265$\footnotesize{ $\pm 0.066$ } & $4.531$\footnotesize{ $\pm 0.207$ } & $-0.549$\footnotesize{ $\pm 0.078$ } & $4.581$\footnotesize{ $\pm 0.685$ } & $-0.753$\footnotesize{ $\pm 0.257$ } \\
HD$\,$53705 & $101.66$\footnotesize{ $\pm 0.14$ } & $1.47$\footnotesize{ $\pm 0.03$ } & $0.254$\footnotesize{ $\pm 0.068$ } & $4.357$\footnotesize{ $\pm 0.095$ } & $-0.728$\footnotesize{ $\pm 0.066$ } & $4.755$\footnotesize{ $\pm 0.258$ } & $-0.762$\footnotesize{ $\pm 0.156$ } \\
16$\,$Cyg$\,$A & $103.24$\footnotesize{ $\pm 0.26$ } & $1.45$\footnotesize{ $\pm 0.05$ } & $0.367$\footnotesize{ $\pm 0.139$ } & $3.886$\footnotesize{ $\pm 0.315$ } & $-0.202$\footnotesize{ $\pm 0.158$ } & $6.697$\footnotesize{ $\pm 0.746$ } & $-0.406$\footnotesize{ $\pm 0.295$ } \\
72$\,$Her & $106.26$\footnotesize{ $\pm 0.14$ } & $1.48$\footnotesize{ $\pm 0.03$ } & $0.264$\footnotesize{ $\pm 0.026$ } & $4.184$\footnotesize{ $\pm 0.216$ } & $-0.247$\footnotesize{ $\pm 0.046$ } & $5.431$\footnotesize{ $\pm 0.595$ } & $-0.434$\footnotesize{ $\pm 0.093$ } \\
19$\,$Dra & $115.29$\footnotesize{ $\pm 0.17$ } & $1.15$\footnotesize{ $\pm 0.03$ } & $0.303$\footnotesize{ $\pm 0.079$ } & $4.544$\footnotesize{ $\pm 0.215$ } & $-0.191$\footnotesize{ $\pm 0.073$ } & $12.540$\footnotesize{ $\pm 0.715$ } & $-0.157$\footnotesize{ $\pm 0.540$ } \\
HD$\,$176051 & $126.93$\footnotesize{ $\pm 0.48$ } & $1.37$\footnotesize{ $\pm 0.08$ } & $0.367$\footnotesize{ $\pm 0.090$ } & $4.111$\footnotesize{ $\pm 0.166$ } & $-0.029$\footnotesize{ $\pm 0.036$ } & $8.812$\footnotesize{ $\pm 0.455$ } & $-0.147$\footnotesize{ $\pm 0.300$ } \\
HD$\,$193664 & $137.20$\footnotesize{ $\pm 0.48$ } & $1.51$\footnotesize{ $\pm 0.08$ } & $-0.069$\footnotesize{ $\pm 0.273$ } & $2.807$\footnotesize{ $\pm 0.151$ } & $-0.520$\footnotesize{ $\pm 0.094$ } & $8.755$\footnotesize{ $\pm 0.689$ } & $-1.019$\footnotesize{ $\pm 0.522$ } \\
26$\,$Dra & $131.22$\footnotesize{ $\pm 1.33$ } & $1.53$\footnotesize{ $\pm 0.25$ } & $-0.912$\footnotesize{ $\pm 0.123$ } & $4.124$\footnotesize{ $\pm 0.512$ } & $0.226$\footnotesize{ $\pm 0.211$ } & -- & -- \\
$\sigma\,$Dra & $182.49$\footnotesize{ $\pm 0.19$ } & $1.50$\footnotesize{ $\pm 0.03$ } & $0.226$\footnotesize{ $\pm 0.032$ } & $4.008$\footnotesize{ $\pm 0.118$ } & $-0.283$\footnotesize{ $\pm 0.037$ } & $11.029$\footnotesize{ $\pm 1.443$ } & $0.105$\footnotesize{ $\pm 0.429$ } \\
\bottomrule
\end{tabular}
\end{table*}

\subsubsection{Asymptotic phase term $\varepsilon$}

\begin{figure}
    \centering
    \includegraphics[width=1\linewidth,trim=0 0 0 0,clip]{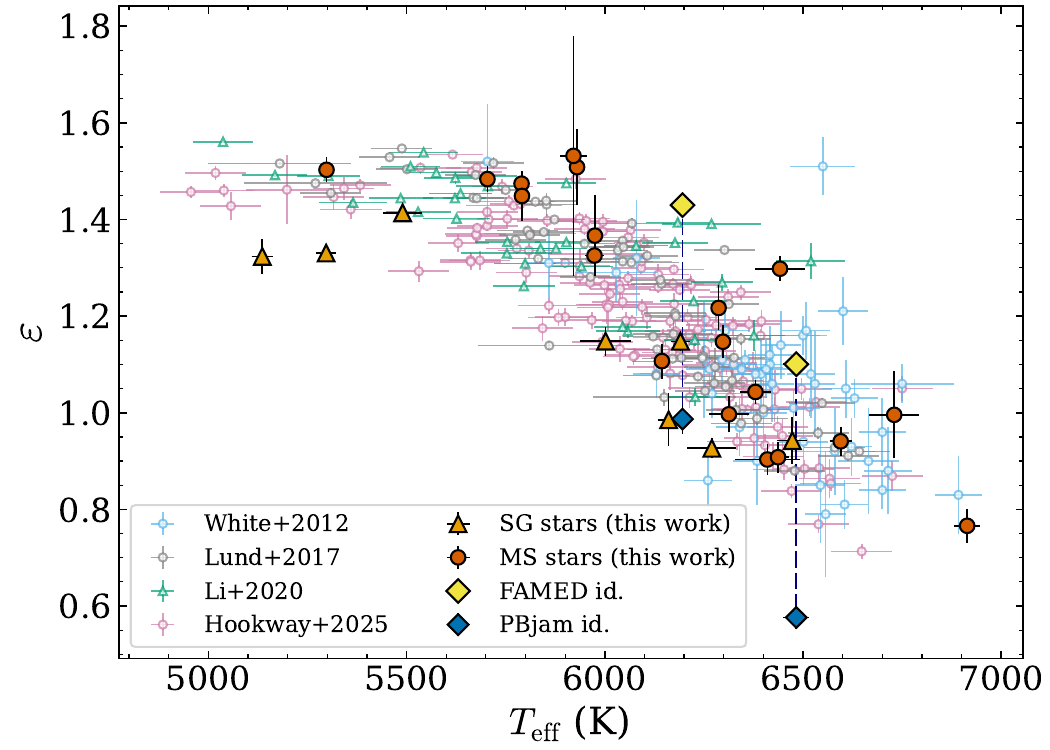}
    \caption{$\varepsilon$ as a function of effective temperature. SG stars are shown as orange triangles, and MS stars as red circles. The two stars for which the fitters did not reach agreement on the mode identification are indicated by diamonds, with FAMED identification in yellow, and \texttt{PBjam}'s one in blue, both symbols belonging to the same star being linked with a dashed blue line. Stars from the four comparison catalogues described in the text are plotted as open symbols: light blue circles \citep{white_2012}, grey circles \citep{lund_2017}, green circles \citep{li_2020}, and pink circles \citep{Hookway_2025}. Error bars are shown where visible and are otherwise smaller than the marker size.}
    \label{fig:eps_teff}
\end{figure}
The resulting $\varepsilon-T_\mathrm{eff}$ relation is shown in Fig.~\ref{fig:eps_teff}, together with measurements from the four comparison catalogues described above. For consistency, the average seismic parameters of the KEYSTONE sample were derived by same fitting procedure described in Sect.~\ref{subsec:fitting_procedure}. For the remaining catalogues, $\varepsilon$ values were taken directly from the corresponding studies. 
For the two stars shown as yellow and blue diamonds ($\theta$ Dra and HR~3220), for which the fitting pipelines did not agree on the mode identification, two estimates of $\varepsilon$ were obtained. The value listed in Table~\ref{tab:freq_sep} corresponds to the pipeline that extracted the largest number of modes from the optimal light curve (FAMED in both cases), while a second estimate was derived using \texttt{PBjam}, which adopted an alternative mode identification.
Both estimates fall within the range defined by the literature samples, consistent with the rest of our targets. However, these stars lie near the edges of the expected $\varepsilon$–$T_\mathrm{eff}$ relation, where mode identification becomes sensitive to small changes in effective temperature and global seismic parameters. This is driven by the strong correlation between $\Delta\nu$ and $\varepsilon$, such that even modest variations in $\Delta\nu$ can propagate into significant changes in $\varepsilon$ and in the resulting ridge assignment in the échelle diagram. This sensitivity is further increased by broad linewidths and, in some cases, mixed modes, which reduce the regularity of the oscillation pattern and can lead to differences in mode identification between pipelines. Only detailed seismic modelling (to be presented in an upcoming paper) can provide a definitive solution in such cases.
Moreover, our sample adds stars in the $T_\mathrm{eff} < 5500$\,K regime, where fewer objects are available, thereby improving the sampling of this region with new TESS observations and supporting the robustness of the derived seismic parameters in this domain.

\subsubsection{Frequency separations}
\begin{figure*}[!ht]
    \centering
    \includegraphics[width=1\linewidth]{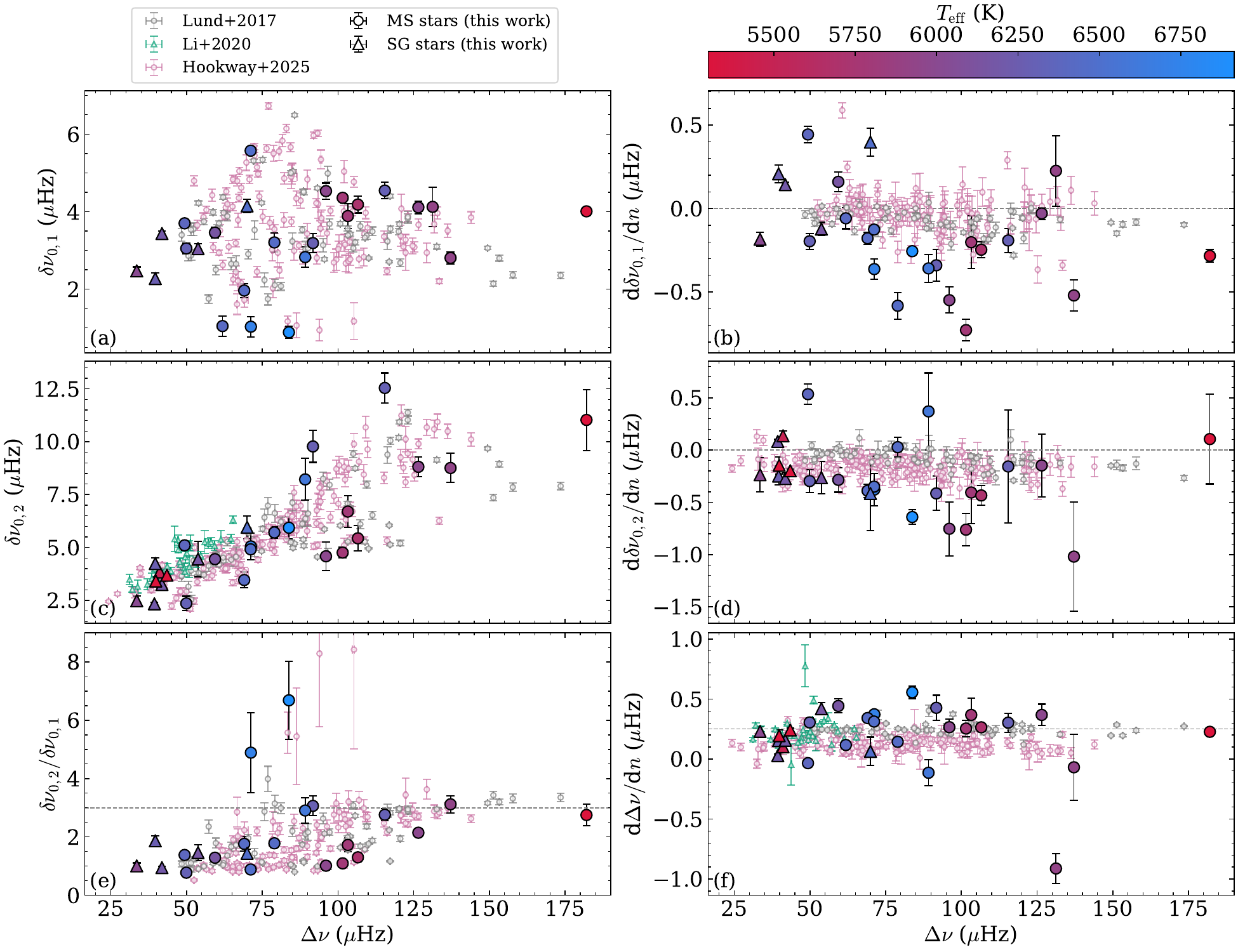}
    \caption{Small frequency separations, $\delta\nu_{0,1}$ (panel a) and $\delta\nu_{0,2}$ (panel c), along with their variation with radial order $n$ (panels b and d), and their ratio, $\delta\nu_{0,1}/\delta\nu_{0,2}$ (panel e), are plotted as a function of $\Delta\nu$ (from Table~\ref{tab:dnu_tls}) for each star. Panel (f) shows the variation in $\Delta\nu$ with $n$ as a function of $\Delta\nu$, with the dashed line indicating the nearly constant value of 0.25\,$\mu$Hz around which most stars appear to lie, as discussed in \citet{lund_2017}. Markers are coloured according to the effective temperature of the stars listed in Table~\ref{tab:fund_params}, and indicated by the colour bar. SG stars are shown as triangles, while MS stars are circles. For comparison, stars from the literature are shown in the background with open markers: grey circles for the LEGACY sample \citep{lund_2017}, green triangles for the SGs of \citet{li_2020}, and pink circles for the KEYSTONE sample \citep{Hookway_2025}. The dashed line in panel (e) indicates the expected value of 3 from the asymptotic relation in Eq.~\eqref{eq:asymp_p_gen}, while those in panels (b) and (d) show the zero value.}
    \label{fig:frequency_separations}
\end{figure*}
The resulting small frequency separations, $\delta\nu_{0,1}$ and $\delta\nu_{0,2}$, and their ratio are shown in Fig.~\ref{fig:frequency_separations}. The values for $\delta\nu_{0,2}$ of the \citet{li_2020} sample are reported from that publication, while those of d$\Delta\nu/$d$n$ were computed from a fit of Eq.~\eqref{eq:fit_separation} ignoring $\ell=1$ and $\ell=2$ modes.
The variations in $\delta\nu_{0,1}$ and $\delta\nu_{0,2}$ with radial order $n$ (panels b and d) are larger than those reported in the LEGACY and KEYSTONE catalogues. However, as these variations are mostly zero or negative, they remain compatible with decreasing functions of frequency, in line with what was noted by \citet{lund_2017}. 
We note a systematic negative offset in $d\delta\nu_{0,2}/dn$ between our sample and the literature stars. This may be due to the shorter observation baseline and lower S/N of TESS compared with \textit{Kepler} or K2, which disproportionately affects the lower-amplitude $\ell=2$ modes and increases the uncertainty in the derived slope.
As noted by \citet{lund_2017}, most stars deviate from the asymptotic regime, and this is also visible in the KEYSTONE sample. Only five stars in our sample (HD 50223, $\omega$ Dra, 19 Dra, HD 193664, and $\sigma$ Dra) are compatible with the asymptotic ratio $\delta\nu_{0,2}/\delta\nu_{0,1}=3$. Therefore, two stars from our sample (HD 191195 and $\theta$ Cyg), as well as four stars from the KEYSTONE sample, depart strongly from the asymptotic regime and are inconsistent with the distribution defined by the LEGACY sample, suggesting structural differences not captured by the asymptotic approximation. 
A revised interpretation of the small frequency separations, extending from the MS to more evolved stages, has recently been proposed, in which $\delta\nu_{0,2}$ is found to scale primarily with global seismic parameters rather than directly tracing the internal sound-speed gradient \citep{ong_2025}.
Finally, $\omega$ Dra and 26 Dra lie below zero in panel (f), which shows $\rm{d}\Delta\nu/\rm{d}n$ as a function of $\Delta\nu$, while all other stars cluster around $\sim$0.25, consistent with \citet{lund_2017} and \citet{Hookway_2025}. These two stars, therefore, appear to exhibit a negative gradient of the large frequency separation with frequency, potentially indicating structural curvature effects. However, this result should be treated with caution, as both stars have relatively low S/N (Table~\ref{tab:dnu_tls}), which may bias the inferred structure.


\section{Conclusion}
\label{sec:conclusion}

In this work, we have performed a detailed seismic characterisation of 29 bright solar-like stars out of a sample of 32 targets from the TESS Luminaries sample that are located in the PLATO LOP fields. Building on the global seismic analysis presented in Paper~I, we now focus on the extraction of individual oscillation mode parameters and provide, for the first time, constraints on individual mode frequencies for 26 of these stars. This allows for more precise determinations of their fundamental properties. A detailed modelling of their masses, radii, and ages will be presented in a forthcoming study (Bétrisey et al., in prep.).
The analysis was performed by combining multiple observing cadences, calibrations, and independent mode-fitting pipelines to ensure robust and consistent results. For the remaining three stars, only the background parameters are reported.
From the extracted mode frequencies, we quantified key seismic diagnostics, including the large and small frequency separations ($\Delta\nu$, $\delta\nu_{0,1}$, $\delta\nu_{0,2}$), the asymptotic phase term $\varepsilon$, as well as mode amplitudes and linewidths. Mixed modes were carefully identified in SG stars, providing insight into their internal structure. Detailed forward modelling of the frequencies will be addressed in a future study.
Our analysis provides several new insights:
\begin{enumerate}
    \item The échelle diagrams and modelled PSD plots are provided for all 29 stars, along with the corresponding lists of extracted mode frequencies.
    \item The mode frequencies extracted for $\sigma$ Dra confirm the results of \citet{hon_2024}, which were obtained using a smaller number of TESS sectors. This star represents one of the coolest MS solar-like stars for which individual oscillation modes have been characterised with TESS, thereby extending the parameter space accessible to asteroseismic studies. Only three other stars in a similar effective temperature range are reported in the \textit{Kepler} LEGACY catalogue \citep{lund_2017}. Studying such low-$T_\mathrm{eff}$ stars is particularly important for exoplanet science: their smaller radii and masses enhance transit and radial velocity signals, enabling more precise detection and characterisation of Earth-sized planets. Additionally, their close-in habitable zones allow more frequent transits, facilitating follow-up observations. Characterising cooler stars seismically provides a critical foundation for understanding the occurrence and properties of potentially habitable planets.
    \item A careful and comparative identification of mixed modes across multiple pipelines is provided, together with a quantification of their agreement and discrepancies. Our work demonstrates the sensitivity of mode identification to pipeline methodology and highlights the need for robust approaches when analysing stars near the MS turn-off. This issue is compounded when modelling the detected mixed modes, as the radial sensitivity of a mode can vary rapidly compared to the frequency differences between stellar models, making detailed seismic modelling particularly challenging \citep[e.g.][]{fellay_2021}.
    \item At least eight stars with double-humped or plateau-like oscillation envelopes were identified, including slightly evolved SG and hotter MS stars. This feature, previously reported only in a few \textit{Kepler} and CoRoT targets, is now observed in TESS data, opening new avenues to study the physical processes shaping envelope morphology.
    \item Comparison with \textit{Kepler} observations for 16 Cyg A and $\theta$ Cyg demonstrates that TESS can reliably recover mode frequencies, even for stars with low S/N or broad F-type modes, while identifying limitations in mode width determination. This emphasises the potential of TESS for expanding the sample of stars with detailed seismic characterisation and identifies prime candidates for future PLATO observations. 
\end{enumerate}
Our results confirm that TESS can provide reliable individual mode frequencies for a broad range of solar-like stars, including F-type MS stars and stars with low S/N, enabling studies of stellar interiors across different evolutionary stages. The observed diversity in mode amplitudes, linewidths, and small separations underscores the importance of combining multiple pipelines and careful mode identification to capture both surface and core properties. Future missions such as PLATO, with higher photometric precision and longer time series, will further improve constraints on mode damping, mixed modes, and stellar interiors, thereby advancing our understanding of stellar structure and evolution.


\section*{Data availability}
The star-by-star resulting mode frequencies obtained from \texttt{apollinaire}, FAMED and \texttt{PBjam} for the optimal light curve are only available in electronic form at the CDS via anonymous ftp to \url{cdsarc.u-strasbg.fr} (130.79.128.5) or via \url{http://cdsweb.u-strasbg.fr/cgi-bin/qcat?J/A+A/}. An example of these tables is shown for HD\,36553 in Appendix~\ref{app:tables_per_star}.

\begin{acknowledgements}
This paper includes data collected with the TESS mission, obtained from the MAST data archive at the Space Telescope Science Institute (STScI). Funding for the TESS mission is provided by the NASA Explorer Program. STScI is operated by the Association of Universities for Research in Astronomy, Inc., under NASA contract NAS 5–26555. EP, RAG, BL, LD, JG, ARGS, and DBP acknowledge financial support from the Centre national d’études spatiales (CNES), France (ROR: \url{https://ror.org/04h1h0y33}), within the framework of the GOLF and PLATO space missions. RAG acknowledges support from the Spanish Ministry of Science and Innovation (MICINN) with the grant No. PID2023-146453NB-I00. SNB and EC acknowledges support from PLATO ASI-INAF agreement no. 2022-28-HH.0 "PLATO Fase D". EC is funded by the European Union NextGenerationEU RRF M4C2 1.1 No. 2022HY2NSX, “CHRONOS: adjusting the clock(s) to unveil the CHRONOchemo-dynamical Structure of the Galaxy” (PI:S.Cassisi). GTH acknowledges the support of the Science and Technology Facilities Council. GTH and MBN acknowledge support from the European Research Council (ERC) under the European Union’s Horizon 2020 research and innovation programme (CartographY G.A. n. 804752). MNL acknowledges support from the ESA PRODEX programme (PEA 4000142995). SM and RAG acknowledge support from the Spanish Ministry of Science and Innovation with the grant no. PID2023-149439NB-C41.
\textit{Softwares}: \texttt{emcee} \citep{foreman-mackey_2013}, \texttt{lmfit} \citep{newville_2025}, Matplotlib \citep{hunter_2007}, NumPy \citep{vanderwalt_2011}, SciPy \citep{jones_2001}, pandas \citep{mckinney_2010, team_2024}. Some portions of the text were revised with the help of OpenAI’s ChatGPT (GPT-4), used for improving clarity and scientific language.
\end{acknowledgements}

\bibliographystyle{aa}
\bibliography{aa59786-26}

@article{anderson_1990,
  title = {Modeling of {{Solar Oscillation Power Spectra}}},
  author = {Anderson, Edwin R. and Duvall, Thomas L. and Jefferies, Stuart M.},
  year = 1990,
  month = dec,
  journal = {ApJ},
  volume = {364},
  pages = {699},
  issn = {0004-637X},
  doi = {10.1086/169452},
  langid = {english}
}

@article{appourchaux_2012a,
  title = {Oscillation Mode Linewidths of Main-Sequence and Subgiant Stars Observed by {{Kepler}}},
  author = {Appourchaux, T. and Benomar, O. and Gruberbauer, M. and Chaplin, W. J. and Garc{\'i}a, R. A. and Handberg, R. and Verner, G. A. and Antia, H. M. and Campante, T. L. and Davies, G. R. and Deheuvels, S. and Hekker, S. and Howe, R. and Salabert, D. and Bedding, T. R. and White, T. R. and Houdek, G. and Silva Aguirre, V. and Elsworth, Y. P. and {van Cleve}, J. and Clarke, B. D. and Hall, J. R. and Kjeldsen, H.},
  year = 2012,
  month = jan,
  journal = {A\&A},
  volume = {537},
  pages = {A134},
  issn = {0004-6361},
  doi = {10.1051/0004-6361/201118496},
  langid = {english}
}

@article{appourchaux_2012,
  title = {Oscillation Mode Frequencies of 61 Main-Sequence and Subgiant Stars Observed by {{Kepler}}},
  author = {Appourchaux, T. and Chaplin, W. J. and Garc{\'i}a, R. A. and Gruberbauer, M. and Verner, G. A. and Antia, H. M. and Benomar, O. and Campante, T. L. and Davies, G. R. and Deheuvels, S. and Handberg, R. and Hekker, S. and Howe, R. and R{\'e}gulo, C. and Salabert, D. and Bedding, T. R. and White, T. R. and Ballot, J. and Mathur, S. and Silva Aguirre, V. and Elsworth, Y. P. and Basu, S. and Gilliland, R. L. and {Christensen-Dalsgaard}, J. and Kjeldsen, H. and Uddin, K. and Stumpe, M. C. and Barclay, T.},
  year = 2012,
  month = jul,
  journal = {A\&A},
  volume = {543},
  pages = {A54},
  issn = {0004-6361},
  doi = {10.1051/0004-6361/201218948},
  langid = {english}
}

@article{appourchaux_2020,
  title = {On Attempting to Automate the Identification of Mixed Dipole Modes for Subgiant Stars},
  author = {Appourchaux, T.},
  year = 2020,
  month = oct,
  journal = {A\&A},
  volume = {642},
  pages = {A226},
  publisher = {EDP Sciences},
  issn = {0004-6361, 1432-0746},
  doi = {10.1051/0004-6361/202038834},
  copyright = {\copyright{} T. Appourchaux et al. 2020},
  langid = {english}
}

@article{barnes_2023a,
  title = {{{DMPP-4}}: Candidate Sub-{{Neptune}} Mass Planets Orbiting a Naked-Eye Star},
  author = {Barnes, J. R. and Standing, M. R. and Haswell, C. A. and Staab, D. and Doherty, J. P. J. and {Waller-Bridge}, M. and Fossati, L. and Soto, M. and {Anglada-Escud{\'e}}, G. and Llama, J. and McCune, C. and Lewis, F. W.},
  year = 2023,
  month = oct,
  journal = {MNRAS},
  volume = {524},
  number = {4},
  pages = {5196--5212},
  issn = {0035-8711},
  doi = {10.1093/mnras/stad2109},
  langid = {english}
}

@article{beck_2011,
  title = {Kepler {{Detected Gravity-Mode Period Spacings}} in a {{Red Giant Star}}},
  author = {Beck, P. G. and Bedding, T. R. and Mosser, B. and Stello, D. and Garcia, R. A. and Kallinger, T. and Hekker, S. and Elsworth, Y. and Frandsen, S. and Carrier, F. and De Ridder, J. and Aerts, C. and White, T. R. and Huber, D. and Dupret, M.-A. and Montalb{\'a}n, J. and Miglio, A. and Noels, A. and Chaplin, W. J. and Kjeldsen, H. and {Christensen-Dalsgaard}, J. and Gilliland, R. L. and Brown, T. M. and Kawaler, S. D. and Mathur, S. and Jenkins, J. M.},
  year = 2011,
  month = apr,
  journal = {Science},
  volume = {332},
  number = {6026},
  pages = {205},
  issn = {0036-8075},
  doi = {10.1126/science.1201939},
  langid = {english}
}

@article{bedding_2005,
  title = {The Non-Detection of Oscillations in {{Procyon}} by {{MOST}}: {{Is}} It Really a Surprise?},
  author = {Bedding, T. R. and Kjeldsen, H. and Bouchy, F. and Bruntt, H. and Butler, R. P. and Buzasi, D. L. and {Christensen-Dalsgaard}, J. and Frandsen, S. and Lebrun, J.-C. and Marti{\'c}, M. and Schou, J.},
  year = 2005,
  month = mar,
  journal = {A\&A},
  volume = {432},
  number = {2},
  pages = {L43-L48},
  issn = {0004-6361},
  doi = {10.1051/0004-6361:200500019},
  langid = {english}
}

@article{belkacem_2008,
  title = {Stochastic Excitation of Non-Radial Modes. {{I}}. {{High-angular-degree}} p Modes},
  author = {Belkacem, K. and Samadi, R. and Goupil, M. -J. and Dupret, M. -A.},
  year = 2008,
  month = jan,
  journal = {A\&A},
  volume = {478},
  pages = {163--174},
  issn = {0004-6361},
  doi = {10.1051/0004-6361:20077775}
}

@article{belkacem_2012,
  title = {Damping Rates of Solar-like Oscillations across the {{HR}} Diagram. {{Theoretical}} Calculations Confronted to {{CoRoT}} and {{Kepler}} Observations},
  author = {Belkacem, K. and Dupret, M. A. and Baudin, F. and Appourchaux, T. and Marques, J. P. and Samadi, R.},
  year = 2012,
  month = apr,
  journal = {A\&A},
  volume = {540},
  pages = {L7},
  publisher = {EDP},
  issn = {0004-6361},
  doi = {10.1051/0004-6361/201218890}
}

@article{bellinger_2021,
  title = {Asteroseismic {{Inference}} of the {{Central Structure}} in a {{Subgiant Star}}},
  author = {Bellinger, Earl P. and Basu, Sarbani and Hekker, Saskia and {Christensen-Dalsgaard}, J{\o}rgen and Ball, Warrick H.},
  year = 2021,
  month = jul,
  journal = {ApJ},
  volume = {915},
  number = {2},
  pages = {100},
  issn = {0004-637X},
  doi = {10.3847/1538-4357/ac0051},
  langid = {english}
}

@article{benomar_2013,
  title = {Properties of {{Oscillation Modes}} in {{Subgiant Stars Observed}} by {{Kepler}}},
  author = {Benomar, O. and Bedding, T. R. and Mosser, B. and Stello, D. and Belkacem, K. and Garcia, R. A. and White, T. R. and Kuehn, C. A. and Deheuvels, S. and {Christensen-Dalsgaard}, J.},
  year = 2013,
  month = apr,
  journal = {ApJ},
  volume = {767},
  number = {2},
  pages = {158},
  issn = {0004-637X},
  doi = {10.1088/0004-637X/767/2/158},
  langid = {english}
}

@article{breton_2022,
  title = {Deciphering Stellar Chorus: Apollinaire, a {{Python}} 3 Module for {{Bayesian}} Peakbagging in Helioseismology and Asteroseismology},
  author = {Breton, S. N. and Garc{\'i}a, R. A. and Ballot, J. and Delsanti, V. and Salabert, D.},
  year = 2022,
  month = jul,
  journal = {A\&A},
  volume = {663},
  pages = {A118},
  issn = {0004-6361},
  doi = {10.1051/0004-6361/202243330}
}

@article{breton_2022a,
  title = {No Swan Song for {{Sun-as-a-star}} Helioseismology: {{Performances}} of the {{Solar-SONG}} Prototype for Individual Mode Characterisation},
  author = {Breton, S. N. and Pall{\'e}, P. L. and Garc{\'i}a, R. A. and Fredslund Andersen, M. and Grundahl, F. and {Christensen-Dalsgaard}, J. and Kjeldsen, H. and Mathur, S.},
  year = 2022,
  month = feb,
  journal = {A\&A},
  volume = {658},
  pages = {A27},
  issn = {0004-6361},
  doi = {10.1051/0004-6361/202141496},
  langid = {english}
}

@article{breton_2023,
  title = {In Search of Gravity Mode Signatures in Main Sequence Solar-Type Stars Observed by {{Kepler}}},
  author = {Breton, S. N. and Dhouib, H. and Garc{\'i}a, R. A. and Brun, A. S. and Mathis, S. and P{\'e}rez Hern{\'a}ndez, F. and Mathur, S. and Dyrek, A. and Santos, A. R. G. and Pall{\'e}, P. L.},
  year = 2023,
  month = nov,
  journal = {A\&A},
  volume = {679},
  pages = {A104},
  issn = {0004-6361},
  doi = {10.1051/0004-6361/202346601}
}

@article{buchele_2025,
  title = {Linearity of {{Structure Kernels}} in {{Main-sequence}} and {{Subgiant Solar-like Oscillators}}},
  author = {Buchele, Lynn and Bellinger, Earl P. and Hekker, Saskia and Basu, Sarbani},
  year = 2025,
  month = aug,
  journal = {ApJ},
  volume = {989},
  number = {2},
  pages = {158},
  issn = {0004-637X},
  doi = {10.3847/1538-4357/adec72},
  langid = {english}
}

@article{buldgen_2024,
  title = {Markov Chain {{Monte Carlo}} Inversions of the Internal Rotation of {{Kepler}} Subgiants},
  author = {Buldgen, G. and Fellay, L. and B{\'e}trisey, J. and Deheuvels, S. and Farnir, M. and Farrell, E.},
  year = 2024,
  month = sep,
  journal = {A\&A},
  volume = {689},
  pages = {A307},
  issn = {0004-6361},
  doi = {10.1051/0004-6361/202450315},
  langid = {english}
}

@article{campante_2011,
  title = {Asteroseismology from Multi-Month {{Kepler}} Photometry: The Evolved {{Sun-like}} Stars {{KIC}} 10273246 and {{KIC}} 10920273},
  author = {Campante, T. L. and Handberg, R. and Mathur, S. and Appourchaux, T. and Bedding, T. R. and Chaplin, W. J. and Garc{\'i}a, R. A. and Mosser, B. and Benomar, O. and Bonanno, A. and Corsaro, E. and Fletcher, S. T. and Gaulme, P. and Hekker, S. and Karoff, C. and R{\'e}gulo, C. and Salabert, D. and Verner, G. A. and White, T. R. and Houdek, G. and Brand{\~a}o, I. M. and Creevey, O. L. and Do{\v g}an, G. and Bazot, M. and {Christensen-Dalsgaard}, J. and Cunha, M. S. and Elsworth, Y. and Huber, D. and Kjeldsen, H. and Lundkvist, M. and {Molenda-{\.Z}akowicz}, J. and Monteiro, M. J. P. F. G. and Stello, D. and Clarke, B. D. and Girouard, F. R. and Hall, J. R.},
  year = 2011,
  month = oct,
  journal = {A\&A},
  volume = {534},
  pages = {A6},
  issn = {0004-6361},
  doi = {10.1051/0004-6361/201116620},
  langid = {english}
}

@article{casagrande_2011,
  title = {New Constraints on the Chemical Evolution of the Solar Neighbourhood and {{Galactic}} Disc(s). {{Improved}} Astrophysical Parameters for the {{Geneva-Copenhagen Survey}}},
  author = {Casagrande, L. and Sch{\"o}nrich, R. and Asplund, M. and Cassisi, S. and Ram{\'i}rez, I. and Mel{\'e}ndez, J. and Bensby, T. and Feltzing, S.},
  year = 2011,
  month = jun,
  journal = {A\&A},
  volume = {530},
  pages = {A138},
  issn = {0004-6361},
  doi = {10.1051/0004-6361/201016276},
  langid = {english}
}

@article{chontos_2022,
  title = {{{pySYD}}: {{Automated}} Measurements of Global Asteroseismic Parameters},
  author = {Chontos, Ashley and Huber, Daniel and Sayeed, Maryum and Yamsiri, Pavadol},
  year = 2022,
  month = nov,
  journal = {JOSS},
  volume = {7},
  number = {79},
  pages = {3331},
  doi = {10.21105/joss.03331},
  langid = {english}
}

@article{compton_2019,
  title = {Asteroseismology of Main-Sequence {{F}} Stars with {{Kepler}}: Overcoming Short Mode Lifetimes},
  author = {Compton, Douglas L. and Bedding, Timothy R. and Stello, Dennis},
  year = 2019,
  month = may,
  journal = {MNRAS},
  volume = {485},
  number = {1},
  pages = {560--569},
  issn = {0035-8711},
  doi = {10.1093/mnras/stz432},
  langid = {english}
}

@article{corsaro_2014,
  title = {{{DIAMONDS}}: {{A}} New {{Bayesian}} Nested Sampling Tool. {{Application}} to Peak Bagging of Solar-like Oscillations},
  author = {Corsaro, E. and De Ridder, J.},
  year = 2014,
  month = nov,
  journal = {A\&A},
  volume = {571},
  pages = {A71},
  issn = {0004-6361},
  doi = {10.1051/0004-6361/201424181},
  langid = {english}
}

@article{Corsaro19,
	adsurl = {https://ui.adsabs.harvard.edu/abs/2019FrASS...6...21C},
	archiveprefix = {arXiv},
	author = {{Corsaro}, Enrico},
	doi = {10.3389/fspas.2019.00021},
	eid = {21},
	eprint = {1903.09409},
	journal = {Frontiers in Astronomy and Space Sciences},
	month = {Apr},
	pages = {21},
	primaryclass = {astro-ph.SR},
	title = {{Fast and automated oscillation frequency extraction using Bayesian multi-modality}},
	volume = {6},
	year = {2019}
    }

@article{corsaro_2020,
  title = {Fast and {{Automated Peak Bagging}} with {{DIAMONDS}} ({{FAMED}})},
  author = {Corsaro, E. and McKeever, J. M. and Kuszlewicz, J. S.},
  year = 2020,
  month = aug,
  journal = {A\&A},
  volume = {640},
  pages = {A130},
  issn = {0004-6361},
  doi = {10.1051/0004-6361/202037930},
  langid = {english}
}

@article{davies_2015,
  title = {Asteroseismic Inference on Rotation, Gyrochronology and Planetary System Dynamics of 16 {{Cygni}}},
  author = {Davies, G. R. and Chaplin, W. J. and Farr, W. M. and Garc{\'i}a, R. A. and Lund, M. N. and Mathis, S. and Metcalfe, T. S. and Appourchaux, T. and Basu, S. and Benomar, O. and Campante, T. L. and Ceillier, T. and Elsworth, Y. and Handberg, R. and Salabert, D. and Stello, D.},
  year = 2015,
  month = jan,
  journal = {MNRAS},
  volume = {446},
  pages = {2959--2966},
  publisher = {OUP},
  issn = {0035-8711},
  doi = {10.1093/mnras/stu2331}
}

@article{deheuvels_2010,
  title = {New Insights on the Interior of Solar-like Pulsators Thanks to {{CoRoT}}: The Case of {{HD}} 49385},
  author = {Deheuvels, S. and Michel, E.},
  year = 2010,
  month = jul,
  journal = {Astrophys. Space Sci.},
  volume = {328},
  number = {1-2},
  pages = {259--263},
  issn = {0004-640X},
  doi = {10.1007/s10509-009-0216-2},
  langid = {english}
}

@article{deheuvels_2011,
  title = {Constraints on the Structure of the Core of Subgiants via Mixed Modes: The Case of {{HD}} 49385},
  author = {Deheuvels, S. and Michel, E.},
  year = 2011,
  month = nov,
  journal = {A\&A},
  volume = {535},
  pages = {A91},
  publisher = {EDP Sciences},
  issn = {0004-6361, 1432-0746},
  doi = {10.1051/0004-6361/201117232},
  copyright = {\copyright{} ESO, 2011},
  langid = {english}
}

@article{deheuvels_2014,
  title = {Seismic Constraints on the Radial Dependence of the Internal Rotation Profiles of Six {{Kepler}} Subgiants and Young Red Giants},
  author = {Deheuvels, S. and Do{\u g}an, G. and Goupil, M. J. and Appourchaux, T. and Benomar, O. and Bruntt, H. and Campante, T. L. and Casagrande, L. and Ceillier, T. and Davies, G. R. and De Cat, P. and Fu, J. N. and Garc{\'i}a, R. A. and Lobel, A. and Mosser, B. and Reese, D. R. and Regulo, C. and Schou, J. and Stahn, T. and Thygesen, A. O. and Yang, X. H. and Chaplin, W. J. and {Christensen-Dalsgaard}, J. and Eggenberger, P. and Gizon, L. and Mathis, S. and {Molenda-{\.Z}akowicz}, J. and Pinsonneault, M.},
  year = 2014,
  month = apr,
  journal = {A\&A},
  volume = {564},
  pages = {A27},
  issn = {0004-6361},
  doi = {10.1051/0004-6361/201322779},
  langid = {english}
}

@article{deheuvels_2020,
  title = {Seismic Evidence for near Solid-Body Rotation in Two {{Kepler}} Subgiants and Implications for Angular Momentum Transport},
  author = {Deheuvels, S. and Ballot, J. and Eggenberger, P. and Spada, F. and Noll, A. and {den Hartogh}, J. W.},
  year = 2020,
  month = sep,
  journal = {A\&A},
  volume = {641},
  pages = {A117},
  issn = {0004-6361},
  doi = {10.1051/0004-6361/202038578},
  langid = {english}
}

@article{desort_2009,
  title = {Extrasolar Planets and Brown Dwarfs around {{A-F}} Type Stars. {{VII}}. {$<$}{{ASTROBJ}}{$>\theta$} {{Cygni}}{$<$}/{{ASTROBJ}}{$>$} Radial Velocity Variations: Planets or Stellar Phenomenon?},
  author = {Desort, M. and Lagrange, A.-M. and Galland, F. and Udry, S. and Montagnier, G. and Beust, H. and Boisse, I. and Bonfils, X. and Bouchy, F. and Delfosse, X. and Eggenberger, A. and Ehrenreich, D. and Forveille, T. and H{\'e}brard, G. and Loeillet, B. and Lovis, C. and Mayor, M. and Meunier, N. and Moutou, C. and Pepe, F. and Perrier, C. and Pont, F. and Queloz, D. and Santos, N. C. and S{\'e}gransan, D. and {Vidal-Madjar}, A.},
  year = 2009,
  month = nov,
  journal = {A\&A},
  volume = {506},
  number = {3},
  pages = {1469--1476},
  issn = {0004-6361},
  doi = {10.1051/0004-6361/200911731},
  langid = {english}
}

@article{fellay_2021,
  title = {Asteroseismology of Evolved Stars to Constrain the Internal Transport of Angular Momentum. {{IV}}. {{Internal}} Rotation of {{Kepler-56}} from an {{MCMC}} Analysis of the Rotational Splittings},
  author = {Fellay, L. and Buldgen, G. and Eggenberger, P. and Khan, S. and Salmon, S. J. a. J. and Miglio, A. and Montalb{\'a}n, J.},
  year = 2021,
  month = oct,
  journal = {A\&A},
  volume = {654},
  pages = {A133},
  issn = {0004-6361},
  doi = {10.1051/0004-6361/202140518},
  langid = {english}
}

@article{foreman-mackey_2013,
  title = {Emcee: {{The MCMC Hammer}}},
  author = {{Foreman-Mackey}, Daniel and Hogg, David W. and Lang, Dustin and Goodman, Jonathan},
  year = 2013,
  month = mar,
  journal = {PASP},
  volume = {125},
  number = {925},
  pages = {306},
  issn = {0004-6280},
  doi = {10.1086/670067},
  langid = {english}
}

@article{garcia_2010,
  title = {{{CoRoT Reveals}} a {{Magnetic Activity Cycle}} in a {{Sun-Like Star}}},
  author = {Garc{\'i}a, Rafael A. and Mathur, Savita and Salabert, David and Ballot, J{\'e}r{\^o}me and R{\'e}gulo, Clara and Metcalfe, Travis S. and Baglin, Annie},
  year = 2010,
  month = aug,
  journal = {Science},
  volume = {329},
  number = {5995},
  pages = {1032},
  issn = {0036-8075},
  doi = {10.1126/science.1191064},
  langid = {english}
}

@article{garcia_2019,
  title = {Asteroseismology of Solar-Type Stars},
  author = {Garc{\'i}a, Rafael A. and Ballot, J{\'e}r{\^o}me},
  year = 2019,
  month = sep,
  journal = {Living Rev. Sol. Phys.},
  volume = {16},
  pages = {4},
  doi = {10.1007/s41116-019-0020-1}
}

@article{garciasaraviaortizdemontellano_2018,
  title = {Automated Asteroseismic Peak Detections},
  author = {{Garc{\'i}a Saravia Ortiz de Montellano}, Andr{\'e}s and Hekker, S. and Theme{\ss}l, N.},
  year = 2018,
  month = may,
  journal = {MNRAS},
  volume = {476},
  number = {2},
  pages = {1470--1496},
  issn = {0035-8711},
  doi = {10.1093/mnras/sty253},
  langid = {english}
}

@article{goldreich_1977,
  title = {Solar Seismology. {{I}}. {{The}} Stability of the Solar p-Modes.},
  author = {Goldreich, P. and Keeley, D. A.},
  year = 1977,
  month = feb,
  journal = {ApJ},
  volume = {211},
  pages = {934--942},
  issn = {0004-637X},
  doi = {10.1086/155005}
}

@article{gould_1855,
  title = {On {{Peirce}}'s {{Criterion}} for the {{Rejection}} of {{Doubtful Observations}}, with Tables for Facilitating Its Application},
  author = {Gould, B. A.},
  year = 1855,
  month = apr,
  journal = {ApJ},
  volume = {4},
  pages = {81--87},
  issn = {0004-6256},
  doi = {10.1086/100480},
  langid = {english}
}

@article{goupil_2024,
  title = {Predicted Asteroseismic Detection Yield for Solar-like Oscillating Stars with {{PLATO}}},
  author = {Goupil, M. J. and Catala, C. and Samadi, R. and Belkacem, K. and Ouazzani, R. M. and Reese, D. R. and Appourchaux, T. and Mathur, S. and Cabrera, J. and B{\"o}rner, A. and Paproth, C. and Moedas, N. and Verma, K. and Lebreton, Y. and Deal, M. and Ballot, J. and Chaplin, W. J. and {Christensen-Dalsgaard}, J. and Cunha, M. and Lanza, A. F. and Miglio, A. and Morel, T. and Serenelli, A. and Mosser, B. and Creevey, O. and Moya, A. and Garcia, R. A. and Nielsen, M. B. and Hatt, E.},
  year = 2024,
  month = mar,
  journal = {A\&A},
  volume = {683},
  pages = {A78},
  issn = {0004-6361},
  doi = {10.1051/0004-6361/202348111}
}

@article{guzik_2016,
  title = {Detection of {{Solar-like Oscillations}}, {{Observational Constraints}}, and {{Stellar Models}} for \texttheta{} {{Cyg}}, the {{Brightest Star Observed By}} the {{Kepler Mission}}},
  author = {Guzik, J. A. and Houdek, G. and Chaplin, W. J. and Smalley, B. and Kurtz, D. W. and Gilliland, R. L. and Mullally, F. and Rowe, J. F. and Bryson, S. T. and Still, M. D. and Antoci, V. and Appourchaux, T. and Basu, S. and Bedding, T. R. and Benomar, O. and Garcia, R. A. and Huber, D. and Kjeldsen, H. and Latham, D. W. and Metcalfe, T. S. and P{\'a}pics, P. I. and White, T. R. and Aerts, C. and Ballot, J. and Boyajian, T. S. and Briquet, M. and Bruntt, H. and Buchhave, L. A. and Campante, T. L. and Catanzaro, G. and {Christensen-Dalsgaard}, J. and Davies, G. R. and Do{\u g}an, G. and Dragomir, D. and Doyle, A. P. and Elsworth, Y. and Frasca, A. and Gaulme, P. and Gruberbauer, M. and Handberg, R. and Hekker, S. and Karoff, C. and Lehmann, H. and Mathias, P. and Mathur, S. and Miglio, A. and {Molenda-{\.Z}akowicz}, J. and Mosser, B. and Murphy, S. J. and R{\'e}gulo, C. and Ripepi, V. and Salabert, D. and Sousa, S. G. and Stello, D. and Uytterhoeven, K.},
  year = 2016,
  month = nov,
  journal = {ApJ},
  volume = {831},
  number = {1},
  pages = {17},
  issn = {0004-637X},
  doi = {10.3847/0004-637X/831/1/17},
  langid = {english}
}

@article{handberg_2014,
  title = {Automated Preparation of {{Kepler}} Time Series of Planet Hosts for Asteroseismic Analysis},
  author = {Handberg, R. and Lund, M. N.},
  year = 2014,
  month = dec,
  journal = {MNRAS},
  volume = {445},
  number = {3},
  pages = {2698--2709},
  issn = {0035-8711},
  doi = {10.1093/mnras/stu1823},
  langid = {english}
}

@article{handberg_2017,
  title = {{{NGC}} 6819: Testing the Asteroseismic Mass Scale, Mass Loss and Evidence for Products of Non-Standard Evolution},
  author = {Handberg, R. and Brogaard, K. and Miglio, A. and Bossini, D. and Elsworth, Y. and Slumstrup, D. and Davies, G. R. and Chaplin, W. J.},
  year = 2017,
  month = nov,
  journal = {MNRAS},
  volume = {472},
  number = {1},
  pages = {979--997},
  issn = {0035-8711},
  doi = {10.1093/mnras/stx1929},
  langid = {english}
}

@article{harvey_1985,
  title = {High-{{Resolution Helioseismology}}},
  author = {Harvey, J.},
  year = 1985,
  month = jun,
  journal = {Future Missions in Solar, Heliospheric \& Space Plasma Physics},
  volume = {235},
  pages = {199},
  issn = {0379-6566},
  langid = {english}
}

@article{hon_2024,
  title = {Asteroseismology of the {{Nearby K Dwarf}} {$\sigma$} {{Draconis Using}} the {{Keck Planet Finder}} and {{TESS}}},
  author = {Hon, Marc and Huber, Daniel and Li, Yaguang and Metcalfe, Travis S. and Bedding, Timothy R. and Ong, Joel and Chontos, Ashley and Rubenzahl, Ryan and Halverson, Samuel and Garc{\'i}a, Rafael A. and Kjeldsen, Hans and Stello, Dennis and Hey, Daniel R. and Campante, Tiago and Howard, Andrew W. and Gibson, Steven R. and Rider, Kodi and Roy, Arpita and Baker, Ashley D. and Edelstein, Jerry and Smith, Chris and Fulton, Benjamin J. and Walawender, Josh and Brodheim, Max and Brown, Matt and Chan, Dwight and Dai, Fei and Deich, William and Gottschalk, Colby and Grillo, Jason and Hale, Dave and Hill, Grant M. and Holden, Bradford and Householder, Aaron and Isaacson, Howard and Ishikawa, Yuzo and Jelinsky, Sharon R. and Kassis, Marc and Kaye, Stephen and Laher, Russ and Lanclos, Kyle and Lee, Chien-Hsiu and Lilley, Scott and McCarney, Ben and Miller, Timothy N. and Payne, Joel and Petigura, Erik A. and Poppett, Claire and Raffanti, Michael and Rockosi, Constance and Sanford, Dale and Schwab, Christian and Shaum, Abby P. and Sirk, Martin M. and Smith, Roger and Thorne, Jim and Valliant, John and Vandenberg, Adam and Wang, Shin Ywan and Wishnow, Edward and Wold, Truman and Yeh, Sherry and Baker, Ashley and Basu, Sarbani and Bedell, Megan and Cegla, Heather M. and Crossfield, Ian and Dressing, Courtney and Dumusque, Xavier and Knutson, Heather and Mawet, Dimitri and O'Meara, John and Stef{\'a}nsson, Gu{\dj}mundur and Teske, Johanna and Vasisht, Gautam and Wang, Sharon Xuesong and Weiss, Lauren M. and Winn, Joshua N. and Wright, Jason T.},
  year = 2024,
  month = nov,
  journal = {The Astrophysical Journal},
  volume = {975},
  number = {1},
  pages = {147},
  issn = {0004-637X},
  doi = {10.3847/1538-4357/ad76a9},
  langid = {english}
}

@article{hookway_2025,
  title = {Peakbagging the {{K2 KEYSTONE}} Sample with {{PBJAM}}: Characterizing the Individual Mode Frequencies in Solar-like Oscillators},
  author = {Hookway, George T. and Nielsen, Martin B. and Davies, Guy R. and Lund, Mikkel N. and Garc{\'i}a, Rafael A. and Mathur, Savita and See, Victor and Stokholm, Amalie},
  year = 2025,
  month = dec,
  journal = {MNRAS},
  volume = {544},
  number = {4},
  pages = {3247--3259},
  issn = {0035-8711},
  doi = {10.1093/mnras/staf1869},
  langid = {english}
}

@article{houdek_2019,
  title = {Damping Rates and Frequency Corrections of {{Kepler LEGACY}} Stars},
  author = {Houdek, G. and Lund, M. N. and Trampedach, R. and {Christensen-Dalsgaard}, J. and Handberg, R. and Appourchaux, T.},
  year = 2019,
  month = jul,
  journal = {MNRAS},
  volume = {487},
  number = {1},
  pages = {595--608},
  issn = {0035-8711},
  doi = {10.1093/mnras/stz1211},
  langid = {english}
}

@article{huber_2011,
  title = {Testing {{Scaling Relations}} for {{Solar-like Oscillations}} from the {{Main Sequence}} to {{Red Giants Using Kepler Data}}},
  author = {Huber, D. and Bedding, T. R. and Stello, D. and Hekker, S. and Mathur, S. and Mosser, B. and Verner, G. A. and Bonanno, A. and Buzasi, D. L. and Campante, T. L. and Elsworth, Y. P. and Hale, S. J. and Kallinger, T. and Silva Aguirre, V. and Chaplin, W. J. and De Ridder, J. and Garc{\'i}a, R. A. and Appourchaux, T. and Frandsen, S. and Houdek, G. and {Molenda-{\.Z}akowicz}, J. and Monteiro, M. J. P. F. G. and {Christensen-Dalsgaard}, J. and Gilliland, R. L. and Kawaler, S. D. and Kjeldsen, H. and Broomhall, A. M. and Corsaro, E. and Salabert, D. and Sanderfer, D. T. and Seader, S. E. and Smith, J. C.},
  year = 2011,
  month = dec,
  journal = {ApJ},
  volume = {743},
  number = {2},
  pages = {143},
  issn = {0004-637X},
  doi = {10.1088/0004-637X/743/2/143},
  langid = {english}
}

@article{huber_2022,
  title = {A 20 {{Second Cadence View}} of {{Solar-type Stars}} and {{Their Planets}} with {{TESS}}: {{Asteroseismology}} of {{Solar Analogs}} and a {{Recharacterization}} of {$\pi$} {{Men}} c},
  author = {Huber, Daniel and White, Timothy R. and Metcalfe, Travis S. and Chontos, Ashley and Fausnaugh, Michael M. and Ho, Cynthia S. K. and Van Eylen, Vincent and Ball, Warrick H. and Basu, Sarbani and Bedding, Timothy R. and Benomar, Othman and Bossini, Diego and Breton, Sylvain and Buzasi, Derek L. and Campante, Tiago L. and Chaplin, William J. and {Christensen-Dalsgaard}, J{\o}rgen and Cunha, Margarida S. and Deal, Morgan and Garc{\'i}a, Rafael A. and Garc{\'i}a Mu{\~n}oz, Antonio and Gehan, Charlotte and {Gonz{\'a}lez-Cuesta}, Luc{\'i}a and Jiang, Chen and Kayhan, Cenk and Kjeldsen, Hans and Lundkvist, Mia S. and Mathis, St{\'e}phane and Mathur, Savita and Monteiro, M{\'a}rio J. P. F. G. and Nsamba, Benard and Ong, Jia Mian Joel and Pak{\v s}tien{\.e}, Erika and Serenelli, Aldo M. and Silva Aguirre, Victor and Stassun, Keivan G. and Stello, Dennis and Norgaard Stilling, Sissel and Lykke Winther, Mark and Wu, Tao and Barclay, Thomas and Daylan, Tansu and G{\"u}nther, Maximilian N. and Hermes, J. J. and Jenkins, Jon M. and Latham, David W. and Levine, Alan M. and Ricker, George R. and Seager, Sara and Shporer, Avi and Twicken, Joseph D. and Vanderspek, Roland K. and Winn, Joshua N.},
  year = 2022,
  month = feb,
  journal = {ApJ},
  volume = {163},
  number = {2},
  pages = {79},
  issn = {0004-6256},
  doi = {10.3847/1538-3881/ac3000},
  langid = {english}
}

@article{hunter_2007,
  title = {Matplotlib: {{A 2D Graphics Environment}}},
  author = {Hunter, John D.},
  year = 2007,
  month = may,
  journal = {CiSE},
  volume = {9},
  pages = {90--95},
  doi = {10.1109/MCSE.2007.55}
}

@article{jones_2001,
  title = {{{SciPy}}: {{Open Source Scientific Tools}} for {{Python}}},
  author = {Jones, Eric and Oliphant, Travis and Peterson, Pearu},
  year = 2001,
  month = jan
}

@article{kallinger_2019,
  title = {Release Note: {{Massive}} Peak Bagging of Red Giants in the {{Kepler}} Field},
  author = {Kallinger, T.},
  year = 2019,
  month = jun,
  journal = {arXiv e-prints},
  pages = {arXiv:1906.09428},
  doi = {10.48550/arXiv.1906.09428},
  langid = {english}
}

@article{kjeldsen_1995,
  title = {Amplitudes of Stellar Oscillations: The Implications for Asteroseismology.},
  author = {Kjeldsen, H. and Bedding, T. R.},
  year = 1995,
  month = jan,
  journal = {A\&A},
  volume = {293},
  pages = {87--106},
  issn = {0004-6361},
  doi = {10.48550/arXiv.astro-ph/9403015},
  langid = {english}
}

@article{koleva_2012,
  title = {Stellar Population Models in the {{UV}}. {{I}}. {{Characterisation}} of the {{New Generation Stellar Library}}},
  author = {Koleva, M. and Vazdekis, A.},
  year = 2012,
  month = feb,
  journal = {A\&A},
  volume = {538},
  pages = {A143},
  issn = {0004-6361},
  doi = {10.1051/0004-6361/201118065},
  langid = {english}
}

@article{li_2020,
  title = {Asteroseismology of 36 {{Kepler}} Subgiants - {{I}}. {{Oscillation}} Frequencies, Linewidths, and Amplitudes},
  author = {Li, Yaguang and Bedding, Timothy R. and Li, Tanda and Bi, Shaolan and Stello, Dennis and Zhou, Yixiao and White, Timothy R.},
  year = 2020,
  month = jun,
  journal = {MNRAS},
  volume = {495},
  number = {2},
  pages = {2363},
  issn = {0035-8711},
  doi = {10.1093/mnras/staa1335},
  langid = {english}
}

@article{lomb_1976,
  title = {Least-{{Squares Frequency Analysis}} of {{Unequally Spaced Data}}},
  author = {Lomb, N. R.},
  year = 1976,
  month = feb,
  journal = {Astrophys. Space Sci.},
  volume = {39},
  number = {2},
  pages = {447--462},
  issn = {0004-640X},
  doi = {10.1007/BF00648343},
  langid = {english}
}

@article{lund_2015,
  title = {{{K2P}}{\textsuperscript{2}}--- {{A Photometry Pipeline}} for the {{K2 Mission}}},
  author = {Lund, Mikkel N. and Handberg, Rasmus and Davies, Guy R. and Chaplin, William J. and Jones, Caitlin D.},
  year = 2015,
  month = jun,
  journal = {ApJ},
  volume = {806},
  number = {1},
  pages = {30},
  issn = {0004-637X},
  doi = {10.1088/0004-637X/806/1/30},
  langid = {english}
}

@article{lund_2017,
  title = {Standing on the Shoulders of Dwarfs: The Kepler Asteroseismic {{LEGACY}} Sample. {{I}}. {{Oscillation}} Mode Parameters},
  author = {Lund, Mikkel N. and Silva Aguirre, V{\'i}ctor and Davies, Guy R. and Chaplin, William J. and {Christensen-Dalsgaard}, J{\o}rgen and Houdek, G{\"u}nter and White, Timothy R. and Bedding, Timothy R. and Ball, Warrick H. and Huber, Daniel and Antia, H. M. and Lebreton, Yveline and Latham, David W. and Handberg, Rasmus and Verma, Kuldeep and Basu, Sarbani and Casagrande, Luca and Justesen, Anders B. and Kjeldsen, Hans and Mosumgaard, Jakob R.},
  year = 2017,
  month = feb,
  journal = {ApJ},
  volume = {835},
  pages = {172},
  publisher = {IOP},
  issn = {0004-637X},
  doi = {10.3847/1538-4357/835/2/172}
}

@article{lund_2024a,
  title = {The {{K2 Asteroseismic KEYSTONE}} Sample of {{Dwarf}} and {{Subgiant Solar-Like Oscillators}} - {{I}}. {{Data}} and {{Asteroseismic}} Parameters},
  author = {Lund, Mikkel N. and Basu, Sarbani and Bieryla, Allyson and Casagrande, Luca and Huber, Daniel and Hekker, Saskia and Viani, Lucas and Davies, Guy R. and Campante, Tiago L. and Chaplin, William J. and Serenelli, Aldo M. and Ong, J. M. Joel and Ball, Warrick H. and Stokholm, Amalie and Bellinger, Earl P. and Bazot, Micha{\"e}l and Stello, Dennis and Latham, David W. and White, Timothy R. and Sayeed, Maryum and {B{\o}rsen-Koch}, V{\'i}ctor Aguirre and Chontos, Ashley},
  year = 2024,
  month = aug,
  journal = {A\&A},
  volume = {688},
  pages = {A13},
  publisher = {EDP Sciences},
  issn = {0004-6361, 1432-0746},
  doi = {10.1051/0004-6361/202450055},
  copyright = {\copyright{} The Authors 2024},
  langid = {english}
}

@article{lund_2025,
  title = {Luminaries in the Sky: {{The TESS}} Legacy Sample of Bright Stars: {{I}}. {{Asteroseismic}} Detections in Naked-Eye Main-Sequence and Subgiant Solar-like Oscillators},
  author = {Lund, Mikkel N. and Chontos, Ashley and Grundahl, Frank and Mathur, Savita and Garc{\'i}a, Rafael A. and Huber, Daniel and Buzasi, Derek and Bedding, Timothy R. and Hon, Marc and Li, Yaguang},
  year = 2025,
  month = sep,
  journal = {A\&A},
  volume = {701},
  pages = {A285},
  issn = {0004-6361},
  doi = {10.1051/0004-6361/202555485},
  langid = {english}
}

@article{mathur_2010,
  title = {Determining Global Parameters of the Oscillations of Solar-like Stars},
  author = {Mathur, S. and Garc{\'i}a, R. A. and R{\'e}gulo, C. and Creevey, O. L. and Ballot, J. and Salabert, D. and Arentoft, T. and Quirion, P.-O. and Chaplin, W. J. and Kjeldsen, H.},
  year = 2010,
  month = feb,
  journal = {A\&A},
  volume = {511},
  pages = {A46},
  issn = {0004-6361},
  doi = {10.1051/0004-6361/200913266},
  langid = {english}
}

@article{mathur_2011a,
  title = {Solar-like {{Oscillations}} in {{KIC}} 11395018 and {{KIC}} 11234888 from 8 {{Months}} of {{Kepler Data}}},
  author = {Mathur, S. and Handberg, R. and Campante, T. L. and Garc{\'i}a, R. A. and Appourchaux, T. and Bedding, T. R. and Mosser, B. and Chaplin, W. J. and Ballot, J. and Benomar, O. and Bonanno, A. and Corsaro, E. and Gaulme, P. and Hekker, S. and R{\'e}gulo, C. and Salabert, D. and Verner, G. and White, T. R. and Brand{\~a}o, I. M. and Creevey, O. L. and Do{\v g}an, G. and Elsworth, Y. and Huber, D. and Hale, S. J. and Houdek, G. and Karoff, C. and Metcalfe, T. S. and {Molenda-{\.Z}akowicz}, J. and Monteiro, M. J. P. F. G. and Thompson, M. J. and {Christensen-Dalsgaard}, J. and Gilliland, R. L. and Kawaler, S. D. and Kjeldsen, H. and Quintana, E. V. and Sanderfer, D. T. and Seader, S. E.},
  year = 2011,
  month = jun,
  journal = {ApJ},
  volume = {733},
  number = {2},
  pages = {95},
  issn = {0004-637X},
  doi = {10.1088/0004-637X/733/2/95},
  langid = {english}
}

@article{mazumdar_2014,
  title = {Measurement of {{Acoustic Glitches}} in {{Solar-type Stars}} from {{Oscillation Frequencies Observed}} by {{Kepler}}},
  author = {Mazumdar, A. and Monteiro, M. J. P. F. G. and Ballot, J. and Antia, H. M. and Basu, S. and Houdek, G. and Mathur, S. and Cunha, M. S. and Silva Aguirre, V. and Garc{\'i}a, R. A. and Salabert, D. and Verner, G. A. and {Christensen-Dalsgaard}, J. and Metcalfe, T. S. and Sanderfer, D. T. and Seader, S. E. and Smith, J. C. and Chaplin, W. J.},
  year = 2014,
  month = feb,
  journal = {ApJ},
  volume = {782},
  number = {1},
  pages = {18},
  issn = {0004-637X},
  doi = {10.1088/0004-637X/782/1/18},
  langid = {english}
}

@article{mckinney_2010,
  title = {Data {{Structures}} for {{Statistical Computing}} in {{Python}}},
  author = {McKinney, Wes},
  year = 2010,
  month = may,
  journal = {scipy},
  doi = {10.25080/Majora-92bf1922-00a},
  langid = {english}
}

@article{metcalfe_2010,
  title = {A {{Precise Asteroseismic Age}} and {{Radius}} for the {{Evolved Sun-like Star KIC}} 11026764},
  author = {Metcalfe, T. S. and Monteiro, M. J. P. F. G. and Thompson, M. J. and {Molenda-{\.Z}akowicz}, J. and Appourchaux, T. and Chaplin, W. J. and Do{\v g}an, G. and Eggenberger, P. and Bedding, T. R. and Bruntt, H. and Creevey, O. L. and Quirion, P.-O. and Stello, D. and Bonanno, A. and Silva Aguirre, V. and Basu, S. and Esch, L. and Gai, N. and Di Mauro, M. P. and Kosovichev, A. G. and Kitiashvili, I. N. and Su{\'a}rez, J. C. and Moya, A. and Piau, L. and Garc{\'i}a, R. A. and Marques, J. P. and Frasca, A. and Biazzo, K. and Sousa, S. G. and Dreizler, S. and Bazot, M. and Karoff, C. and Frandsen, S. and Wilson, P. A. and Brown, T. M. and {Christensen-Dalsgaard}, J. and Gilliland, R. L. and Kjeldsen, H. and Campante, T. L. and Fletcher, S. T. and Handberg, R. and R{\'e}gulo, C. and Salabert, D. and Schou, J. and Verner, G. A. and Ballot, J. and Broomhall, A.-M. and Elsworth, Y. and Hekker, S. and Huber, D. and Mathur, S. and New, R. and Roxburgh, I. W. and Sato, K. H. and White, T. R. and Borucki, W. J. and Koch, D. G. and Jenkins, J. M.},
  year = 2010,
  month = nov,
  journal = {ApJ},
  volume = {723},
  number = {2},
  pages = {1583--1598},
  issn = {0004-637X},
  doi = {10.1088/0004-637X/723/2/1583},
  langid = {english}
}

@article{metcalfe_2012,
  title = {Asteroseismology of the {{Solar Analogs}} 16 {{Cyg A}} and {{B}} from {{Kepler Observations}}},
  author = {Metcalfe, T. S. and Chaplin, W. J. and Appourchaux, T. and Garc{\'i}a, R. A. and Basu, S. and Brand{\~a}o, I. and Creevey, O. L. and Deheuvels, S. and Do{\v g}an, G. and Eggenberger, P. and Karoff, C. and Miglio, A. and Stello, D. and Y{\i}ld{\i}z, M. and {\c C}elik, Z. and Antia, H. M. and Benomar, O. and Howe, R. and R{\'e}gulo, C. and Salabert, D. and Stahn, T. and Bedding, T. R. and Davies, G. R. and Elsworth, Y. and Gizon, L. and Hekker, S. and Mathur, S. and Mosser, B. and Bryson, S. T. and Still, M. D. and {Christensen-Dalsgaard}, J. and Gilliland, R. L. and Kawaler, S. D. and Kjeldsen, H. and Ibrahim, K. A. and Klaus, T. C. and Li, J.},
  year = 2012,
  month = mar,
  journal = {ApJ},
  volume = {748},
  number = {1},
  pages = {L10},
  issn = {0004-637X},
  doi = {10.1088/2041-8205/748/1/L10},
  langid = {english}
}

@article{montalto_2021,
  title = {The All-Sky {{PLATO}} Input Catalogue},
  author = {Montalto, M. and Piotto, G. and Marrese, P. M. and Nascimbeni, V. and Prisinzano, L. and Granata, V. and Marinoni, S. and Desidera, S. and Ortolani, S. and Aerts, C. and Alei, E. and Altavilla, G. and Benatti, S. and B{\"o}rner, A. and Cabrera, J. and Claudi, R. and Deleuil, M. and Fabrizio, M. and Gizon, L. and Goupil, M. J. and Heras, A. M. and Magrin, D. and Malavolta, L. and {Mas-Hesse}, J. M. and Pagano, I. and Paproth, C. and Pertenais, M. and Pollacco, D. and Ragazzoni, R. and Ramsay, G. and Rauer, H. and Udry, S.},
  year = 2021,
  month = sep,
  journal = {A\&A},
  volume = {653},
  pages = {A98},
  issn = {0004-6361},
  doi = {10.1051/0004-6361/202140717},
  langid = {english}
}

@misc{montalto_2026,
  title = {The {{PLATO Input Catalogue}} of Targets ({{tPIC}}) for the First {{Long Pointing Field}}},
  author = {Montalto, M. and Piotto, G. and Marrese, P. M. and Prisinzano, L. and Marinoni, S. and Granata, V. and Cabrera, J. and Nascimbeni, V. and Desidera, S. and Adibekyan, V. and Ortolani, S. and Alei, E. and Aerts, C. and Altavilla, G. and Belkacem, K. and Benatti, S. and B{\"o}rner, A. and Deleuil, M. and Fabrizio, M. and Gizon, L. and Goupil, M. J. and G{\"u}nther, M. and Heras, A. M. and Magrin, D. and Malavolta, L. and {Mas-Hesse}, J. M. and Pagano, I. and Paproth, C. and Pollacco, D. and Ragazzoni, R. and Ramsay, G. and Rauer, H. and Udry, S.},
  year = 2026,
  month = apr,
  number = {arXiv:2604.03369},
  eprint = {2604.03369},
  primaryclass = {astro-ph},
  publisher = {arXiv},
  doi = {10.48550/arXiv.2604.03369},
  archiveprefix = {arXiv}
}

@article{mosser_2011,
  title = {The Universal Red-Giant Oscillation Pattern. {{An}} Automated Determination with {{CoRoT}} Data},
  author = {Mosser, B. and Belkacem, K. and Goupil, M. J. and Michel, E. and Elsworth, Y. and Barban, C. and Kallinger, T. and Hekker, S. and De Ridder, J. and Samadi, R. and Baudin, F. and Pinheiro, F. J. G. and Auvergne, M. and Baglin, A. and Catala, C.},
  year = 2011,
  month = jan,
  journal = {A\&A},
  volume = {525},
  pages = {L9},
  issn = {0004-6361},
  doi = {10.1051/0004-6361/201015440},
  langid = {english}
}

@article{muterspaugh_2010,
  title = {The {{Phases Differential Astrometry Data Archive}}. {{V}}. {{Candidate Substellar Companions}} to {{Binary Systems}}},
  author = {Muterspaugh, Matthew W. and Lane, Benjamin F. and Kulkarni, S. R. and Konacki, Maciej and Burke, Bernard F. and Colavita, M. M. and Shao, M. and Hartkopf, William I. and Boss, Alan P. and Williamson, M.},
  year = 2010,
  month = dec,
  journal = {ApJ},
  volume = {140},
  number = {6},
  pages = {1657--1671},
  issn = {0004-6256},
  doi = {10.1088/0004-6256/140/6/1657},
  langid = {english}
}

@article{nascimbeni_2022,
  title = {The {{PLATO}} Field Selection Process. {{I}}. {{Identification}} and Content of the Long-Pointing Fields},
  author = {Nascimbeni, V. and Piotto, G. and B{\"o}rner, A. and Montalto, M. and Marrese, P. M. and Cabrera, J. and Marinoni, S. and Aerts, C. and Altavilla, G. and Benatti, S. and Claudi, R. and Deleuil, M. and Desidera, S. and Fabrizio, M. and Gizon, L. and Goupil, M. J. and Granata, V. and Heras, A. M. and Magrin, D. and Malavolta, L. and {Mas-Hesse}, J. M. and Ortolani, S. and Pagano, I. and Pollacco, D. and Prisinzano, L. and Ragazzoni, R. and Ramsay, G. and Rauer, H. and Udry, S.},
  year = 2022,
  month = feb,
  journal = {A\&A},
  volume = {658},
  pages = {A31},
  issn = {0004-6361},
  doi = {10.1051/0004-6361/202142256},
  langid = {english}
}

@article{nascimbeni_2025,
  title = {The {{PLATO}} Field Selection Process: {{II}}. {{Characterization}} of {{LOPS2}}, the First Long-Pointing Field},
  author = {Nascimbeni, V. and Piotto, G. and Cabrera, J. and Montalto, M. and Marinoni, S. and Marrese, P. M. and Aerts, C. and Altavilla, G. and Benatti, S. and B{\"o}rner, A. and Deleuil, M. and Desidera, S. and Gizon, L. and Goupil, M. J. and Granata, V. and Heras, A. M. and Magrin, D. and Malavolta, L. and {Mas-Hesse}, J. M. and Osborn, H. P. and Pagano, I. and Paproth, C. and Pollacco, D. and Prisinzano, L. and Ragazzoni, R. and Ramsay, G. and Rauer, H. and Tkachenko, A. and Udry, S.},
  year = 2025,
  month = feb,
  journal = {A\&A},
  volume = {694},
  pages = {A313},
  issn = {0004-6361},
  doi = {10.1051/0004-6361/202452325},
  langid = {english}
}

@article{nascimbeni_2026,
  title = {The {{PLATO}} Field Selection Process {{III}}. {{Selection}} of the {{Prime Sample}} for the {{LOPS2}} Field},
  author = {Nascimbeni, V. and Piotto, G. and Granata, V. and Marinoni, S. and Marrese, P. M. and Montalto, M. and Cabrera, J. and Aerts, C. and Altavilla, G. and Belkacem, K. and Benatti, S. and Bergemann, M. and B{\"o}rner, A. and Covone, G. and Deleuil, M. and Desidera, S. and Gizon, L. and Goupil, M. J. and G{\"u}nther, M. and Heras, A. M. and Malavolta, L. and {Mas-Hesse}, J. M. and Nardiello, D. and Osborn, H. P. and Pagano, I. and Paproth, C. and Pollacco, D. and Prisinzano, L. and Ragazzoni, R. and Ramsay, G. and Rauer, H. and Udry, S. and Zingales, T.},
  year = 2026,
  month = apr,
  journal = {arXiv e-prints},
  pages = {arXiv:2604.03365},
  doi = {10.48550/arXiv.2604.03365},
  langid = {english}
}

@misc{newville_2025,
  title = {{{LMFIT}}: {{Non-Linear Least-Squares Minimization}} and {{Curve-Fitting}} for {{Python}}},
  author = {Newville, Matthew and Otten, Renee and Nelson, Andrew and Stensitzki, Till and Ingargiola, Antonino and Allan, Daniel and Fox, Austin and Carter, Faustin and Rawlik, Michal},
  year = 2025,
  month = jul,
  doi = {10.5281/zenodo.16175987},
  howpublished = {Zenodo}
}

@article{nielsen_2021,
  title = {{{PBjam}}: {{A Python Package}} for {{Automating Asteroseismology}} of {{Solar-like Oscillators}}},
  author = {Nielsen, M. B. and Davies, G. R. and Ball, W. H. and Lyttle, A. J. and Li, T. and Hall, O. J. and Chaplin, W. J. and Gaulme, P. and Carboneau, L. and Ong, J. M. J. and Garc{\'i}a, R. A. and Mosser, B. and Roxburgh, I. W. and Corsaro, E. and Benomar, O. and Moya, A. and Lund, M. N.},
  year = 2021,
  month = feb,
  journal = {ApJ},
  volume = {161},
  number = {2},
  pages = {62},
  issn = {0004-6256},
  doi = {10.3847/1538-3881/abcd39},
  langid = {english}
}

@article{nielsen_2025,
  title = {Asteroseismology with {{PBjam}} 2.0: {{Measuring Dipole Mode Frequencies}} in {{Coupling Regimes}} from {{Main-sequence}} to {{Low-luminosity Red Giant Stars}}},
  author = {Nielsen, M. B. and Ong, J. M. J. and Hatt, E. J. and Davies, G. R. and Chaplin, W. J. and Hookway, G. T. and Stokholm, A. and Scutt, O. J. and Lund, M. N. and Garc{\'i}a, R. A.},
  year = 2025,
  month = jun,
  journal = {ApJ},
  volume = {169},
  number = {6},
  pages = {322},
  issn = {0004-6256},
  doi = {10.3847/1538-3881/adcb37},
  langid = {english}
}

@article{ong_2020,
  title = {Semianalytic {{Expressions}} for the {{Isolation}} and {{Coupling}} of {{Mixed Modes}}},
  author = {Ong, J. M. Joel and Basu, Sarbani},
  year = 2020,
  month = aug,
  journal = {ApJ},
  volume = {898},
  number = {2},
  pages = {127},
  issn = {0004-637X},
  doi = {10.3847/1538-4357/ab9ffb},
  langid = {english}
}

@article{ong_2023,
  title = {Mode {{Mixing}} and {{Rotational Splittings}}. {{II}}. {{Reconciling Different Approaches}} to {{Mode Coupling}}},
  author = {Ong, J. M. Joel and Gehan, Charlotte},
  year = 2023,
  month = apr,
  journal = {ApJ},
  volume = {946},
  number = {2},
  pages = {92},
  issn = {0004-637X},
  doi = {10.3847/1538-4357/acbf2f},
  langid = {english}
}

@article{ong_2025,
  title = {Resolving an {{Asteroseismic Catastrophe}}: {{Structural Diagnostics}} from p-Mode {{Phase Functions}} off the {{Main Sequence}}},
  author = {Ong, J. M. Joel and Lindsay, Christopher J. and Reyes, Claudia and Stello, Dennis and Roxburgh, Ian W.},
  year = 2025,
  month = feb,
  journal = {ApJ},
  volume = {980},
  number = {2},
  pages = {199},
  publisher = {The American Astronomical Society},
  issn = {0004-637X},
  doi = {10.3847/1538-4357/ada949},
  langid = {english}
}

@article{peirce_1852,
  title = {Criterion for the Rejection of Doubtful Observations},
  author = {Peirce, Benjamin},
  year = 1852,
  month = jul,
  journal = {ApJ},
  volume = {2},
  pages = {161--163},
  issn = {0004-6256},
  doi = {10.1086/100259},
  langid = {english}
}

@article{perdelwitz_2024,
  title = {Analysis of the Public {{HARPS}}/{{ESO}} Spectroscopic Archive. {{Ca II H}}\&amp;{{K}} Time Series for the {{HARPS}} Radial Velocity Database},
  author = {Perdelwitz, V. and Trifonov, T. and Teklu, J. T. and Sreenivas, K. R. and {Tal-Or}, L.},
  year = 2024,
  month = mar,
  journal = {A\&A},
  volume = {683},
  pages = {A125},
  issn = {0004-6361},
  doi = {10.1051/0004-6361/202348263},
  langid = {english}
}

@article{rauer_2025,
  title = {The {{PLATO}} Mission},
  author = {Rauer, Heike and Aerts, Conny and Cabrera, Juan and Deleuil, Magali and Erikson, Anders and Gizon, Laurent and Goupil, Mariejo and Heras, Ana and Walloschek, Thomas and {Lorenzo-Alvarez}, Jose and Marliani, Filippo and {Martin-Garcia}, C{\'e}sar and {Mas-Hesse}, J. Miguel and O'Rourke, Laurence and Osborn, Hugh and Pagano, Isabella and Piotto, Giampaolo and Pollacco, Don and Ragazzoni, Roberto and Ramsay, Gavin and Udry, St{\'e}phane and Appourchaux, Thierry and Benz, Willy and Brandeker, Alexis and G{\"u}del, Manuel and {Janot-Pacheco}, Eduardo and Kabath, Petr and Kjeldsen, Hans and Min, Michiel and Santos, Nuno and Smith, Alan and Suarez, Juan-Carlos and Werner, Stephanie C. and Aboudan, Alessio and Abreu, Manuel and Acu{\~n}a, Lorena and Adams, Moritz and Adibekyan, Vardan and Affer, Laura and Agneray, Fran{\c c}ois and Agnor, Craig and {Aguirre B{\o}rsen-Koch}, Victor and Ahmed, Saad and Aigrain, Suzanne and {Al-Bahlawan}, Ashraf and Alcacera Gil, Ma de los Angeles and Alei, Eleonora and Alencar, Silvia and Alexander, Richard and {Alfonso-Garz{\'o}n}, Julia and Alibert, Yann and Allende Prieto, Carlos and Almeida, Leonardo and Alonso Sobrino, Roi and Altavilla, Giuseppe and Althaus, Christian and Alvarez Trujillo, Luis Alonso and Amarsi, Anish and {Ammler-von Eiff}, Matthias and Am{\^o}res, Eduardo and Andrade, Laerte and {Antoniadis-Karnavas}, Alexandros and Ant{\'o}nio, Carlos and {Aparicio del Moral}, Beatriz and Appolloni, Matteo and Arena, Claudio and Armstrong, David and Aroca Aliaga, Jose and Asplund, Martin and Audenaert, Jeroen and Auricchio, Natalia and Avelino, Pedro and Baeke, Ann and Bailli{\'e}, Kevin and Balado, Ana and Ballber Balaguer{\'o}, Pau and Balestra, Andrea and Ball, Warrick and Ballans, Herve and Ballot, Jerome and Barban, Caroline and Barbary, Ga{\"e}le and Barbieri, Mauro and Barcel{\'o} Forteza, Sebasti{\`a} and Barker, Adrian and Barklem, Paul and Barnes, Sydney and Barrado Navascues, David and Barragan, Oscar and Baruteau, Cl{\'e}ment and Basu, Sarbani and Baudin, Frederic and Baumeister, Philipp and Bayliss, Daniel and Bazot, Michael and Beck, Paul G. and Belkacem, Kevin and Bellinger, Earl and Benatti, Serena and Benomar, Othman and B{\'e}rard, Diane and Bergemann, Maria and Bergomi, Maria and Bernardo, Pierre and Biazzo, Katia and Bignamini, Andrea and Bigot, Lionel and Billot, Nicolas and Binet, Martin and Biondi, David and Biondi, Federico and Birch, Aaron C. and Bitsch, Bertram and Bluhm Ceballos, Paz Victoria and B{\'o}di, Attila and Bogn{\'a}r, Zs{\'o}fia and Boisse, Isabelle and Bolmont, Emeline and Bonanno, Alfio and Bonavita, Mariangela and Bonfanti, Andrea and Bonfils, Xavier and Bonito, Rosaria and Bonomo, Aldo Stefano and B{\"o}rner, Anko and Boro Saikia, Sudeshna and Borreguero Mart{\'i}n, Elisa and Borsa, Francesco and Borsato, Luca and Bossini, Diego and Bouchy, Francois and Bou{\'e}, Gwena{\"e}l and Boufleur, Rodrigo and Boumier, Patrick and Bourrier, Vincent and Bowman, Dominic M. and Bozzo, Enrico and Bradley, Louisa and Bray, John and Bressan, Alessandro and Breton, Sylvain and Brienza, Daniele and Brito, Ana and Brogi, Matteo and Brown, Beverly and Brown, David J. A. and Brun, Allan Sacha and Bruno, Giovanni and Bruns, Michael and Buchhave, Lars A. and Bugnet, Lisa and Buldgen, Ga{\"e}l and Burgess, Patrick and Busatta, Andrea and Busso, Giorgia and Buzasi, Derek and Caballero, Jos{\'e} A. and Cabral, Alexandre and Cabrero Gomez, Juan-Francisco and Calderone, Flavia and Cameron, Robert and Cameron, Andrew and Campante, Tiago and Campos Gestal, N{\'e}stor and Canto Martins, Bruno Leonardo and Cara, Christophe and Carone, Ludmila and Carrasco, Josep Manel and Casagrande, Luca and Casewell, Sarah L. and Cassisi, Santi and Castellani, Marco and Castro, Matthieu and Catala, Claude and Catal{\'a}n Fern{\'a}ndez, Irene and Catelan, M{\'a}rcio and Cegla, Heather and Cerruti, Chiara and Cessa, Virginie and Chadid, Merieme and Chaplin, William and Charpinet, Stephane and Chiappini, Cristina and Chiarucci, Simone and Chiavassa, Andrea and Chinellato, Simonetta and Chirulli, Giovanni and {Christensen-Dalsgaard}, J{\o}rgen and Church, Ross and Claret, Antonio and Clarke, Cathie and Claudi, Riccardo and Clermont, Lionel and Coelho, Hugo and Coelho, Joao and Cogato, Fabrizio and Colom{\'e}, Josep and Condamin, Mathieu and Conde Garc{\'i}a, Fernando and Conseil, Simon and Corbard, Thierry and Correia, Alexandre C. M. and Corsaro, Enrico and Cosentino, Rosario and Costes, Jean and Cottinelli, Andrea and Covone, Giovanni and Creevey, Orlagh L. and Crida, Aurelien and Csizmadia, Szilard and Cunha, Margarida and Curry, Patrick and {da Costa}, Jefferson and {da Silva}, Francys and Dalal, Shweta and Damasso, Mario and Damiani, Cilia and Damiani, Francesco and {das Chagas}, Maria Liduina and Davies, Melvyn and Davies, Guy and Davies, Ben and Davison, Gary and {de Almeida}, Leandro and {de Angeli}, Francesca and {de Barros}, Susana Cristina Cabral and {de CastroLe{\~a}o}, Izan and {de Freitas}, Daniel Brito and {de Freitas}, Marcia Cristina and De Martino, Domitilla and {de Medeiros}, Jos{\'e} Renan and {de Paula}, Luiz Alberto and {de Pedraza G{\'o}mez}, {\'A}lvaro and {de Plaa}, Jelle and De Ridder, Joris and Deal, Morgan and Decin, Leen and Deeg, Hans and Degl'Innocenti, Scilla and Deheuvels, Sebastien and {del Burgo}, Carlos and Del Sordo, Fabio and {Delgado-Mena}, Elisa and Demangeon, Olivier and Denk, Tilmann and Derekas, Aliz and Desert, Jean-Michel and Desidera, Silvano and Dexet, Marc and Di Criscienzo, Marcella and Di Giorgio, Anna Maria and Di Mauro, Maria Pia and Diaz Rial, Federico Jose and {D{\'i}az-Garc{\'i}a}, Jos{\'e}-Javier and Dima, Marco and Dinuzzi, Giacomo and Dionatos, Odysseas and Distefano, Elisa and {do Nascimento Jr.}, Jose-Dias and Domingo, Albert and D'Orazi, Valentina and Dorn, Caroline and Doyle, Lauren and Duarte, Elena and Ducellier, Florent and Dumaye, Luc and Dumusque, Xavier and Dupret, Marc-Antoine and Eggenberger, Patrick and Ehrenreich, David and Eigm{\"u}ller, Philipp and Eising, Johannes and Emilio, Marcelo and Eriksson, Kjell and Ermocida, Marco and Escate Giribaldi, Riano Isidoro and Eschen, Yoshi and Espinosa Y{\'a}{\~n}ez, Luc{\'i}a and Estrela, In{\^e}s and Evans, Dafydd Wyn and Fabbian, Damian and Fabrizio, Michele and Faria, Jo{\~a}o Pedro and Farina, Maria and Farinato, Jacopo and Feliz, Dax and Feltzing, Sofia and Fenouillet, Thomas and Fern{\'a}ndez, Miguel and Ferrari, Lorenza and {Ferraz-Mello}, Sylvio and Fialho, Fabio and Fienga, Agnes and Figueira, Pedro and Fiori, Laura and Flaccomio, Ettore and Focardi, Mauro and Foley, Steve and Fontignie, Jean and Ford, Dominic},
  year = 2025,
  month = apr,
  journal = {Exp. Astron.},
  volume = {59},
  number = {3},
  pages = {26},
  issn = {1572-9508},
  doi = {10.1007/s10686-025-09985-9},
  langid = {english}
}

@article{regulo_2016,
  title = {Magnetic Activity Cycles in Solar-like Stars: {{The}} Cross-Correlation Technique of p-Mode Frequency Shifts},
  author = {R{\'e}gulo, C. and Garc{\'i}a, R. A. and Ballot, J.},
  year = 2016,
  month = may,
  journal = {A\&A},
  volume = {589},
  pages = {A103},
  publisher = {EDP Sciences},
  issn = {0004-6361, 1432-0746},
  doi = {10.1051/0004-6361/201425408},
  copyright = {\copyright{} ESO, 2016},
  langid = {english}
}

@article{ricker_2014,
  title = {Transiting {{Exoplanet Survey Satellite}} ({{TESS}})},
  author = {Ricker, George R. and Winn, Joshua N. and Vanderspek, Roland and Latham, David W. and Bakos, G{\'a}sp{\'a}r {\'A} and Bean, Jacob L. and {Berta-Thompson}, Zachory K. and Brown, Timothy M. and Buchhave, Lars and Butler, Nathaniel R. and Butler, R. Paul and Chaplin, William J. and Charbonneau, David and {Christensen-Dalsgaard}, J{\o}rgen and Clampin, Mark and Deming, Drake and Doty, John and De Lee, Nathan and Dressing, Courtney and Dunham, E. W. and Endl, Michael and Fressin, Francois and Ge, Jian and Henning, Thomas and Holman, Matthew J. and Howard, Andrew W. and Ida, Shigeru and Jenkins, Jon and Jernigan, Garrett and Johnson, John A. and Kaltenegger, Lisa and Kawai, Nobuyuki and Kjeldsen, Hans and Laughlin, Gregory and Levine, Alan M. and Lin, Douglas and Lissauer, Jack J. and MacQueen, Phillip and Marcy, Geoffrey and McCullough, P. R. and Morton, Timothy D. and Narita, Norio and Paegert, Martin and Palle, Enric and Pepe, Francesco and Pepper, Joshua and Quirrenbach, Andreas and Rinehart, S. A. and Sasselov, Dimitar and Sato, Bun'ei and Seager, Sara and Sozzetti, Alessandro and Stassun, Keivan G. and Sullivan, Peter and Szentgyorgyi, Andrew and Torres, Guillermo and Udry, Stephane and Villasenor, Joel},
  year = 2014,
  month = aug,
  journal = {Proceedings of the SPIE},
  volume = {9143},
  pages = {914320},
  issn = {0277-786X},
  doi = {10.1117/12.2063489},
  langid = {english}
}

@article{samadi_2001,
  title = {Excitation of Stellar P-Modes by Turbulent Convection. {{I}}. {{Theoretical}} Formulation},
  author = {Samadi, R. and Goupil, M. -J.},
  year = 2001,
  month = apr,
  journal = {A\&A},
  volume = {370},
  pages = {136--146},
  issn = {0004-6361},
  doi = {10.1051/0004-6361:20010212}
}

@article{santos_2019,
  title = {Signatures of Magnetic Activity: {{On}} the Relation between Stellar Properties and p-Mode Frequency Variations},
  author = {Santos, A. R. G. and Campante, T. L. and Chaplin, W. J. and Cunha, M. S. and {van Saders}, J. L. and Karoff, C. and Metcalfe, T. S. and Mathur, S. and Garc{\'i}a, R. A. and Lund, M. N. and Kiefer, R. and Silva Aguirre, V. and Davies, G. R. and Howe, R. and Elsworth, Y.},
  year = 2019,
  month = sep,
  journal = {ApJ},
  volume = {883},
  pages = {65},
  publisher = {IOP},
  issn = {0004-637X},
  doi = {10.3847/1538-4357/ab397a}
}

@article{scargle_1982,
  title = {Studies in Astronomical Time Series Analysis. {{II}}. {{Statistical}} Aspects of Spectral Analysis of Unevenly Spaced Data.},
  author = {Scargle, J. D.},
  year = 1982,
  month = dec,
  journal = {ApJ},
  volume = {263},
  pages = {835--853},
  issn = {0004-637X},
  doi = {10.1086/160554},
  langid = {english}
}

@article{shibahashi_1979,
  title = {Modal {{Analysis}} of {{Stellar Nonradial Oscillations}} by an {{Asymptotic Method}}},
  author = {Shibahashi, Hiromoto},
  year = 1979,
  month = mar,
  journal = {PASJ},
  volume = {31},
  number = {1},
  pages = {87--104},
  issn = {0004-6264},
  doi = {10.1093/pasj/31.1.87},
  langid = {english}
}

@article{Skilling04,
	author = {Skilling, J.},
	doi = {http://dx.doi.org/10.1063/1.1835238},
	journal = {AIP Conference Proceedings},
	number = {1},
	pages = {395 (SK04)-405},
	title = {Nested Sampling},
	url = {http://scitation.aip.org/content/aip/proceeding/aipcp/10.1063/1.1835238},
	volume = {735},
	year = {2004}
    }

@article{silvaaguirre_2017,
  title = {Standing on the {{Shoulders}} of {{Dwarfs}}: The {{Kepler Asteroseismic LEGACY Sample}}. {{II}}.{{Radii}}, {{Masses}}, and {{Ages}}},
  author = {Silva Aguirre, V{\'i}ctor and Lund, Mikkel N. and Antia, H. M. and Ball, Warrick H. and Basu, Sarbani and {Christensen-Dalsgaard}, J{\o}rgen and Lebreton, Yveline and Reese, Daniel R. and Verma, Kuldeep and Casagrande, Luca and Justesen, Anders B. and Mosumgaard, Jakob R. and Chaplin, William J. and Bedding, Timothy R. and Davies, Guy R. and Handberg, Rasmus and Houdek, G{\"u}nter and Huber, Daniel and Kjeldsen, Hans and Latham, David W. and White, Timothy R. and Coelho, Hugo R. and Miglio, Andrea and Rendle, Ben},
  year = 2017,
  month = feb,
  journal = {ApJ},
  volume = {835},
  pages = {173},
  publisher = {IOP},
  issn = {0004-637X},
  doi = {10.3847/1538-4357/835/2/173}
}

@article{soubiran_2022,
  title = {Assessment of [{{Fe}}/{{H}}] Determinations for {{FGK}} Stars in Spectroscopic Surveys},
  author = {Soubiran, C. and Brouillet, N. and Casamiquela, L.},
  year = 2022,
  month = jul,
  journal = {A\&A},
  volume = {663},
  pages = {A4},
  issn = {0004-6361},
  doi = {10.1051/0004-6361/202142409},
  langid = {english}
}

@article{soubiran_2024,
  title = {Gaia {{FGK}} Benchmark Stars: {{Fundamental T}}{\textsubscript{eff}} and Log g of the Third Version},
  author = {Soubiran, C. and Creevey, O. L. and Lagarde, N. and Brouillet, N. and Jofr{\'e}, P. and Casamiquela, L. and Heiter, U. and {Aguilera-G{\'o}mez}, C. and Vitali, S. and Worley, C. and {de Brito Silva}, D.},
  year = 2024,
  month = feb,
  journal = {A\&A},
  volume = {682},
  pages = {A145},
  issn = {0004-6361},
  doi = {10.1051/0004-6361/202347136},
  langid = {english}
}

@article{speagle_2020,
  title = {{{DYNESTY}}: A Dynamic Nested Sampling Package for Estimating {{Bayesian}} Posteriors and Evidences},
  author = {Speagle, Joshua S.},
  year = 2020,
  month = apr,
  journal = {MNRAS},
  volume = {493},
  number = {3},
  pages = {3132--3158},
  issn = {0035-8711},
  doi = {10.1093/mnras/staa278},
  langid = {english}
}

@article{tassoul_1980,
  title = {Asymptotic Approximations for Stellar Nonradial Pulsations.},
  author = {Tassoul, M.},
  year = 1980,
  month = aug,
  journal = {ApJS},
  volume = {43},
  pages = {469--490},
  issn = {0067-0049},
  doi = {10.1086/190678},
  langid = {english}
}

@article{themessl_2018,
  title = {Oscillating Red Giants in Eclipsing Binary Systems: Empirical Reference Value for Asteroseismic Scaling Relation},
  author = {Theme{\ss}l, N. and Hekker, S. and Southworth, J. and Beck, P. G. and Pavlovski, K. and Tkachenko, A. and Angelou, G. C. and Ball, W. H. and Barban, C. and Corsaro, E. and Elsworth, Y. and Handberg, R. and Kallinger, T.},
  year = 2018,
  month = aug,
  journal = {MNRAS},
  volume = {478},
  number = {4},
  pages = {4669--4696},
  issn = {0035-8711},
  doi = {10.1093/mnras/sty1113},
  langid = {english}
}

@misc{team_2024,
  title = {Pandas-Dev/Pandas: {{Pandas}}},
  author = {{The pandas development team}},
  year = 2024,
  month = sep,
  doi = {10.5281/zenodo.13819579},
  howpublished = {Zenodo}
}

@article{vanderwalt_2011,
  title = {The {{NumPy Array}}: {{A Structure}} for {{Efficient Numerical Computation}}},
  author = {{van der Walt}, St{\'e}fan and Colbert, S. Chris and Varoquaux, Ga{\"e}l},
  year = 2011,
  month = mar,
  journal = {CiSE},
  volume = {13},
  pages = {22--30},
  doi = {10.1109/MCSE.2011.37}
}

@article{white_2011,
  title = {Calculating {{Asteroseismic Diagrams}} for {{Solar-like Oscillations}}},
  author = {White, Timothy R. and Bedding, Timothy R. and Stello, Dennis and {Christensen-Dalsgaard}, J{\o}rgen and Huber, Daniel and Kjeldsen, Hans},
  year = 2011,
  month = dec,
  journal = {ApJ},
  volume = {743},
  number = {2},
  pages = {161},
  issn = {0004-637X},
  doi = {10.1088/0004-637X/743/2/161},
  langid = {english}
}

@article{white_2012,
  title = {Solving the {{Mode Identification Problem}} in {{Asteroseismology}} of {{F Stars Observed}} with {{Kepler}}},
  author = {White, Timothy R. and Bedding, Timothy R. and Gruberbauer, Michael and Benomar, Othman and Stello, Dennis and Appourchaux, Thierry and Chaplin, William J. and {Christensen-Dalsgaard}, J{\o}rgen and Elsworth, Yvonne P. and Garc{\'i}a, Rafael A. and Hekker, Saskia and Huber, Daniel and Kjeldsen, Hans and Mosser, Beno{\^i}t and Kinemuchi, Karen and Mullally, Fergal and Still, Martin},
  year = 2012,
  month = jun,
  journal = {ApJL},
  volume = {751},
  number = {2},
  pages = {L36},
  issn = {0004-637X},
  doi = {10.1088/2041-8205/751/2/L36},
  langid = {english}
}

\begin{appendix}

\onecolumn
\section{List of the 34 TLS in the PLATO LOP fields}
\begin{table}[H]
\centering
\caption{List of the 34 TLS in the PLATO fields.}
\label{tab:fund_params}
\setlength{\tabcolsep}{5pt} 
\begin{tabular}{llccccccccc}
\toprule
Name & TIC & $T_{\rm{eff}}$ [K] & $\log g$ & [Fe/H] & Ref. & $\nu_{\rm{max}}$ [$\mu$Hz] (I) & $\Delta \nu$ [$\mu$Hz] (I) & \#\,120s & \#\,20s & Note\\
\toprule
\multicolumn{10}{c}{Inside PLATO LOPS 2}\\
\midrule
HR$\,$3220 & 308844962 & $6483\pm33$ & $4.18\pm0.05$ & $-0.29\pm0.03$ & (7) & $1386.8 \pm 27.0$ & $71.6 \pm 0.6$ & 18 & 7 & SB \\
HD$\,$62644 & 123699670 & $5490\pm50$ & $3.82\pm0.07$ & $0.01\pm0.04$ & (6) & $708.4 \pm 6.7$ & $41.1 \pm 0.3$ & 8 & 4 & BS\\ 
HD$\,$50223 & 170225363 & $6437\pm27$ & $4.10\pm0.07$ & $-0.22\pm0.04$ & (7) & $1279.7 \pm 66.0$ & $69.0 \pm 0.9$ & 8 & 3 & BS \\ 
$\nu^2\,$Col & 32500750 & $6373\pm70$ & $3.94\pm0.10$ & $-0.11\pm0.05$ & (7) & $691.3 \pm 43.5$ & $38.9 \pm 1.3$ & 4 & 0 & \\ 
171$\,$Pup & 149672905 & $5754\pm22$ & $4.11\pm0.06$ & $-0.85\pm0.02$ & (7) & $2107.7 \pm 59.8$ & $104.4 \pm 0.9$ & 5 & 3 & SB\\ 
$\zeta\,$Pic & 219420836 & $6380\pm38$ & $4.00\pm0.07$ & $0.07\pm0.04$ & (7) & $853.0 \pm 42.4$ & $49.5 \pm 0.4$ & 7 & 2 & \\ 
HD$\,$36553 & 354552931 & $6002\pm65$ & $3.82\pm0.06$ & $0.35\pm0.03$ & (7) & $544.7 \pm 16.1$ & $33.7 \pm 0.2$ & 7 & 1 & \\ 
HD$\,$53705 & 130645536 & $5790\pm15$ & $4.33\pm0.04$ & $-0.22\pm0.03$ & (7) & $1989.9 \pm 98.4$ & $101.8 \pm 0.7$ & 8 & 5 & \\ 
HD$\,$46569 & 255630992 & $6313\pm50$ & $4.03\pm0.06$ & $0.04\pm0.05$ & (6) & $932.2 \pm 20.5$ & $49.7 \pm 0.3$ & 13 & 5 & SB\\ 
HD$\,$65907 & 372914091 & $5997\pm16$ & $4.52\pm0.03$ & $-0.31\pm0.02$ & (7) & $3006.9 \pm 165.7$ & $128.5 \pm 1.2$ & 19 & 13 & \\ 
\midrule
\multicolumn{10}{c}{Inside PLATO LOPN 1}\\
\midrule
$\chi\,$Dra & 341873045 & $6101\pm25$ & $4.29\pm0.04$ & $-0.64\pm0.04$ & (7) & $2314.7 \pm 24.4$ & $108.4 \pm 0.1$ & 38 & 16 & SB\\ 
$\theta\,$Dra & 161825882 & $6196\pm24$ & $3.96\pm0.17$ & $0.20\pm0.03$ & (7) & $723.0 \pm 15.8$ & $40.2 \pm 0.3$ & 14 & 1 & SB \\ 
$\theta\,$Cyg & 27014182 & $6914\pm33$ & $4.24\pm0.01$ & $-0.03\pm0.03$ & (4) & $1759.1 \pm 67.1$ & $82.8 \pm 1.2$ & 13 & 7 & BS, EHC (3) \\ 
$\upsilon\,$Cep & 421444084 & $6161\pm27$ & $3.86\pm0.08$  & $0.12\pm0.03$ & (7) & $958.7 \pm 22.3$ & $53.6 \pm 0.4$ & 12 & 4 & SB \\ 
$\psi^1\,$Dra$\,$A & 441804568 & $6410\pm82$ & $4.00\pm0.05$ & $0.01\pm0.03$ & (7) & $1232.4 \pm 19.8$ & $61.8 \pm 0.3$ & 39 & 18 & SB \\ 
$\sigma\,$Dra & 259237827 & $5298\pm14$ & $4.52\pm0.02$ & $-0.21\pm0.01$ & (7) & $4217.9 \pm 122.6$ & $182.2 \pm 0.5$ & 41 & 28 & \\ 
$\omega\,$Dra & 233195546 & $6595\pm28$ & $4.19\pm0.11$ & $-0.03\pm0.06$ & (2) & $1999.5 \pm 55.0$ & $88.3 \pm 2.0$ & 39 & 19 & SB \\ 
19$\,$Dra & 289622310 & $6298\pm20$ & $4.27\pm0.07$ & $-0.14\pm0.02$ & (7) & $2313.1 \pm 493.9$ & $104.8 \pm 3.5$ & 41 & 28 & SB \\ 
36$\,$Dra & 233121747 & $6473\pm38$ & $4.09\pm0.04$ & $-0.29\pm0.02$ & (7) & $1312.0 \pm 16.9$ & $69.6 \pm 0.2$ & 36 & 14 & \\ 
17$\,$Cyg & 58445695 & $6442\pm63$ & $4.17\pm0.02$ & $-0.07\pm0.05$ & (1) & $1484.4 \pm 36.1$ & $78.6 \pm 1.5$ & 5 & 3 & \\ 
35$\,$Dra & 441813918 & $6191\pm25$ & $3.82\pm0.05$ & $-0.13\pm0.04$ & (7) & $705.3 \pm 7.0$ & $42.1 \pm 0.1$ & 40 & 13 & \\ 
99$\,$Her & 22516402 & $5974\pm23$ & $4.20\pm0.03$ & $-0.55\pm0.02$ & (7) & $1950.7 \pm 41.0$ & $96.2 \pm 0.6$ & 5 & 4 & SB \\ 
HD$\,$136064 & 232563914 & $6144\pm20$ & $4.02\pm0.04$ & $-0.02\pm0.02$ & (7) & $1081.9 \pm 23.8$ &$ 59.1 \pm 0.2$ & 18 & 4 & \\ 
HD$\,$175225 & 48194330 & $5297\pm26$ & $3.77\pm0.08$ & $0.14\pm0.05$ & (7) & $752.1 \pm 11.3$ & $43.5 \pm 0.3$ & 23 & 5 & \\ 
26$\,$Dra & 219777482 & $5921\pm33$ & $4.44\pm0.04$ & $-0.04\pm0.03$ & (7) & $3059 \pm 176.8$ & $132.9 \pm 0.9$ & 40 & 27 & SB \\ 
27$\,$Cyg & 41195655 & $5136\pm27$ & $3.66\pm0.06$ & $-0.02\pm0.02$ & (7) & $702.8 \pm 44.4$ & $40.6 \pm 1.2$ & 9 & 2 & RS CVn \\ 
72$\,$Her & 9728611 & $5704\pm13$ & $4.33\pm0.03$ & $-0.39\pm0.01$ & (7) & $2241.4 \pm 85.1$ & $106.2 \pm 1.0$ & 5 & 3 &  \\ 
HD$\,$176051 & 20601206 & $5975\pm20$ & $4.54\pm0.06$ & $-0.10\pm0.02$ & (7) & $2902.2 \pm 132.1$ & $127.4 \pm 1.7$ & 8 & 6 & BS, 1 EH* (5) \\ 
68$\,$Dra & 236871353 & $6270\pm62$ & $4.00\pm0.13$ & $-0.03\pm0.04$ & (7) & $716.2 \pm 14.7$ & $40.2 \pm 0.6$ & 20 & 5 & \\ 
HD$\,$184960 & 26884478 & $6287\pm19$ & $4.29\pm0.06$ & $-0.08\pm0.03$ & (7) & $1870.6 \pm 73.9$ & $91.9 \pm 0.9$ & 13 & 11 & 1 EH (1)\\ 
HD$\,$191195 & 405902259 & $6730\pm63$ & $4.18\pm0.02$ & $-0.05\pm0.05$ & (1) & $1353.0 \pm 146.7$ & $69.2 \pm 2.3$ & 11 & 6 & \\  
HD$\,$193664 & 403585118 & $5930\pm18$ & $4.48\pm0.03$ & $-0.10\pm0.02$ & (8) & $3174.6 \pm 201.7$ & $138.5 \pm 2.8$ & 26 & 17 & \\ 
16$\,$Cyg$\,$A & 27533341 & $5791\pm9$ & $4.29\pm0.02$ & $0.08\pm0.01$ & (7) & $2236.5 \pm 97.9$ & $103.5 \pm 1.1$ & 12 & 10 & BS \\ 
HD$\,$152303 & 233503400 & $6573\pm80$ & - & - & (2) & $1689.5 \pm 49.6$ & $82.2 \pm 1.5$ & 34 & 14 & \\
\bottomrule
\end{tabular}
\tablebib{
(1) \citet{barnes_2023a}; 
(2) \citet{casagrande_2011}; 
(3) \citet{desort_2009}; 
(4) \citet{koleva_2012}; 
(5) \citet{muterspaugh_2010}; 
(6) \citet{perdelwitz_2024}; 
(7) \citet{soubiran_2022} ; 
(8) \citet{soubiran_2024}; 
}
\tablefoot{Spectroscopic parameters are compiled from the literature (sources being listed in the Ref. column), while global seismic parameters are taken from Paper~I. The number of TESS observing sectors is reported separately for each cadence, and the notes indicate system multiplicity and some stellar properties. Values of $\nu_\mathrm{max}$ and $\Delta\nu$ were computed with the \texttt{pySYD} pipeline in Paper~I \citep{chontos_2022, lund_2025}. Abbreviations: RS CVn: RS Canum Venaticorum–type variable; BS: binary or multiple system; SB: spectroscopic binary; EH: confirmed exoplanet host (* indicates a confirmed planet in the system but not necessarily orbiting the target star); EHC: exoplanet host candidate. 
\#\,120s: number of sectors observed at 120-s cadence; \#\,20s: number of sectors observed at 20-s cadence.}
\end{table}

\section{Background parameters from the \texttt{apollinaire} fitting procedure}

\begin{sidewaystable}
\centering
\caption{Background parameters from the \texttt{apollinaire} fitting procedure. }
\label{tab:back_params}
\setlength{\tabcolsep}{3pt}
\begin{tabular}{lccccccccccc}
\toprule
\multirow{2}{*}{Name} & \multirow{2}{*}{version} & \multirow{2}{*}{$n_{\mathcal{H}}$} & low cut & $A_{\mathcal{H}_1}$ & $\nu_{\rm{c,\mathcal{H}}_1}$ & $A_{\mathcal{H}_2}$ & $\nu_{\rm{c,\mathcal{H}}_2}$ & $H_{\mathrm{max}}$ & $\nu_{\mathrm{max}}$ & $W_{\mathrm{env}}$ & $P_\mathrm{n}$ \\
 & & & [$\mu$Hz] & [ppm$^2$/$\mu$Hz] & [$\mu$Hz] & [ppm$^2$/$\mu$Hz] & [$\mu$Hz] & [ppm$^2$/$\mu$Hz] & [$\mu$Hz] & [$\mu$Hz] & [ppm$^2$/$\mu$Hz] \\
\midrule
HD$\,$36553 & 20s \& 120s (K2P$^2$) & 2 & 80 & 12.81$\pm$0.79 & 203$\pm$15.63 & 4.4$\pm$0.77 & 602.72$\pm$33.94 & 8.08$\pm$0.57 & 573.42$\pm$6.2 & 116.25$\pm$12.71 & 3.77$\pm$0.02 \\
$\theta\,$Dra & 20s \& 120s (K2P$^2$) & 1 & 200 & 14.37$\pm$2.48 & 157.7$\pm$16.82 & - & - & 3.69$\pm$0.14 & 616.66$\pm$14.37 & 334.53$\pm$14.94 & 1.04$\pm$0.01 \\
$\nu^2\,$Col & 120s (K2P$^2$) & 1 & 80 & 20.65$\pm$1.9 & 167.05$\pm$12.66 & - & - & 3.69$\pm$0.35 & 670.01$\pm$14.92 & 207.43$\pm$26.67 & 3.4$\pm$0.02 \\
HD$\,$62644 & 20s \& 120s (K2P$^2$) & 2 & 120 & 21.24$\pm$2.61 & 165.55$\pm$9.22 & 4.77$\pm$0.23 & 637.36$\pm$14.37 & 10.03$\pm$0.49 & 704.17$\pm$2.64 & 84.65$\pm$4.32 & 1.84$\pm$0.01 \\
68$\,$Dra & 20s \& 120s (K2P$^2$) & 2 & 80 & 12.18$\pm$0.35 & 224.04$\pm$7.33 & 2.34$\pm$0.26 & 750.72$\pm$33.91 & 4$\pm$0.24 & 711.83$\pm$7.26 & 157.44$\pm$13.89 & 4.42$\pm$0.01 \\
35$\,$Dra & 20s (K2P$^2$) & 2 & 80 & 6.87$\pm$0.25 & 247.13$\pm$9.85 & 1.51$\pm$0.27 & 811.48$\pm$43.61 & 4.16$\pm$0.2 & 712.86$\pm$7.49 & 202.06$\pm$9.76 & 0.99$\pm$0 \\
27$\,$Cyg & 20s \& 120s (K2P$^2$) & 2 & 100 & 16.57$\pm$1.02 & 184.19$\pm$8.06 & 4.32$\pm$0.32 & 587.6$\pm$23.92 & 1.77$\pm$0.27 & 730.25$\pm$10.47 & 99.29$\pm$16.09 & 2.78$\pm$0.01 \\
HD$\,$175225 & 20s \& 120s (K2P$^2$) & 2 & 150 & 16.57$\pm$1.36 & 194.79$\pm$7.81 & 4.2$\pm$0.16 & 681.55$\pm$14.57 & 3.93$\pm$0.19 & 776.55$\pm$3.65 & 108.76$\pm$6.45 & 2.87$\pm$0.01 \\
$\zeta\,$Pic & 20s \& 120s (SPOC) & 1 & 100 & 8.7$\pm$0.47 & 253.48$\pm$13.12 & - & - & 2.79$\pm$0.17 & 875.65$\pm$10.61 & 232.31$\pm$16.11 & 2.86$\pm$0.01 \\
HD$\,$46569 & 20s (K2P$^2$) & 2 & 80 & 4.74$\pm$0.23 & 329.54$\pm$17.52 & 0.86$\pm$0.22 & 1115.27$\pm$112.38 & 2.84$\pm$0.21 & 929.85$\pm$10.82 & 204.93$\pm$17.81 & 1.7$\pm$0.01 \\
$\upsilon\,$Cep & 20s \& 120s (K2P$^2$) & 1 & 100 & 3.72$\pm$0.14 & 267.62$\pm$10.3 & - & - & 0.81$\pm$0.04 & 1024.59$\pm$13.09 & 333.82$\pm$22.61 & 1.04$\pm$0 \\
HD$\,$136064 & 20s (K2P$^2$) & 1 & 80 & 6.94$\pm$0.34 & 274.13$\pm$12.23 & - & - & 2.2$\pm$0.11 & 1079.18$\pm$10.76 & 279.65$\pm$15.37 & 1.13$\pm$0.01 \\
$\psi^1\,$Dra$\,$A & 20s (K2P$^2$) & 2 & 100 & 1.5$\pm$0.04 & 405.91$\pm$10.95 & 0.42$\pm$0.04 & 1370.37$\pm$42.39 & 0.51$\pm$0.03 & 1248.93$\pm$15.63 & 281.1$\pm$24.38 & 0.6$\pm$0 \\
HD$\,$191195 & 20s (K2P$^2$) & 2 & 80 & 1.38$\pm$0.77 & 156.27$\pm$58.61 & 2$\pm$0.22 & 538.58$\pm$46.02 & 1.26$\pm$0.07 & 1330.31$\pm$63.18 & 701.97$\pm$89.25 & 2.2$\pm$0.01 \\
HD 50223 & 20s \& 120s (K2P$^2$) & 1 & 100 & 3.21$\pm$0.13 & 417.61$\pm$19.6 & - & - & 1.08$\pm$0.09 & 1339.29$\pm$19.33 & 315.7$\pm$34.89 & 2.05$\pm$0.01 \\
36$\,$Dra & 20s \& 120s (K2P$^2$) & 1 & 150 & 3.41$\pm$0.07 & 439.96$\pm$10.39 & - & - & 1.5$\pm$0.04 & 1354.15$\pm$6.61 & 348.77$\pm$10.2 & 1.84$\pm$0.01 \\
HR$\,$3220 & 20s \& 120s (SPOC) & 2 & 50 & 11.27$\pm$1.15 & 77.63$\pm$3.84 & 2.77$\pm$0.07 & 501.22$\pm$9.94 & 0.92$\pm$0.03 & 1368.51$\pm$19.23 & 529.94$\pm$29.16 & 1.58$\pm$0.01 \\
HD$\,$152303 & 20s (K2P$^2$) & 2 & 55 & 5.06$\pm$0.49 & 102.7$\pm$6.56 & 1.54$\pm$0.08 & 545.14$\pm$29.12 & 0.63$\pm$0.04 & 1544.53$\pm$113.29 & 986.96$\pm$180.56 & 2.63$\pm$0.01 \\
17$\,$Cyg & 20s (K2P$^2$) & 1 & 50 & 2.75$\pm$0.12 & 423$\pm$21.11 & - & - & 0.78$\pm$0.07 & 1584.21$\pm$28.46 & 408.98$\pm$43.87 & 1.16$\pm$0.01 \\
$\theta\,$Cyg & 20s (K2P$^2$) & 2 & 80 & 3.38$\pm$0.28 & 119.57$\pm$5.1 & 1.66$\pm$0.03 & 651.46$\pm$9.12 & 0.71$\pm$0.01 & 1657.55$\pm$15.6 & 764.84$\pm$22.27 & 0.75$\pm$0 \\
HD$\,$184960 & 20s (SPOC) & 2 & 150 & 1.21$\pm$0.06 & 600.87$\pm$33.95 & 0.3$\pm$0.06 & 2178.79$\pm$304.45 & 0.68$\pm$0.08 & 1857.74$\pm$34.34 & 392.68$\pm$63.82 & 2.05$\pm$0.02 \\
$\omega\,$Dra & 20s (K2P$^2$) & 1 & 150 & 1.04$\pm$0.03 & 576.89$\pm$19.65 & - & - & 0.27$\pm$0.02 & 1936.96$\pm$26.64 & 539.44$\pm$42.35 & 0.9$\pm$0 \\
99$\,$Her & 20s (K2P$^2$) & 2 & 120 & 1.44$\pm$0.07 & 483.11$\pm$20.74 & 0.29$\pm$0.03 & 2243.33$\pm$319.22 & 0.51$\pm$0.07 & 1998.75$\pm$30.43 & 306.28$\pm$53.06 & 1.1$\pm$0.02 \\
HD$\,$53705 & 20s \& 120s (K2P$^2$) & 2 & 100 & 0.99$\pm$0.11 & 501.02$\pm$41.67 & 0.36$\pm$0.11 & 1239.21$\pm$194.67 & 0.37$\pm$0.06 & 2037.73$\pm$41.69 & 303.96$\pm$54.09 & 1.75$\pm$0.01 \\
171$\,$Pup & 20s (K2P$^2$) & 2 & 100 & 1.09$\pm$0.11 & 519.8$\pm$54.62 & 0.28$\pm$0.12 & 1612.01$\pm$456.68 & 0.63$\pm$0.09 & 2133.05$\pm$33.66 & 316.5$\pm$59.86 & 1.37$\pm$0.02 \\
16$\,$Cyg$\,$A & 20s (K2P$^2$) & 1 & 150 & 1.34$\pm$0.06 & 485.2$\pm$24.23 & - & - & 0.28$\pm$0.03 & 2159.18$\pm$44.16 & 473.23$\pm$65.73 & 1.34$\pm$0.01 \\
72$\,$Her & 20s (SPOC) & 1 & 200 & 3.1$\pm$0.23 & 484.6$\pm$32.79 & - & - & 0.5$\pm$0.11 & 2186.25$\pm$40.77 & 242.2$\pm$48.81 & 1.53$\pm$0.01 \\
$\chi\,$Dra & 20s (K2P$^2$) & 2 & 80 & 0.65$\pm$0.02 & 598.57$\pm$14.38 & 0.16$\pm$0.01 & 1966.68$\pm$212.16 & 0.35$\pm$0.02 & 2336.97$\pm$10.83 & 437.2$\pm$28.17 & 0.36$\pm$0 \\
19$\,$Dra & 20s (K2P$^2$) & 1 & 100 & 0.99$\pm$0.02 & 615.67$\pm$15.08 & - & - & 0.14$\pm$0.01 & 2706.51$\pm$49.8 & 763.94$\pm$99.3 & 0.97$\pm$0 \\
HD$\,$176051 & 20s (K2P$^2$) & 1 & 100 & 1.06$\pm$0.05 & 564.68$\pm$38.56 & - & - & 0.18$\pm$0.04 & 2834.23$\pm$173.05 & 749.66$\pm$387.7 & 1.37$\pm$0.02 \\
HD$\,$65907 & 20s (K2P$^2$) & 2 & 80 & 0.61$\pm$0.04 & 564.2$\pm$40.72 & 0.25$\pm$0.04 & 1895.19$\pm$224.65 & 0.19$\pm$0.05 & 3055.34$\pm$84.01 & 412.51$\pm$146.14 & 1.71$\pm$0.01 \\
HD$\,$193664 & 20s (K2P$^2$) & 2 & 80 & 0.53$\pm$0.17 & 686.15$\pm$234.43 & 0.25$\pm$0.18 & 1529.04$\pm$799.34 & 0.19$\pm$0.07 & 3062.17$\pm$338.21 & 1310.4$\pm$641.45 & 2.13$\pm$0.04 \\
26$\,$Dra & 20s (K2P$^2$) & 1 & 80 & 0.83$\pm$0.02 & 649.78$\pm$23.75 & - & - & 0.04$\pm$0.02 & 3272.71$\pm$251.63 & 899.35$\pm$405.33 & 1.18$\pm$0.01 \\
$\sigma\,$Dra & 20s (SPOC) & 2 & 80 & 0.56$\pm$0.01 & 646.82$\pm$15.29 & 0.18$\pm$0.02 & 3058.52$\pm$113.3 & 0.09$\pm$0.02 & 4611.55$\pm$118.33 & 934.5$\pm$89.85 & 0.64$\pm$0.02 \\
\bottomrule
\end{tabular}
\tablefoot{The table lists the amplitudes and central frequencies of the Harvey components, the Gaussian parameters of the oscillation global pattern, and the noise level. Results correspond to the optimal light curve indicated in the version column. The number of Harvey profiles ($n_\mathrm{\mathcal{H}}$) and the low-frequency cut-off applied before fitting are also reported.}
\end{sidewaystable}

\twocolumn
\section{Mode identification}
\label{app:modes_id}

The identification of oscillation modes is a crucial step in asteroseismic analyses, as it establishes the link between observed spectral features and their physical interpretation. Accurate mode identification enables the assignment of angular degree and radial order to individual oscillation modes, which is essential for subsequent seismic modelling and inference of stellar properties. Depending on the evolutionary stage and oscillation characteristics of the star, different strategies may be required to reliably identify modes, particularly in the presence of mixed modes, broadened linewidths, or overlapping ridges in the power spectrum.

\subsection{Pure p modes}

For MS stars with clearly resolved ridges in the power spectrum, all pipelines adopt the standard asymptotic expression \citep[e.g.][]{tassoul_1980, mosser_2011, lund_2017, breton_2022}. The frequencies of $\ell=0$ and $\ell=2$ modes are given by
\begin{equation}
\label{eq:asymp_p_gen}
    \nu_\mathrm{p_{n,\ell}}
    = \left[n + \frac{\ell}{2} + \varepsilon + \frac{\alpha}{2}\left(n - \frac{\nu_{\rm max}}{\Delta\nu}\right)^2 \right]\Delta\nu - \delta\nu_{0,\ell} \,,
\end{equation}
where $n$ and $\ell$ denote the radial order and angular degree, $\varepsilon$ is the phase offset, $\delta\nu_{0,\ell}$ the small separation, and $\alpha$ accounts for the curvature of the large separation. Global seismic parameters $\Delta\nu$ and $\nu_{\rm max}$ are first measured from the power spectrum, then predicted mode frequencies are generated from Eq.~\eqref{eq:asymp_p_gen}. In this scheme, quadrupole modes are offset from radial modes by $\delta\nu_{0,2}$:
\begin{equation}
    \nu_{n_{\mathrm p}-1,2} = \nu_{n_{\mathrm p},0}-\delta\nu_{0,2}\,.
    \label{eq:l=2}
\end{equation}
and dipole ($\ell=1$) modes, when not mixed, are offset by
\begin{equation}
    \nu_{n_{\mathrm p},1} = \nu_{n_{\mathrm p},0}-\delta\nu_{0,1}+\frac{\Delta\nu}{2}\,.
    \label{eq:l=1_p-modes}
\end{equation}
Observed peaks are then matched to the nearest predicted frequencies \citep{corsaro_2014, corsaro_2020, nielsen_2021, nielsen_2025, breton_2022}, and modes are labelled according to their angular degree. This approach works well for stars with narrow, regular ridges but becomes challenging for mixed modes, F-like stars, or stars with broad linewidths, where ridge distortion and mode blending can obscure the asymptotic pattern.

\subsection{Mixed-mode formulation}
\label{subsec:mix_modes_form}

For stars exhibiting signatures of mode coupling, particularly dipole mixed modes, mode identification requires a formalism that accounts for the interaction between the p- and g- mode cavities.
In this formulation, pure pressure modes are computed using Eq.~\eqref{eq:asymp_p_gen}, while pure gravity modes follow \citet{tassoul_1980}:
\begin{equation}
    \nu_{{\rm g}} = \frac{1}{\Delta\Pi_\mathrm{1} \left(n_\mathrm{g} + \frac{1}{2} + \varepsilon_\mathrm{g}\right)}\,,
    \label{eq:g_mode}
\end{equation}
where $\varepsilon_{\rm g}$ is the gravity-mode phase offset and $\Delta\Pi_\ell$ is the asymptotic period spacing. We note that, in the case of SGs, the gravity-mode component of mixed modes does not strictly follow the asymptotic regime, as the g-mode cavity is relatively small and the mode pattern may deviate significantly from this simple relation \citep[e.g.][]{ong_2020}. The asymptotic description adopted here should therefore be regarded as an approximation in this evolutionary phase. 

Two complementary approaches were applied using \texttt{apollinaire} and \texttt{PBjam}.
For \texttt{apollinaire} results, an a posteriori mixed mode-identification method was implemented. First, the $\ell=0$ and $\ell=2$ modes are identified using the standard asymptotic relation of Eq.~\eqref{eq:asymp_p_gen} for pure pressure modes. Remaining significant peaks are provisionally associated with dipole mixed modes, which are then compared to the frequencies predicted by the best-fitting asymptotic mixed-mode model. Peaks that do not correspond to model-predicted modes are treated as noise and excluded. The mixed-mode frequencies are computed by solving the implicit asymptotic relation \citep{shibahashi_1979}:
\begin{equation}
    \label{asymp_mix}
    \cot\!\left(\Theta_{\rm g}\right)\tan\!\left(\Theta_{\rm p}\right) = q\,,
\end{equation}
with phases defined as \citep{ong_2023}:
\begin{align}
\label{eq:phases}
    \Theta_{\rm p} &= \frac{\pi}{\Delta\nu}\left(\nu - \nu_{\rm p}\right)\,,\\
    \Theta_{\rm g} &= \frac{\pi}{2} - \frac{\pi}{\Delta\Pi_\ell} \left(\frac{1}{\nu_{\rm g}} - \frac{1}{\nu}\right)\,,
\end{align}
where $\nu_\mathrm{p}$ and $\nu_\mathrm{g}$ are the closest pure pressure and gravity mode frequencies to $\nu$, respectively.
The asymptotic parameters $\Delta\nu$, $\alpha$, $\delta\nu_{0,1}$, $\varepsilon$, $\varepsilon_{\mathrm{g}}$, $q$, and $\Delta\Pi_1$ are adjusted within a Bayesian MCMC framework using \texttt{emcee}, with a Gaussian log-likelihood:
\begin{equation}
    \ln \mathcal{L}
    = -\frac{1}{2}\sum_{\rm modes}
      \frac{(\nu_{\rm obs} - \nu_{\rm model})^2}{\sigma_{\rm obs}^2}\,,
\end{equation}
where $\nu_{\rm obs}$ are the observed peak-bagged frequencies, $\nu_{\rm model}$ the model frequencies, and $\sigma_{\rm obs}$ the observational uncertainties, with the larger of asymmetric bounds adopted conservatively. This procedure enables a clear separation between genuine oscillation modes and spurious noise features. Although the MCMC convergence yields asymptotic parameters describing the oscillation pattern, here it is employed solely as a mode-identification tool.

\texttt{PBjam}, in contrast, follows an a priori non-asymptotic approach, particularly suited for SG stars with mixed modes. Radial and quadrupole modes are modelled with standard asymptotic relations of Eqs.~\eqref{eq:asymp_p_gen} and \eqref{eq:l=2}, whereas dipole modes are treated as combinations of pure pressure and gravity modes with Eq.~\eqref{eq:g_mode}, and described by a non-asymptotic relation \citep{deheuvels_2010, ong_2020}, which explicitly accounts for the coupling strength between the p- and g-mode cavities \citep{nielsen_2025}.

\subsection{Data driven approach}
\label{subsec:data_driven}

Unlike purely asymptotic approaches, which rely primarily on theoretical mode-spacing relations, the method adopted by FAMED identifies oscillation modes using a data-driven, multi-step analysis of the power spectrum. Candidate peaks are first detected from the background-corrected power spectrum using a significance criterion based on their signal-to-noise ratio (S/N), without imposing a fixed asymptotic pattern a priori. This allows oscillation modes, including dipole modes affected by mixed-mode behaviour, to be identified based on their statistical significance rather than their proximity to expected asymptotic locations.
Once candidate frequencies are identified, global asteroseismic parameters are refined using asymptotic descriptions of Eqs.~\eqref{eq:asymp_p_gen} and \eqref{eq:l=2}. These well-behaved modes define the oscillation pattern, enabling a clearer separation of dipole modes from both noise and neighbouring ridges.

For each candidate dipole peak, Bayesian model comparison is employed to assess whether the feature is statistically significant relative to the background and whether it is best described by a single Lorentzian profile, multiple components, noise, or more complex structures such as rotational splitting. This step allows genuine mixed dipole modes to be distinguished from spurious detections and noise-driven artefacts. The identification process is iterative, with ambiguous peaks discarded and mode classifications refined based on Bayesian evidence and consistency with the evolving global pattern.
In this framework, dipole mixed modes are identified primarily through Bayesian model comparison, based on their statistical evidence relative to background-only or alternative multi-component models. Asymptotic relations are applied only at a later stage to refine mode labelling and spacing.

\section{Échelle diagrams}
\label{app:results-per-star}

We present here the star-by-star échelle diagram for the optimal light curve with observed modes as extracted by the three pipelines used in this work, and the fitted model with \texttt{apollinaire}. Stars are listed by increasing $\nu_\mathrm{max}$.

\begin{figure}[H]
    \centering
    \includegraphics[width=1\linewidth,trim={0 0 0 0.5cm},clip]{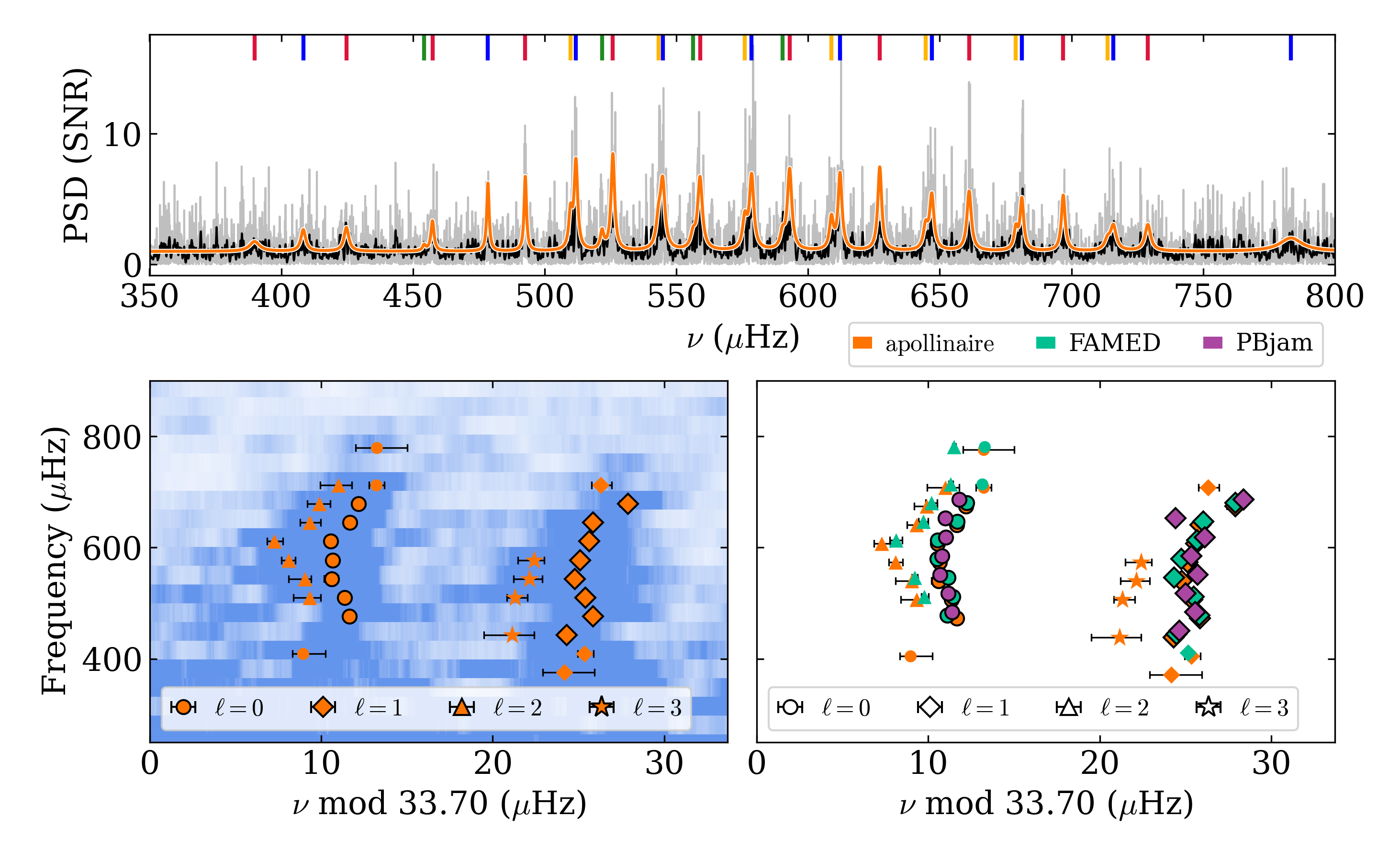}
    \caption{Top panel: The PSD computed from the optimal light curve of TIC 354552931 (HD 36553) is shown in grey, over a frequency range centred on $\nu_\mathrm{max}$. Its moving mean (10-point window) is shown in black. The orange curve represents the model fitted by \texttt{apollinaire}, with the corresponding frequencies listed in Table~\ref{tab:354552931_freqs}. Identified peaks are indicated by vertical coloured lines above the PSD: blue for $\ell=0$, red for $\ell=1$, yellow for $\ell=2$ and green for $\ell=3$. Bottom panels: The left panel shows an échelle diagram with the selected reference frequency dataset: circles ($\ell=0$), diamonds ($\ell=1$), triangles ($\ell=2$), and stars ($\ell=3$). The right panel compares these frequencies with those obtained by the other pipelines, where the markers of each pipeline was slightly vertically shifted within $\Delta\nu$ to improve readability. Markers are the same as in the left panel, with \texttt{apollinaire} in orange, FAMED in green, and \texttt{PBjam} in violet. Markers with black edges correspond to modes included in the minimal list, while markers without black edges correspond to modes in the maximal list or excluded from any list. For clarity, the échelle diagram has been offset by -5$\,\mu$Hz to improve ridges visibility.}
    \label{fig:354552931_echelle}
\end{figure}

\begin{figure}[H]
    \centering
    \includegraphics[width=1\linewidth,trim={0 0 0 0.5cm},clip]{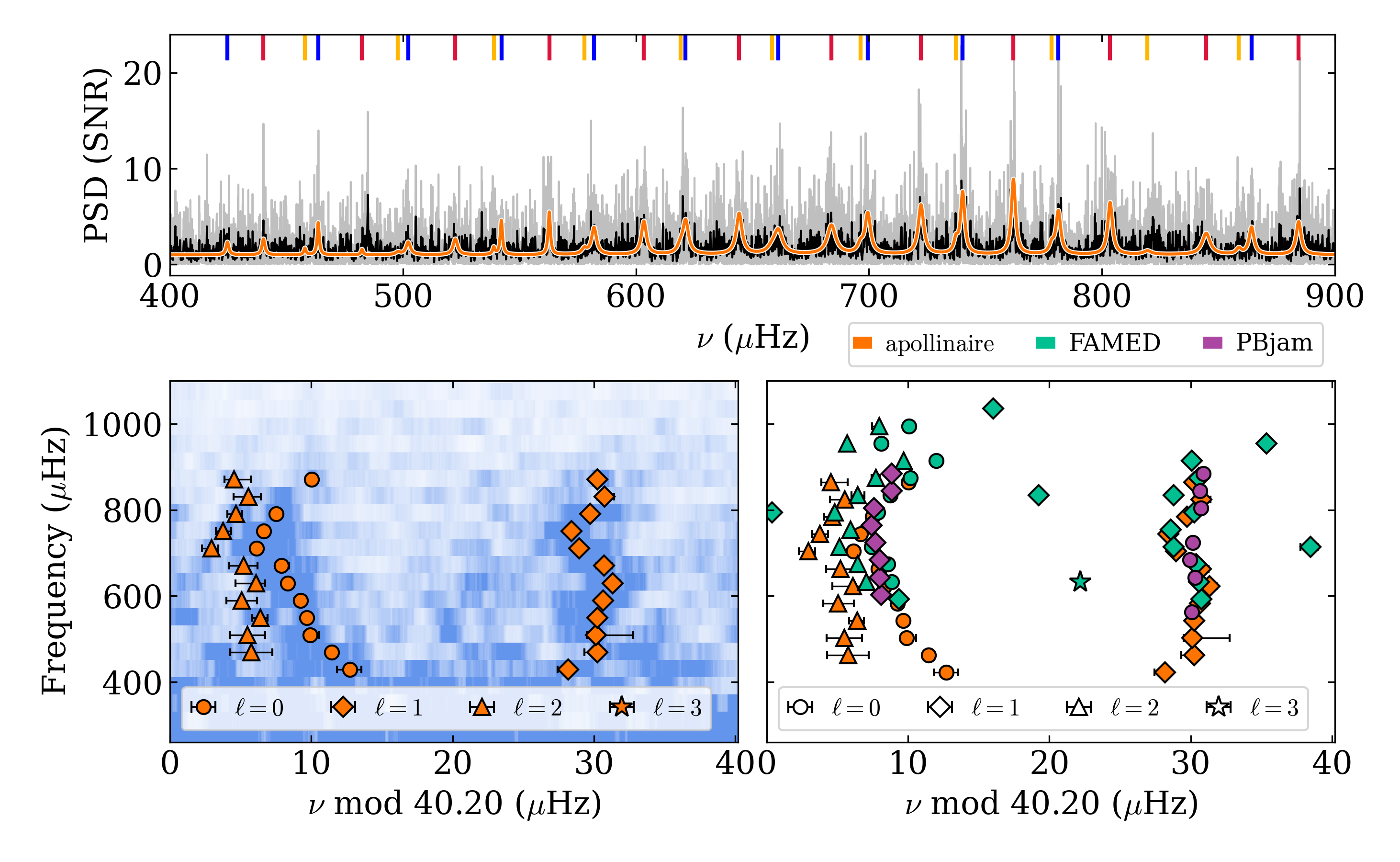}
    \caption{Same as Figure~\ref{fig:354552931_echelle}, but for TIC 161825882 ($\theta$ Dra). The échelle diagram has been offset by 10$\,\mu$Hz to improve ridges visibility. Modes for this star have not been flagged, as the pipelines do not agree on their identification; all are shown with black edges markers.}
    \label{fig:161825882_echelle}
\end{figure}

\begin{figure}[H]
    \centering
    \includegraphics[width=1\linewidth,trim={0 0 0 0.5cm},clip]{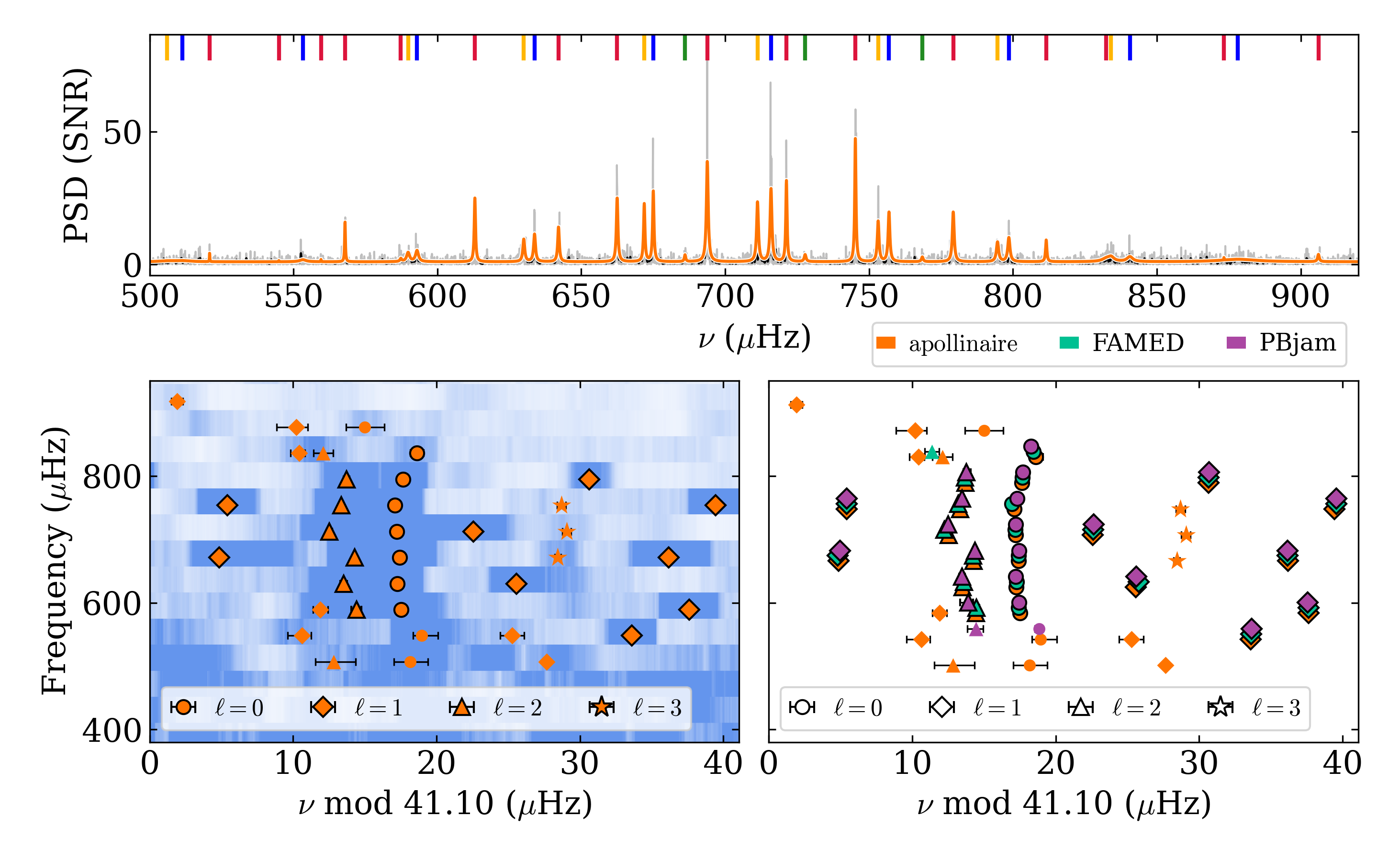}
    \caption{Same as Figure~\ref{fig:354552931_echelle}, but for TIC 123699670 (HD 62644). The échelle diagram has not been offset.}
    \label{fig:123699670_echelle}
\end{figure}

\begin{figure}[H]
    \centering
    \includegraphics[width=1\linewidth,trim={0 0 0 0.5cm},clip]{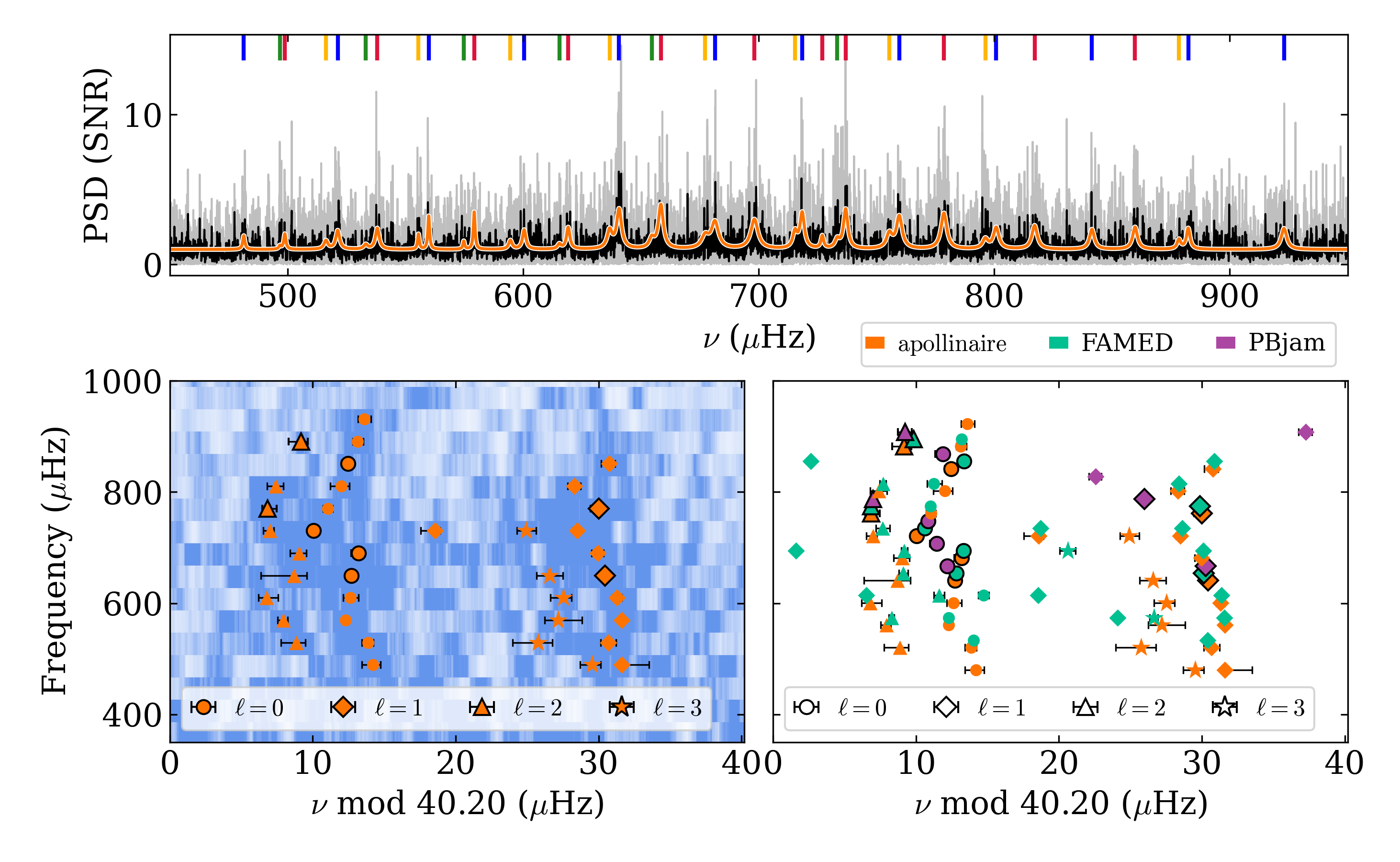}
    \caption{Same as Figure~\ref{fig:354552931_echelle}, but for TIC 236871353 (68 Dra). The échelle diagram has been offset by 25$\,\mu$Hz to improve ridges visibility.}
    \label{fig:236871353_echelle}
\end{figure}

\begin{figure}[H]
    \centering
    \includegraphics[width=1\linewidth,trim={0 0 0 0.5cm},clip]{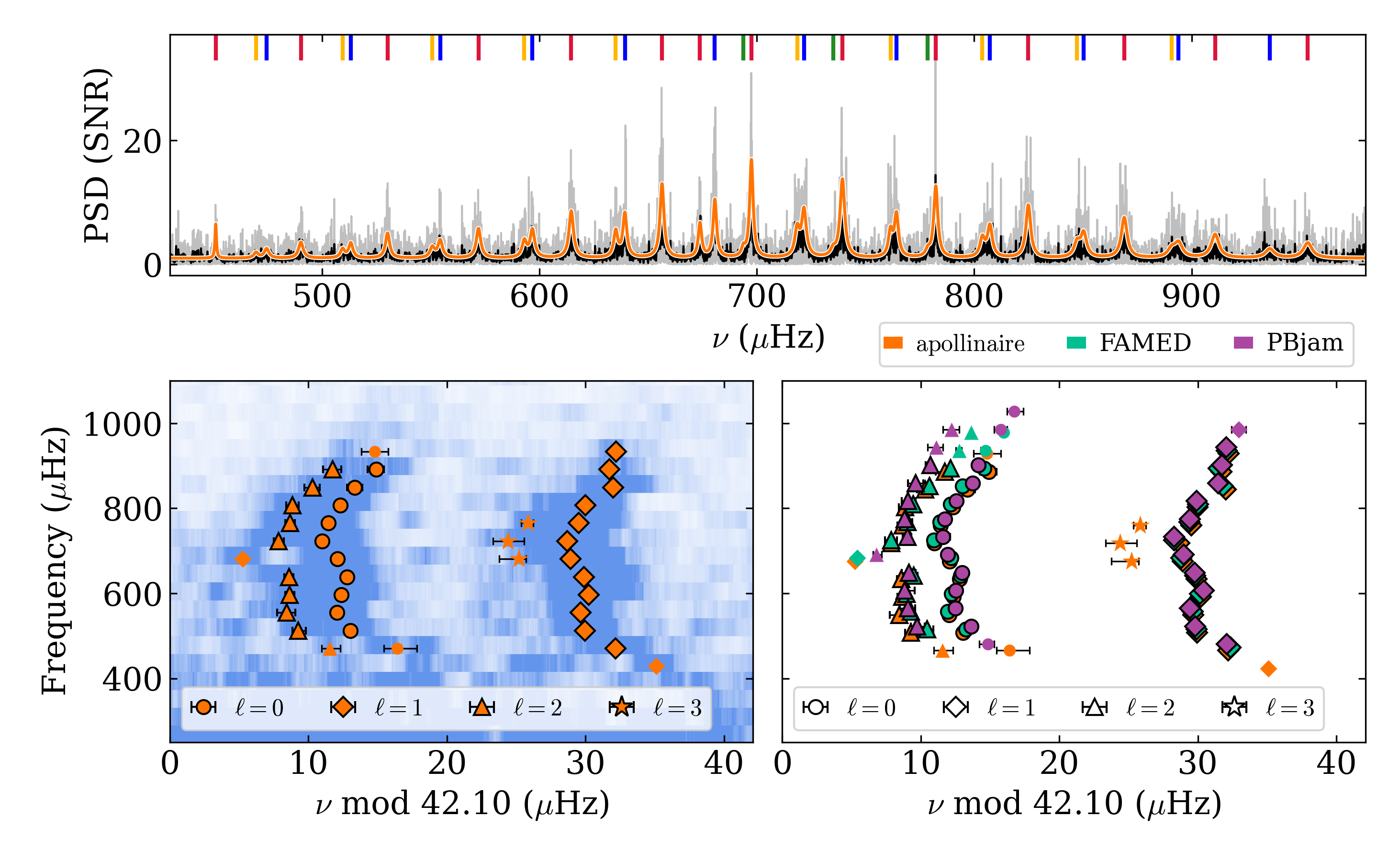}
    \caption{Same as Figure~\ref{fig:354552931_echelle}, but for TIC 441813918 (35 Dra). The échelle diagram has been offset by -5$\,\mu$Hz to improve ridges visibility.}
    \label{fig: 441813918_echelle}
\end{figure}

\begin{figure}[H]
    \centering
    \includegraphics[width=1\linewidth,trim={0 0 0 0.5cm},clip]{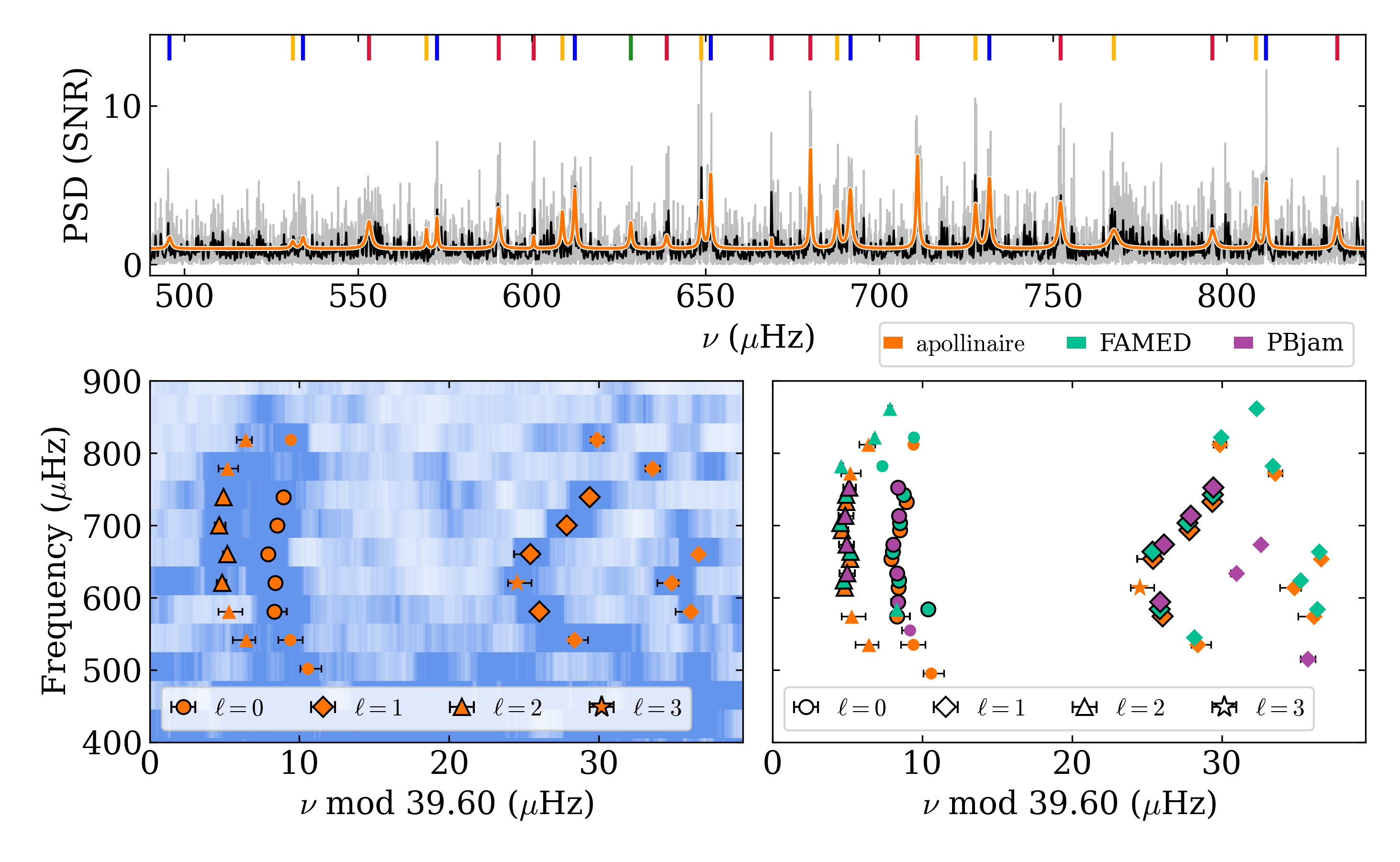}
    \caption{Same as Figure~\ref{fig:354552931_echelle}, but for TIC 41195655 (27\,Cyg). The échelle diagram has been offset by 10$\,\mu$Hz to improve ridges visibility.}
    \label{fig:41195655_echelle}
\end{figure}

\begin{figure}[H]
    \centering
    \includegraphics[width=1\linewidth,trim={0 0 0 0.5cm},clip]{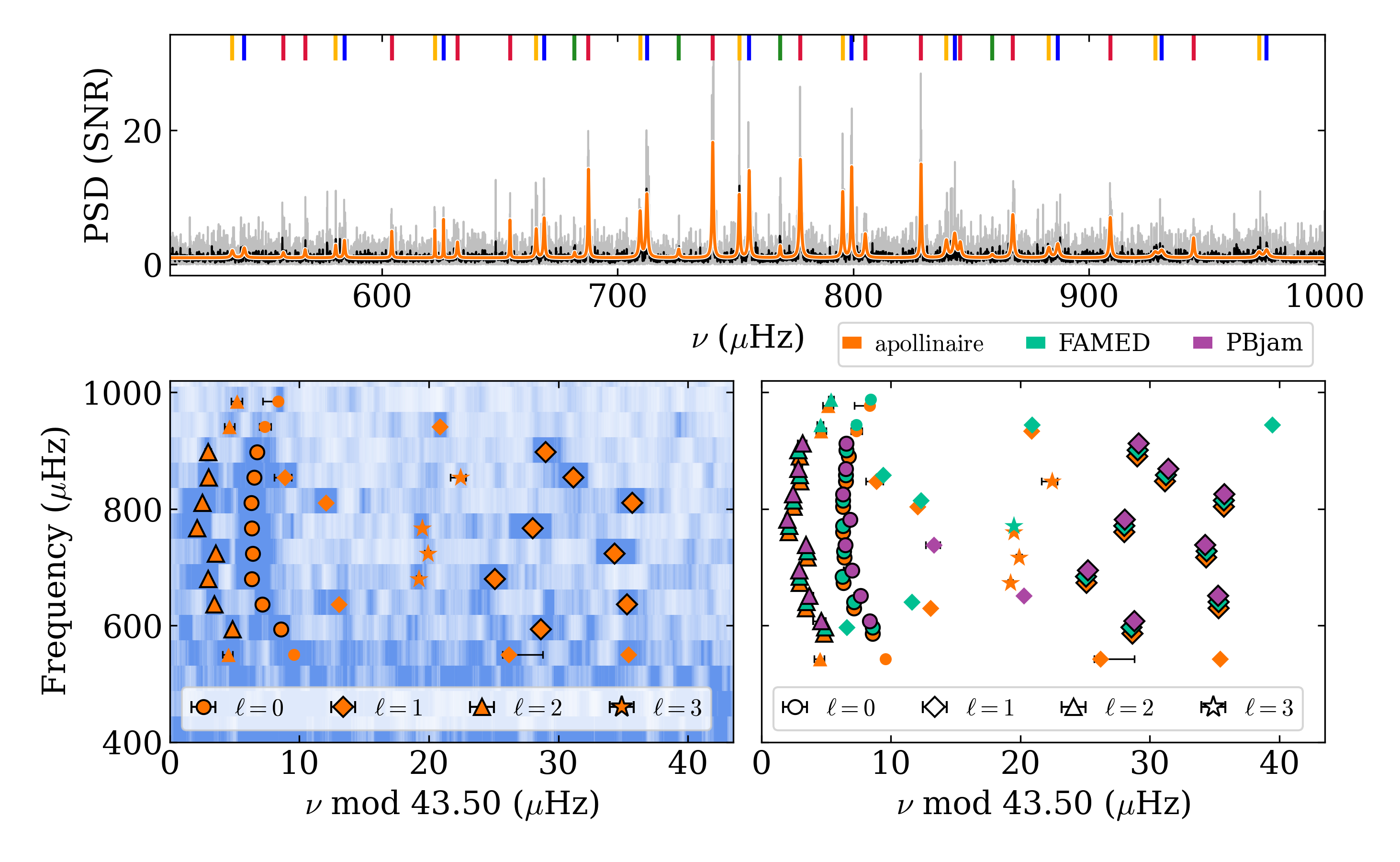}
    \caption{Same as Figure~\ref{fig:354552931_echelle}, but for TIC 48194330 (HD 175225). The échelle diagram has been offset by 10$\,\mu$Hz to improve ridges visibility.}
    \label{fig:48194330_echelle}
\end{figure}

\begin{figure}[H]
    \centering
    \includegraphics[width=1\linewidth,trim={0 0 0 0.5cm},clip]{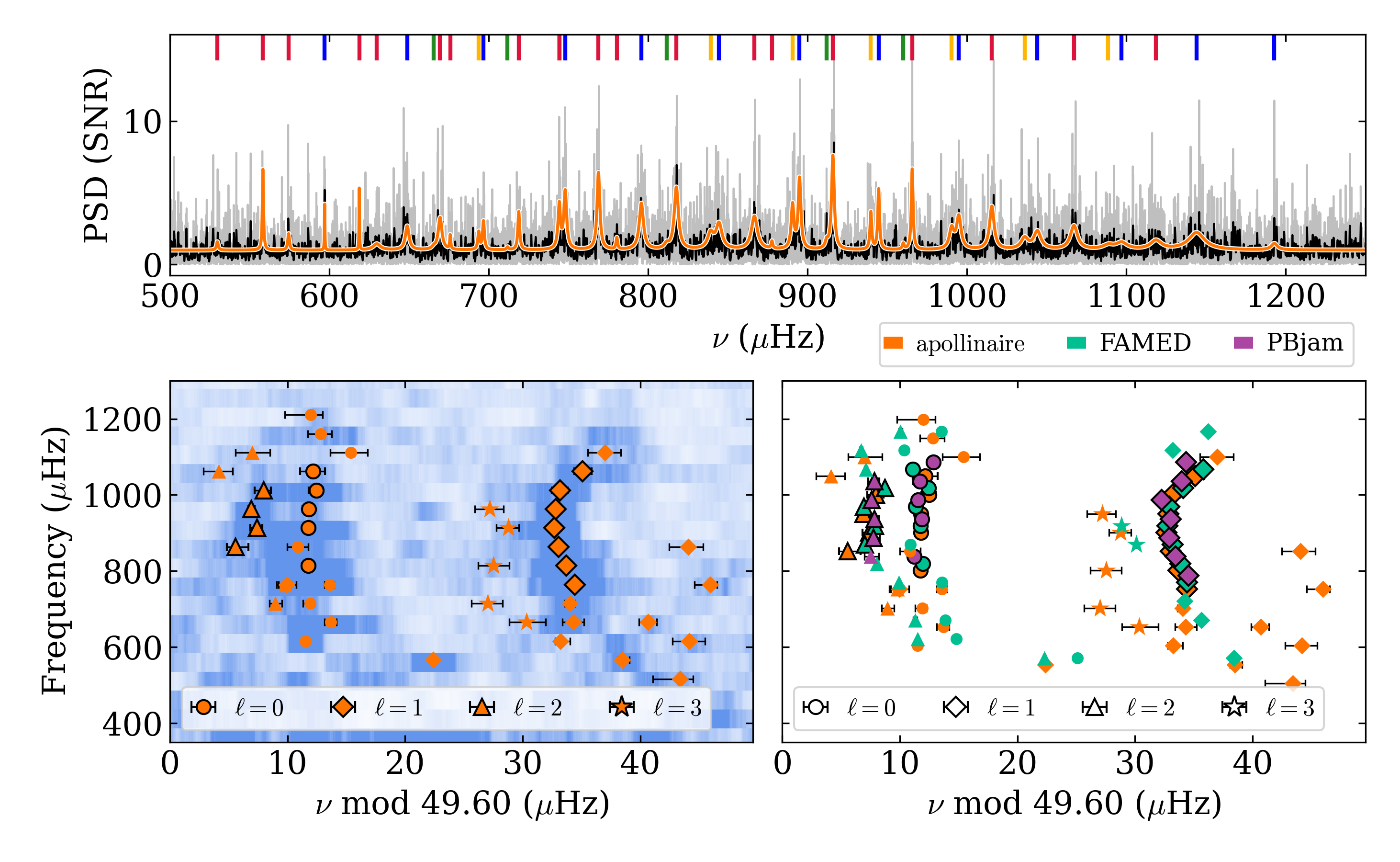}
    \caption{Same as Figure~\ref{fig:354552931_echelle}, but for TIC 219420836 ($\zeta$ Pic). The échelle diagram has been offset by 40$\,\mu$Hz to improve ridges visibility.}
    \label{fig:219420836_echelle}
\end{figure}

\begin{figure}[H]
    \centering
    \includegraphics[width=1\linewidth,trim={0 0 0 0.5cm},clip]{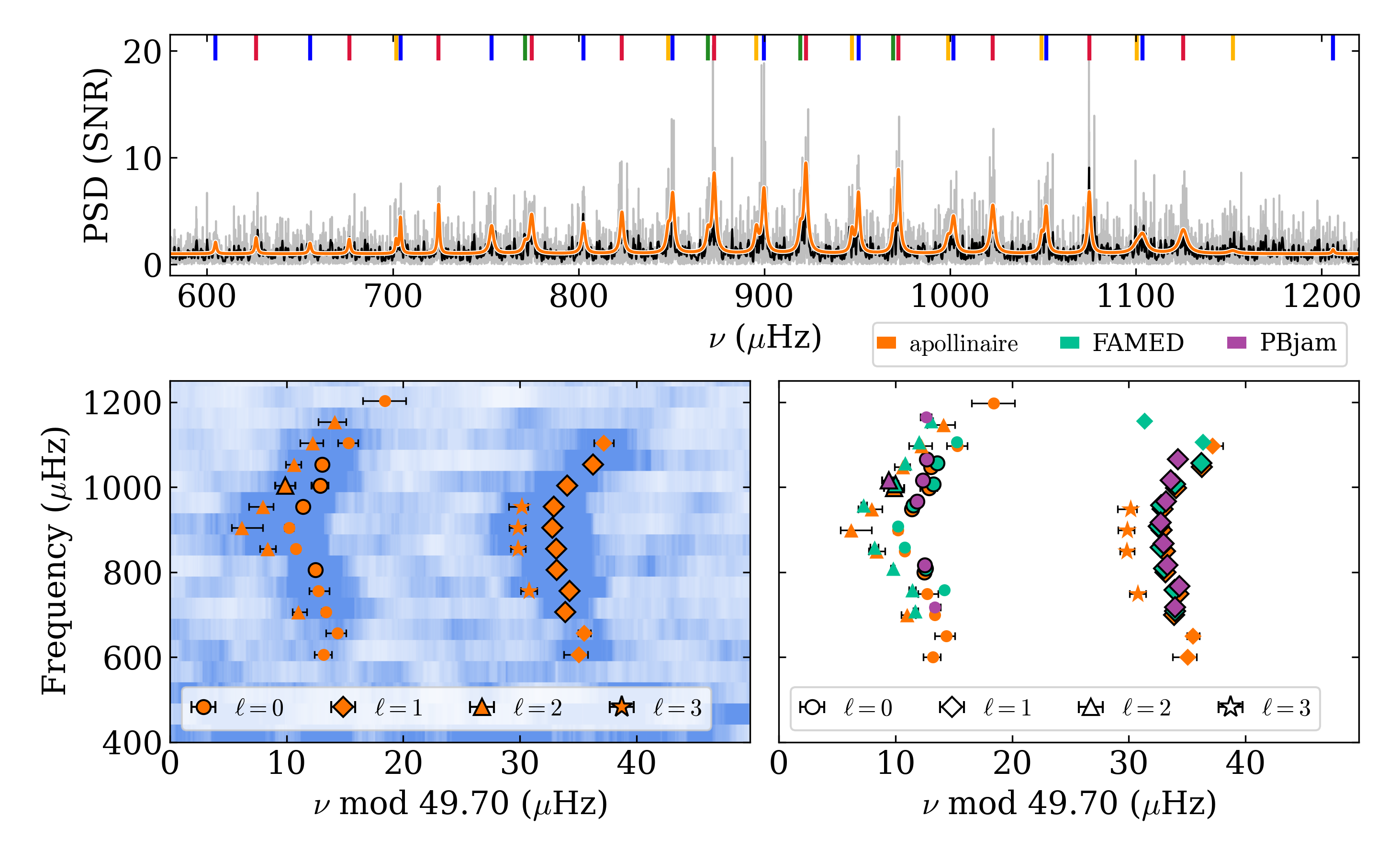}
    \caption{Same as Figure~\ref{fig:354552931_echelle}, but for TIC 255630992 (HD 46569). The échelle diagram has been offset by -5$\,\mu$Hz to improve ridges visibility.}
    \label{fig:255630992_echelle}
\end{figure}

\begin{figure}[H]
    \centering
    \includegraphics[width=1\linewidth,trim={0 0 0 0.5cm},clip]{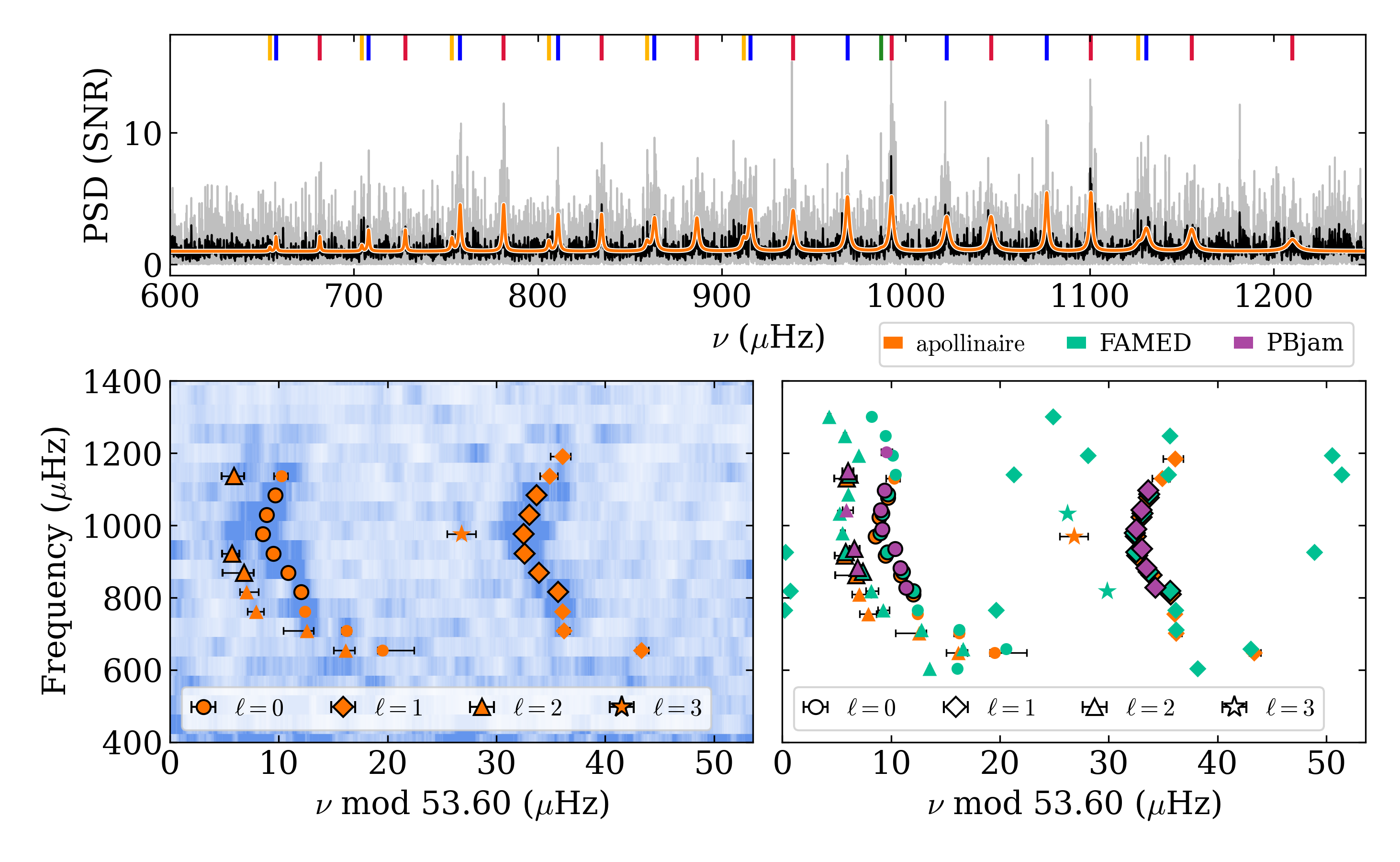}
    \caption{Same as Figure~\ref{fig:354552931_echelle}, but for TIC 421444084 ($\upsilon$ Cep). The échelle diagram has been offset by -5$\,\mu$Hz to improve ridges visibility.}
    \label{fig:421444084_echelle}
\end{figure}

\begin{figure}[H]
    \centering
    \includegraphics[width=1\linewidth,trim={0 0 0 0.5cm},clip]{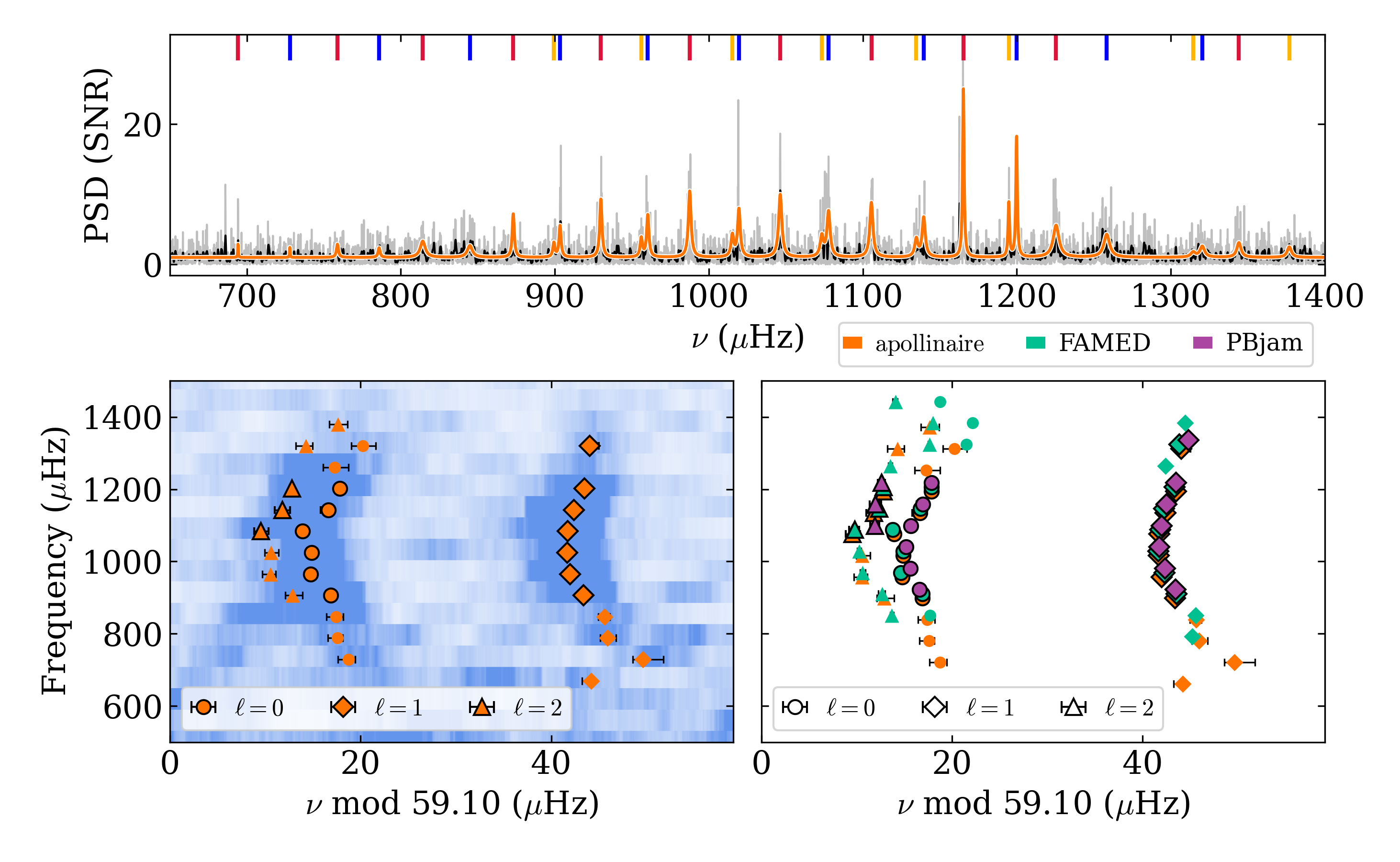}
    \caption{Same as Figure~\ref{fig:354552931_echelle}, but for TIC 232563914 (HD 136064). The échelle diagram has not been offset.}
    \label{fig:232563914_echelle}
\end{figure}

\begin{figure}[H]
    \centering
    \includegraphics[width=1\linewidth,trim={0 0 0 0.5cm},clip]{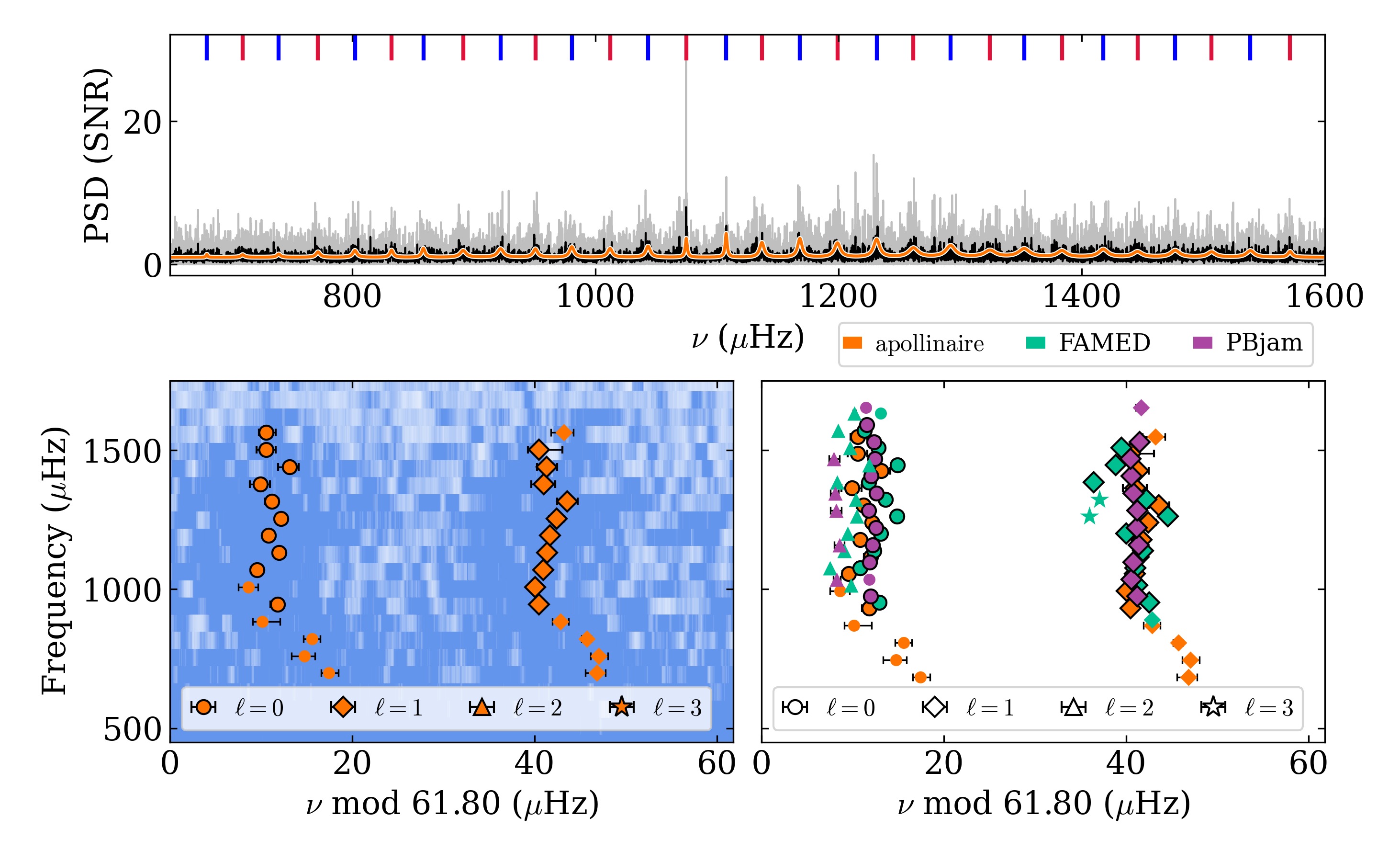}
    \caption{Same as Figure~\ref{fig:354552931_echelle}, but for TIC 441804568 ($\psi^1$ Dra A). The échelle diagram has been offset by 5$\,\mu$Hz to improve ridges visibility.}
    \label{fig:441804568_echelle}
\end{figure}

\begin{figure}[H]
    \centering
    \includegraphics[width=1\linewidth,trim={0 0 0 0.5cm},clip]{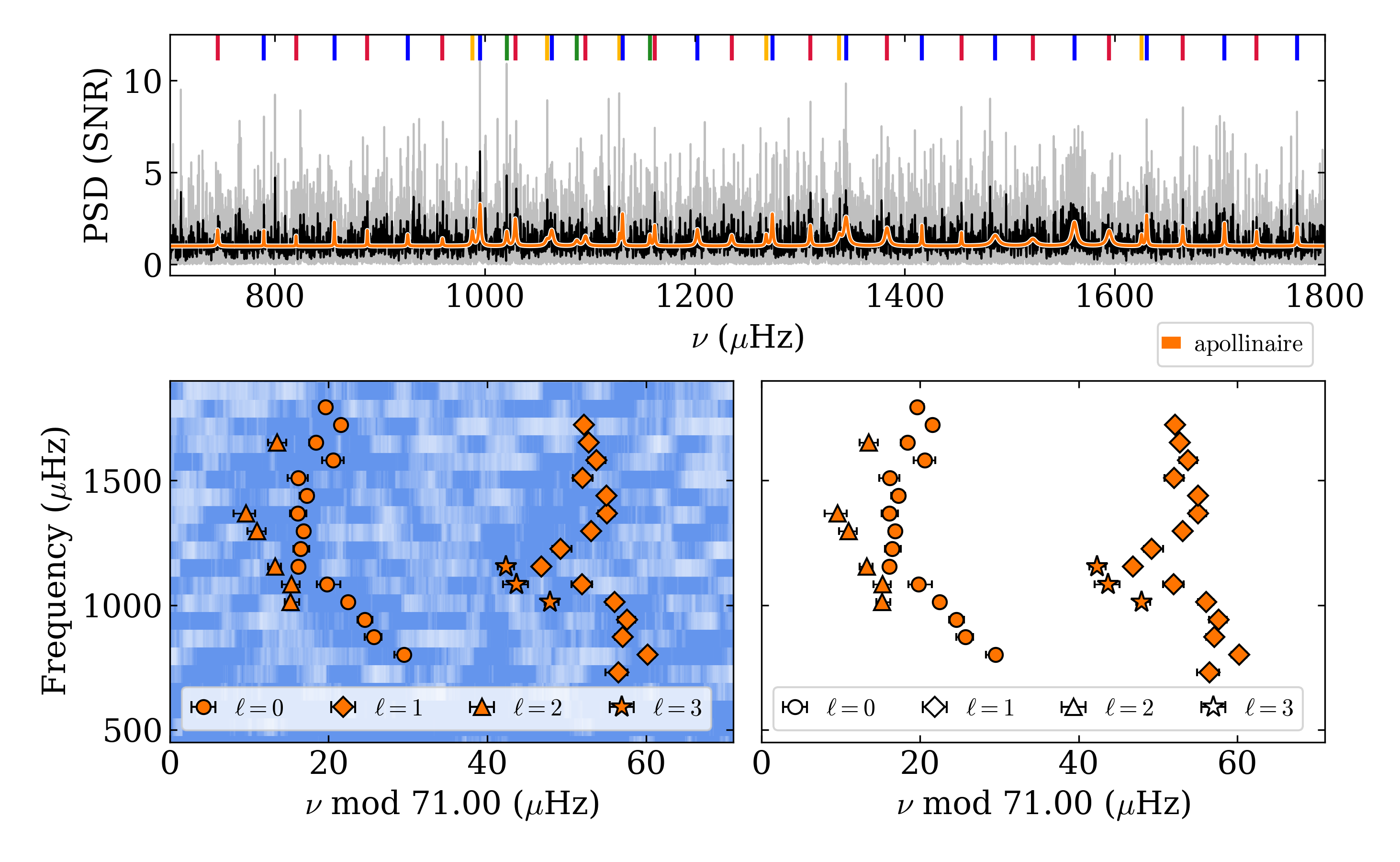}
    \caption{Same as Figure~\ref{fig:354552931_echelle}, but for TIC 405902259 (HD 191195). The échelle diagram has been offset by 50$\,\mu$Hz to improve ridges visibility.}
    \label{fig:405902259_echelle}
\end{figure}

\begin{figure}[H]
    \centering
    \includegraphics[width=1\linewidth,trim={0 0 0 0.5cm},clip]{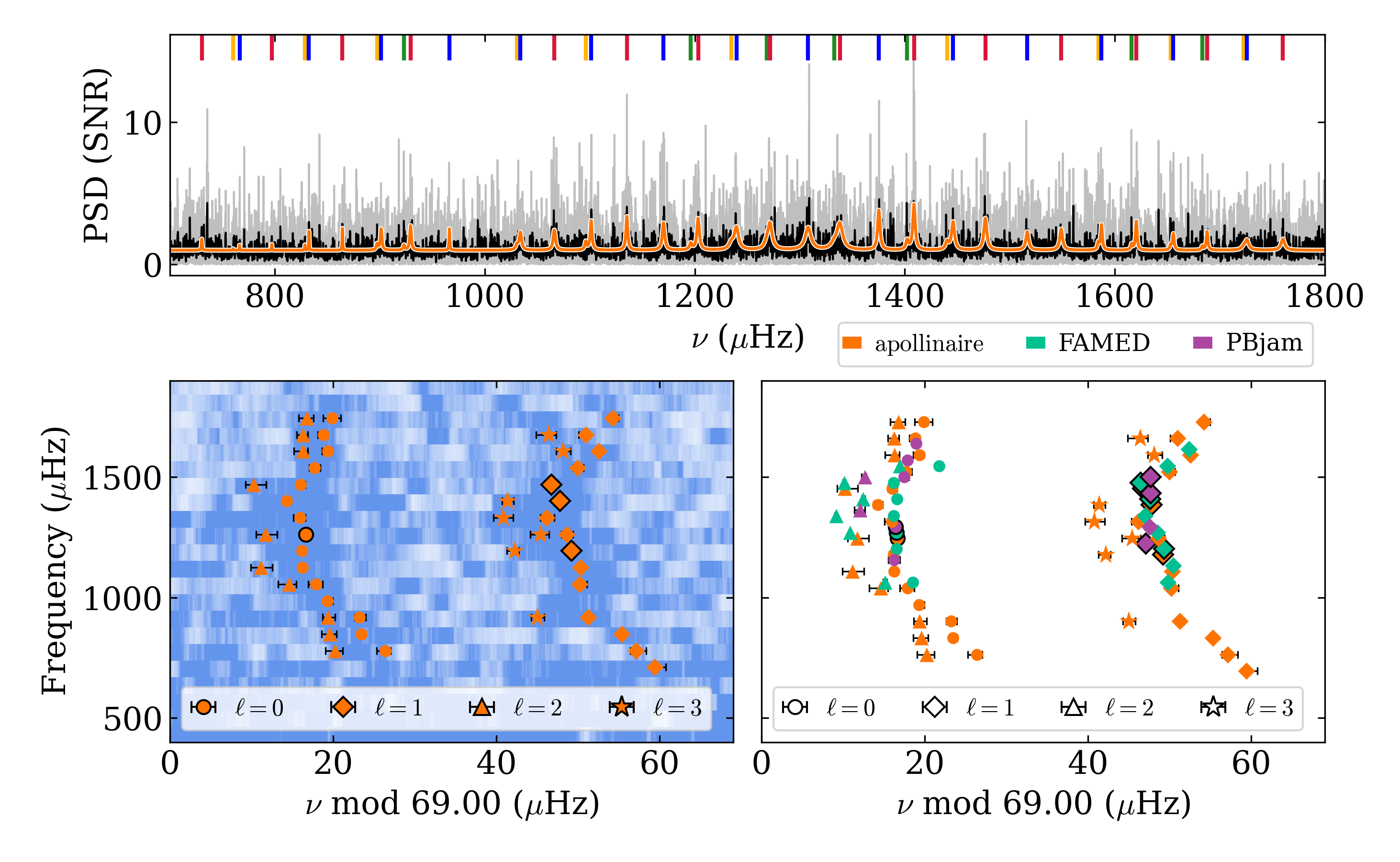}
    \caption{Same as Figure~\ref{fig:354552931_echelle}, but for TIC 170225363 (HD 50223). The échelle diagram has been offset by 50$\,\mu$Hz to improve ridges visibility.}
    \label{fig:170225363_echelle}
\end{figure}

\begin{figure}[H]
    \centering
    \includegraphics[width=1\linewidth,trim={0 0 0 0.5cm},clip]{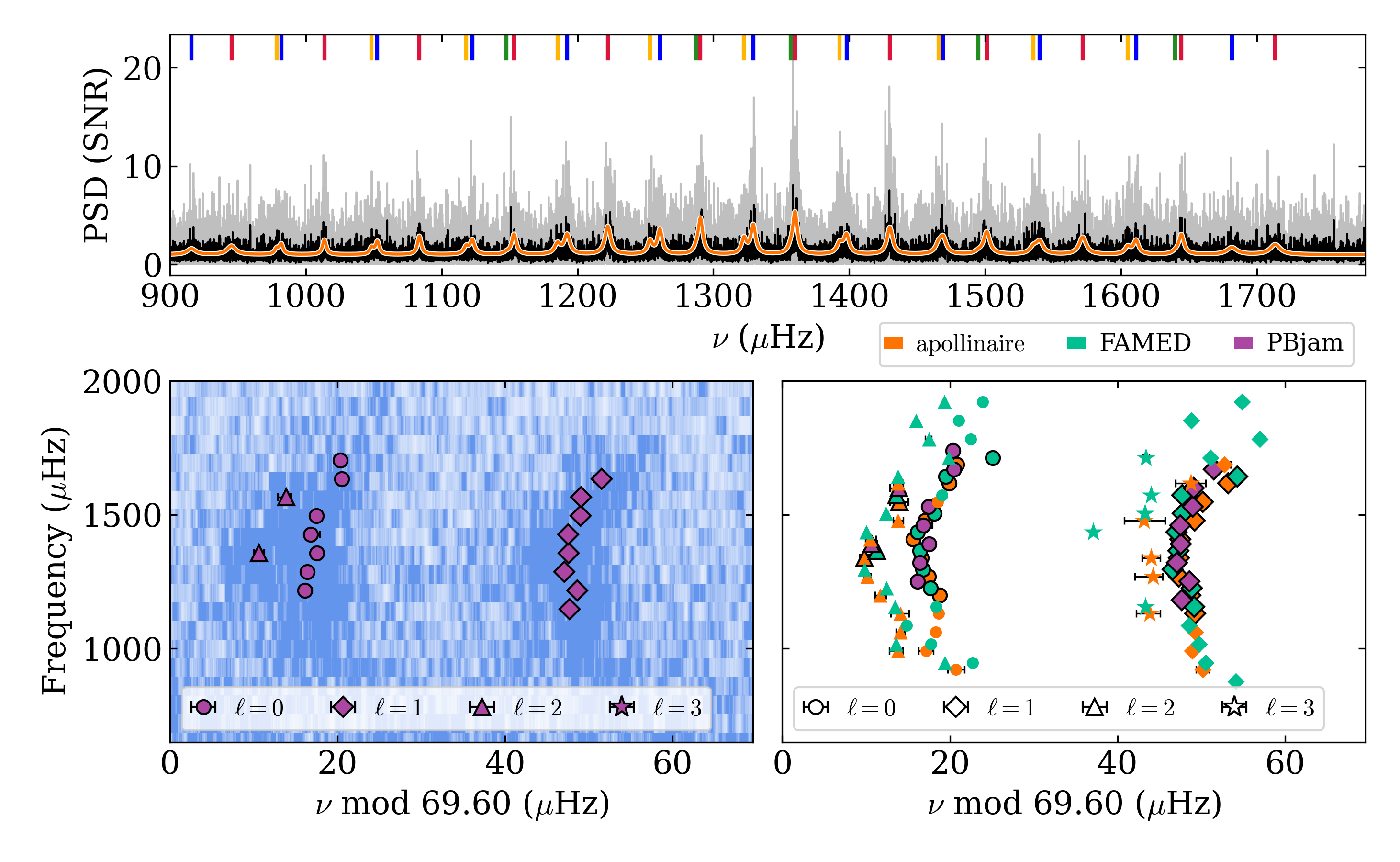}
    \caption{Same as Figure~\ref{fig:354552931_echelle}, but for TIC 233121747 (36 Dra). The échelle diagram has been offset by 60$\,\mu$Hz to improve ridges visibility.}
    \label{fig:233121747_echelle}
\end{figure}

\begin{figure}[H]
    \centering
    \includegraphics[width=1\linewidth,trim={0 0 0 0.5cm},clip]{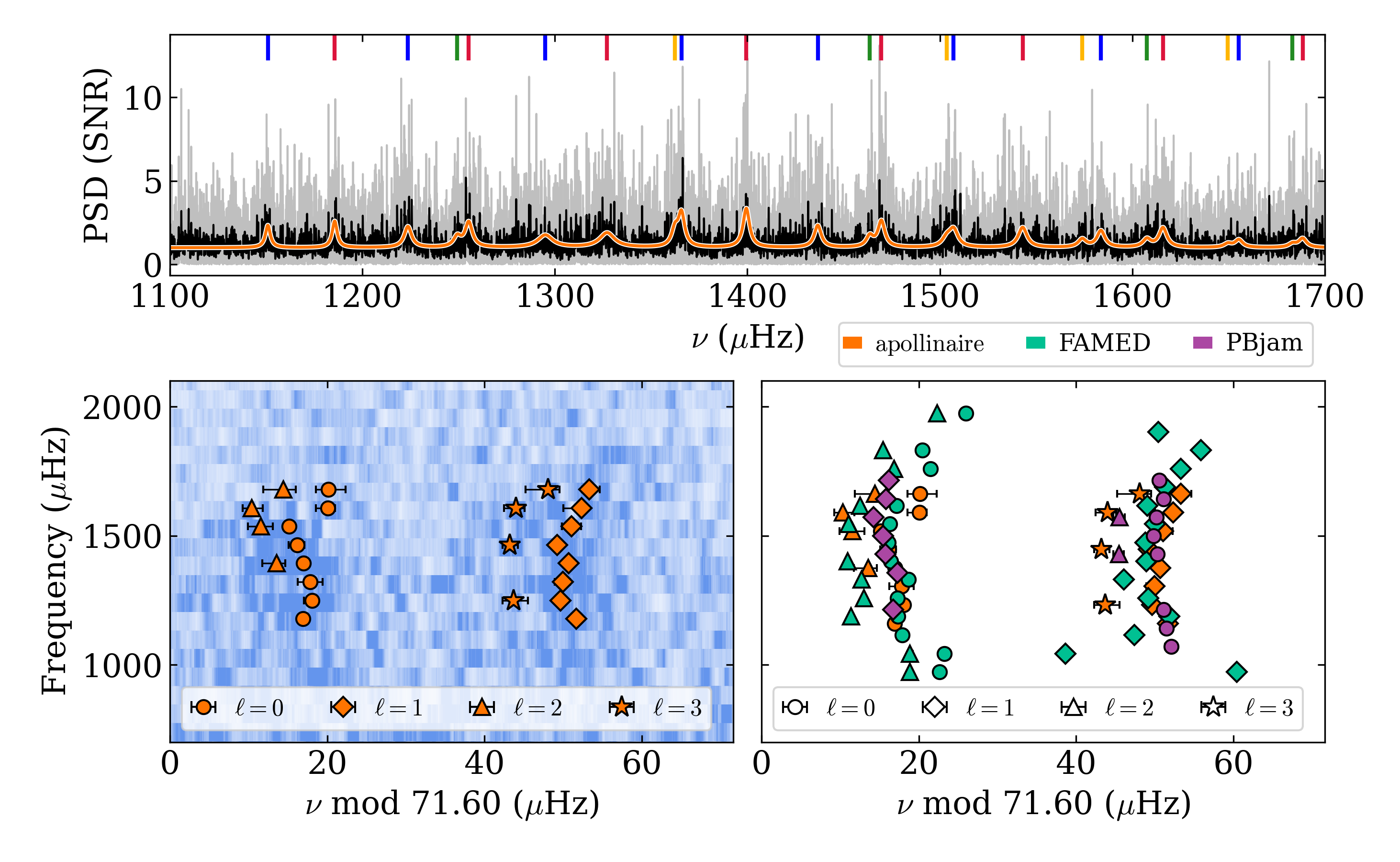}
    \caption{Same as Figure~\ref{fig:354552931_echelle}, but for TIC 308844962 (HR 3220). The échelle diagram has been offset by 60$\,\mu$Hz to improve ridges visibility. Modes for this star have not been flagged, as the pipelines do not agree on their identification; all are shown with black edges markers.}
    \label{fig:308844962_echelle}
\end{figure}

\begin{figure}[H]
    \centering
    \includegraphics[width=1\linewidth,trim={0 0 0 0.5cm},clip]{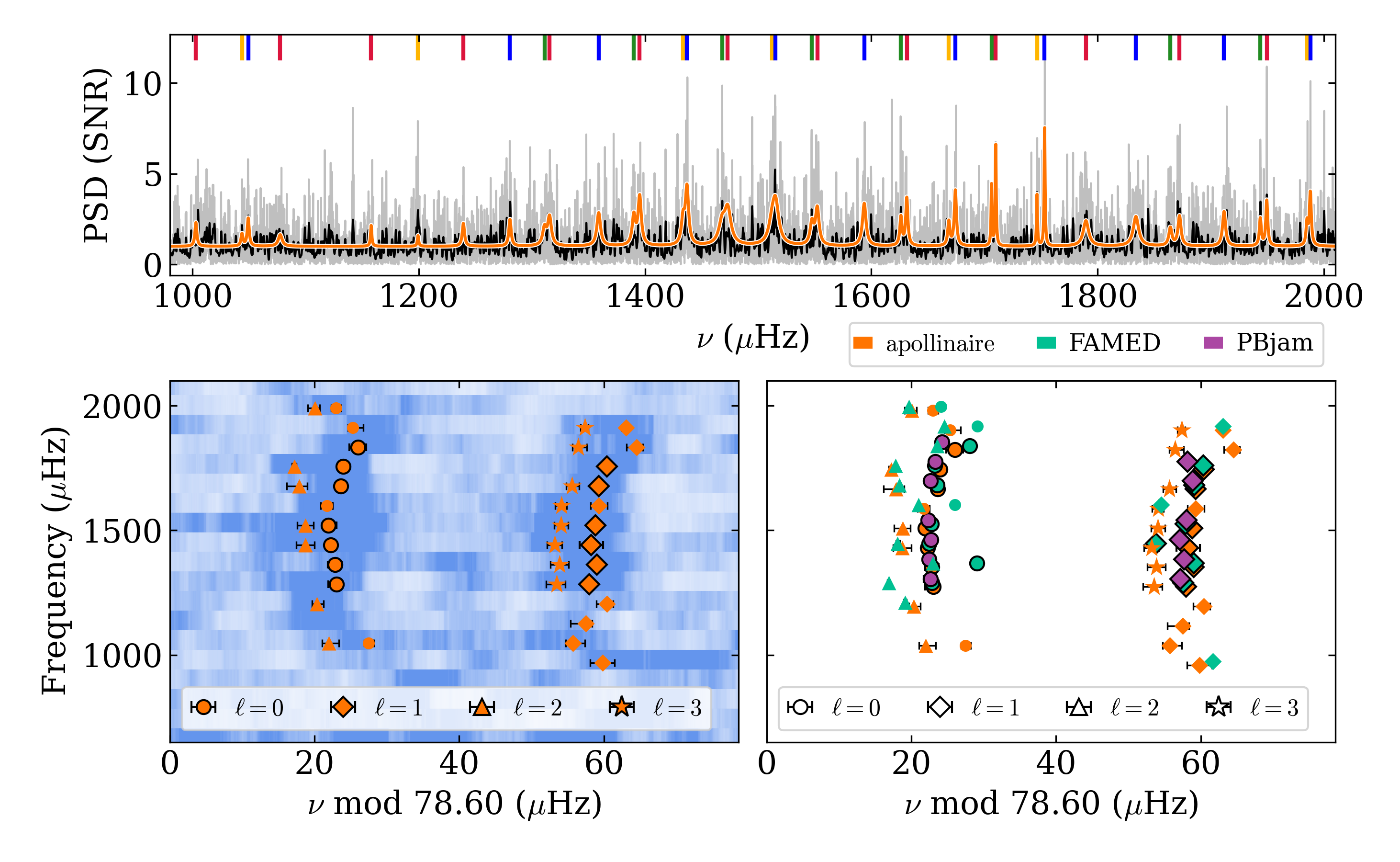}
    \caption{Same as Figure~\ref{fig:354552931_echelle}, but for TIC 58445695 (17 Cyg). The échelle diagram has not been offset.}
    \label{fig:58445695_echelle}
\end{figure}

\begin{figure}[H]
    \centering
    \includegraphics[width=1\linewidth,trim={0 0 0 0.5cm},clip]{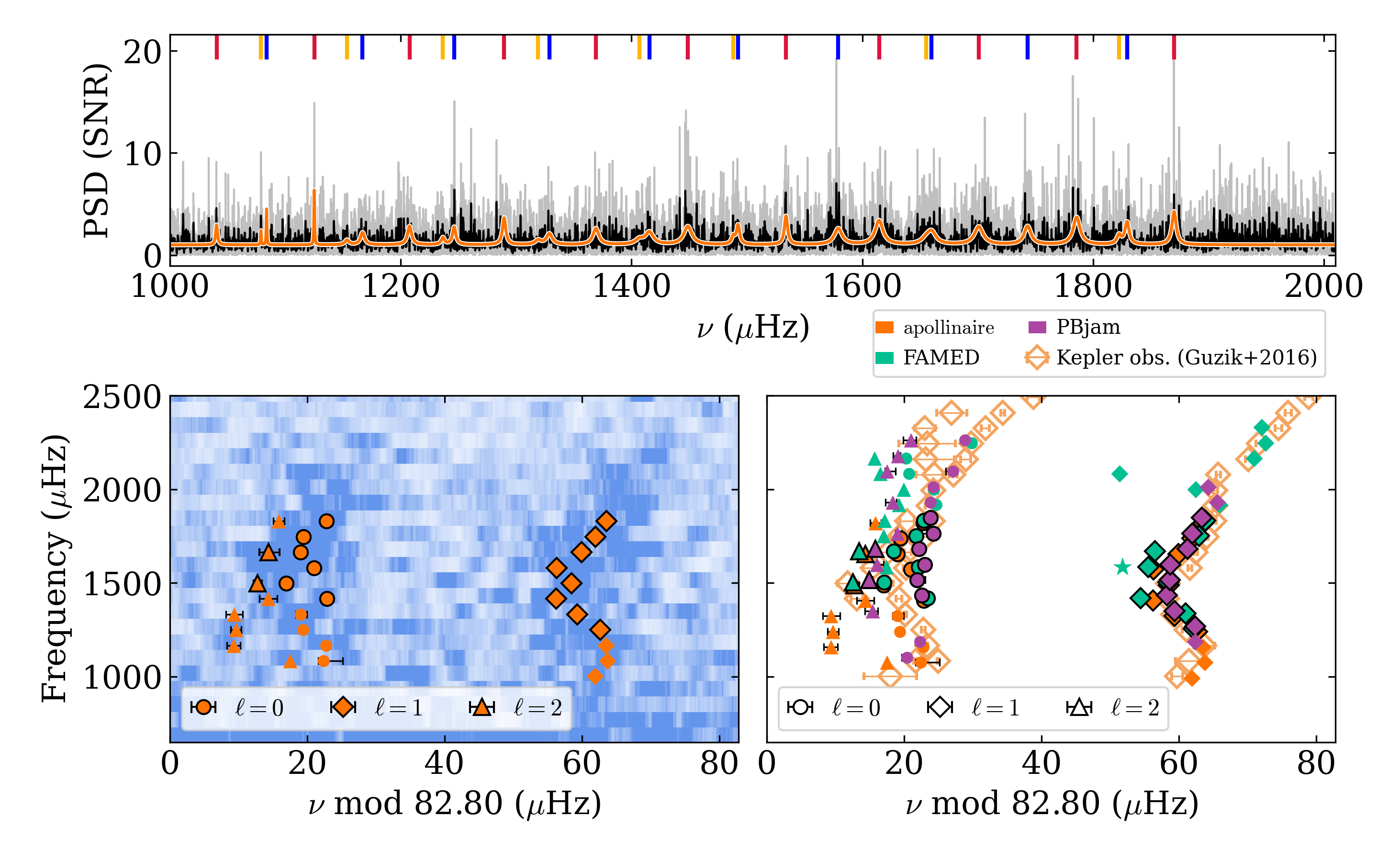}
    \caption{Same as Figure~\ref{fig:354552931_echelle}, but for TIC 27014182 ($\theta$ Cyg). The échelle diagram has been offset by -15$\,\mu$Hz to improve ridges visibility.}
    \label{fig:27014182_echelle}
\end{figure}

\begin{figure}[H]
    \centering
    \includegraphics[width=1\linewidth,trim={0 0 0 0.5cm},clip]{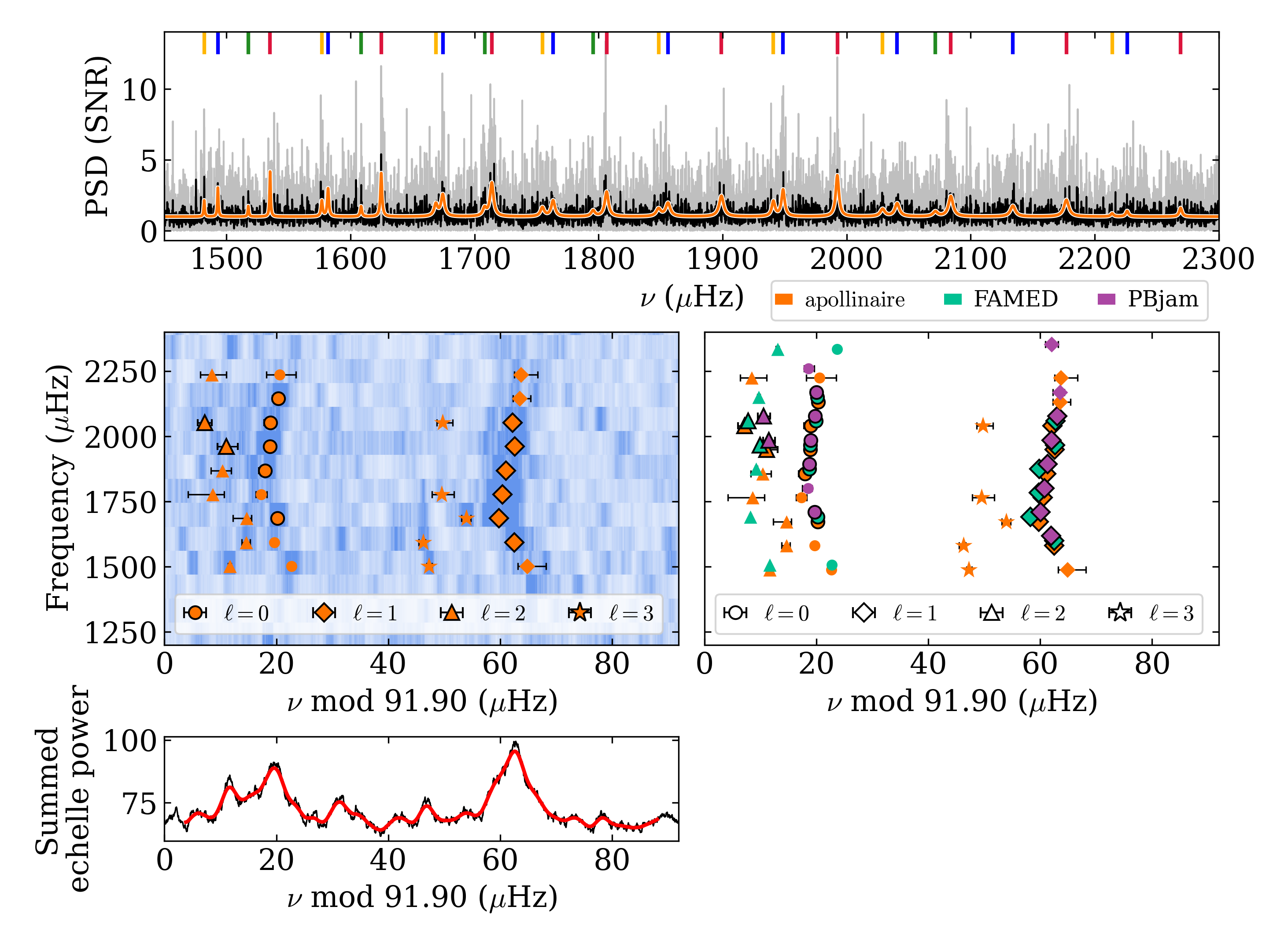}
    \caption{Same as Figure~\ref{fig:354552931_echelle}, but for TIC 26884478 (HD 184960). The échelle diagram has not been offset. The bottom panel shows the sum of the échelle diagram over frequency as a function of $\Delta\nu$ in black, with its moving mean (window = 1000 points) in red.}
    \label{fig:26884478_echelle}
\end{figure}

\begin{figure}[H]
    \centering
    \includegraphics[width=1\linewidth,trim={0 0 0 0.5cm},clip]{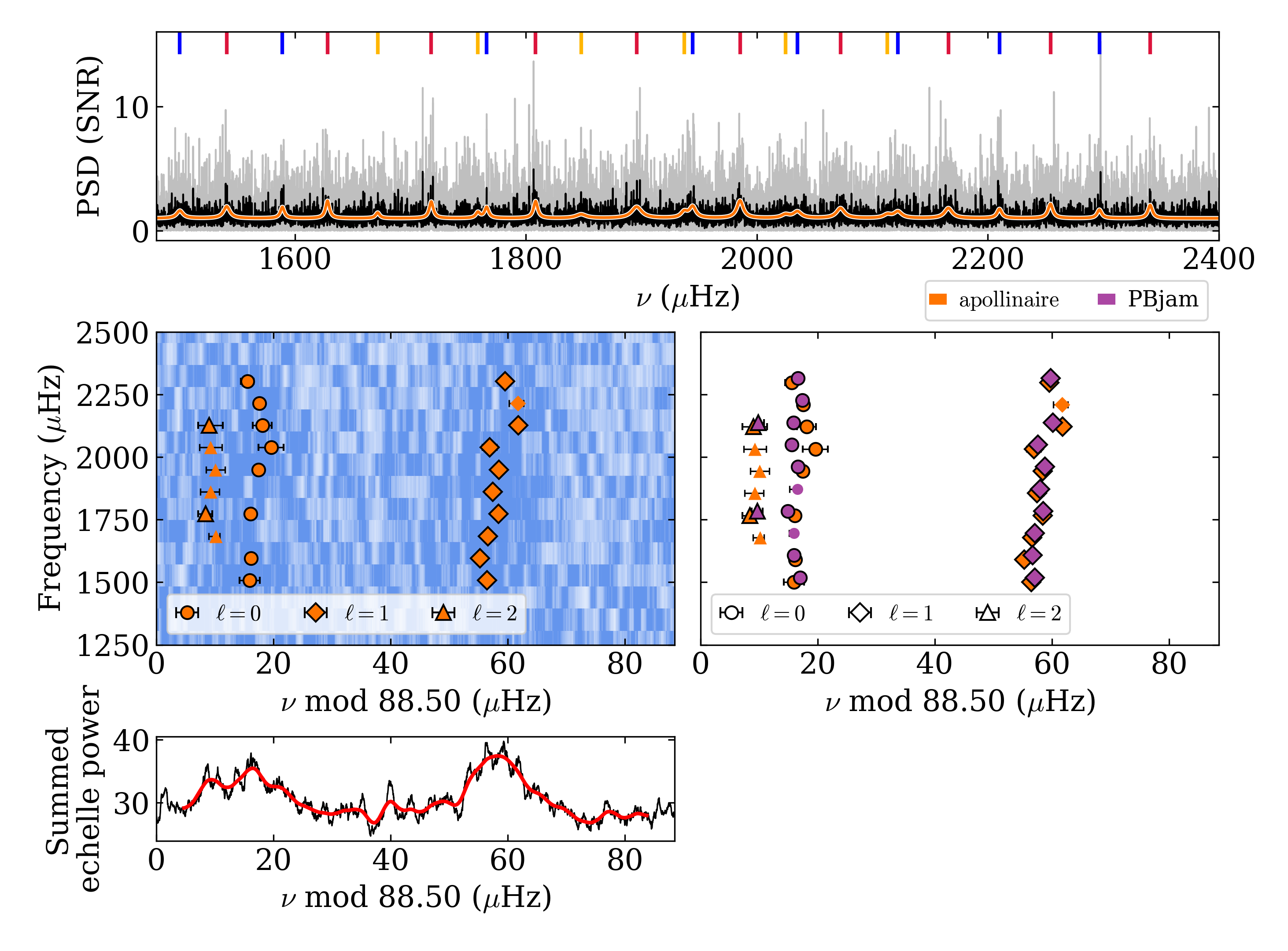}
    \caption{Same as Figure~\ref{fig:354552931_echelle}, but for TIC 233195546 ($\omega$ Dra). The échelle diagram has been offset by -20$\,\mu$Hz to improve ridges visibility. The bottom panel shows the sum of the échelle diagram over frequency as a function of $\Delta\nu$ in black, with its moving mean (window = 2000 points) in red.}
    \label{fig:233195546_echelle}
\end{figure}

\begin{figure}[H]
    \centering
    \includegraphics[width=1\linewidth,trim={0 0 0 0.5cm},clip]{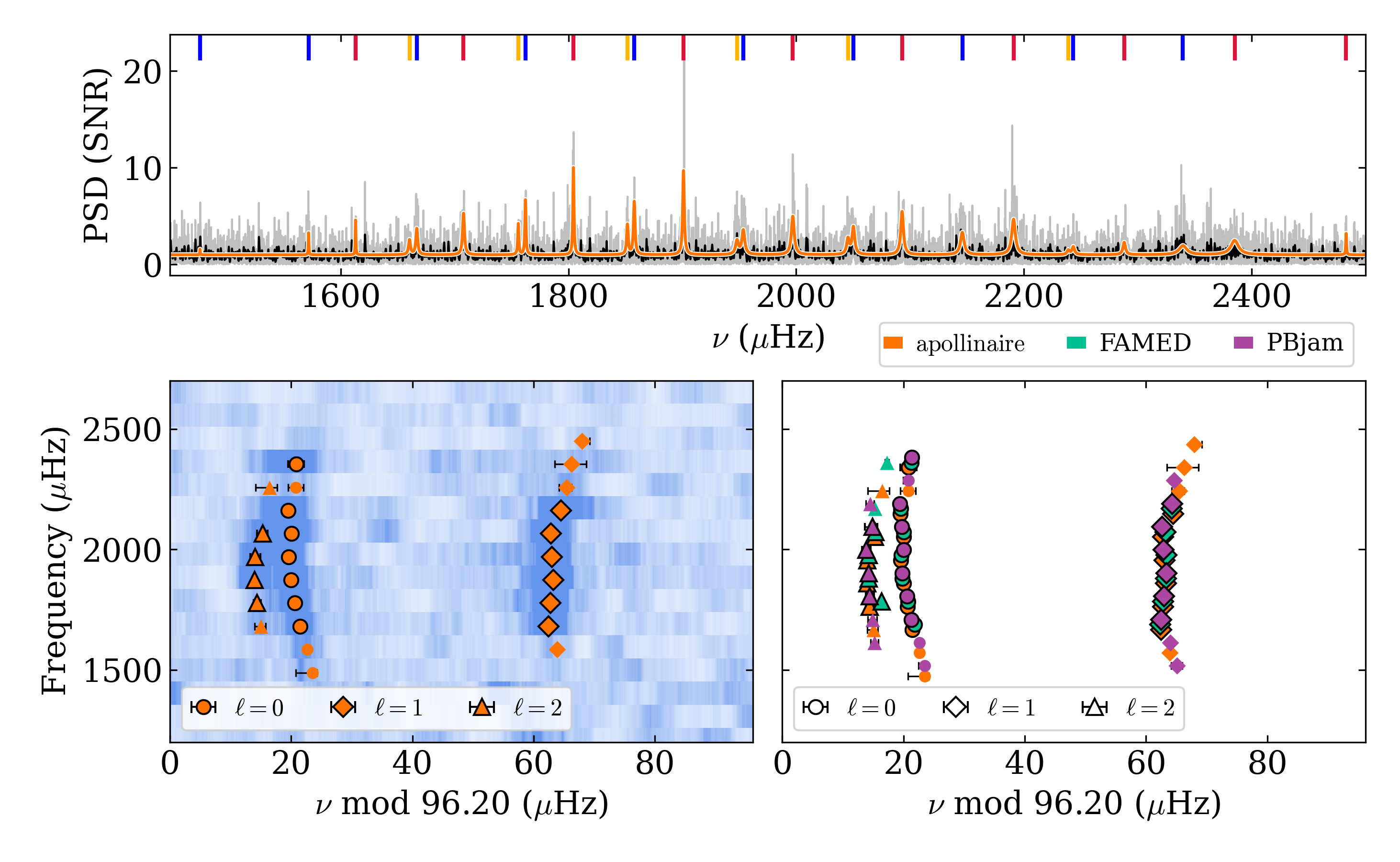}
    \caption{Same as Figure~\ref{fig:354552931_echelle}, but for TIC 22516402 (99 Her). The échelle diagram has been offset by 10$\,\mu$Hz to improve ridges visibility.}
    \label{fig:22516402_echelle}
\end{figure}

\begin{figure}[H]
    \centering
    \includegraphics[width=1\linewidth,trim={0 0 0 0.5cm},clip]{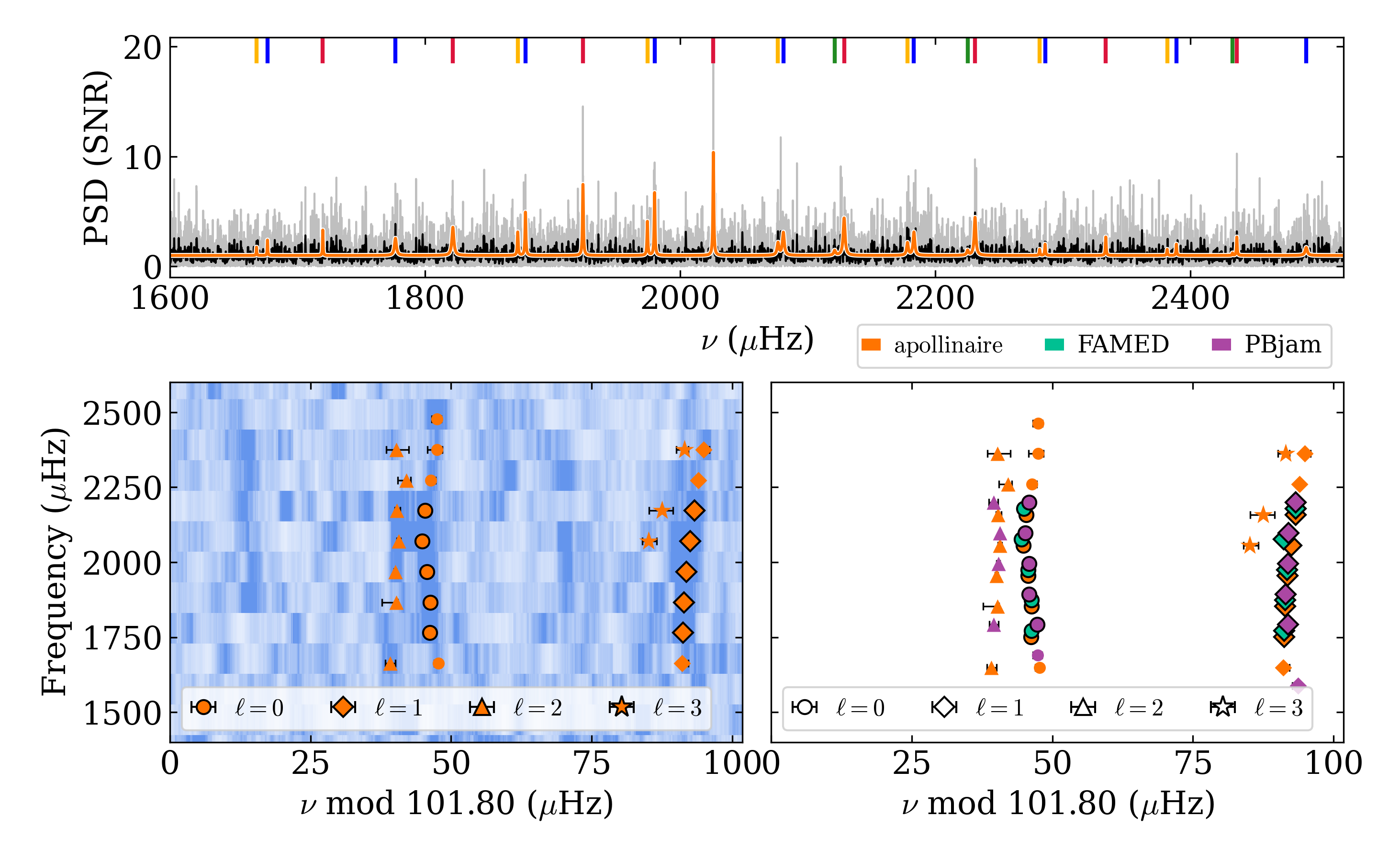}
    \caption{Same as Figure~\ref{fig:354552931_echelle}, but for TIC 130645536 (HD 53705). The échelle diagram has not been offset.}
    \label{fig:130645536_echelle}
\end{figure}

\begin{figure}[H]
    \centering
    \includegraphics[width=1\linewidth,trim={0 0 0 0.5cm},clip]{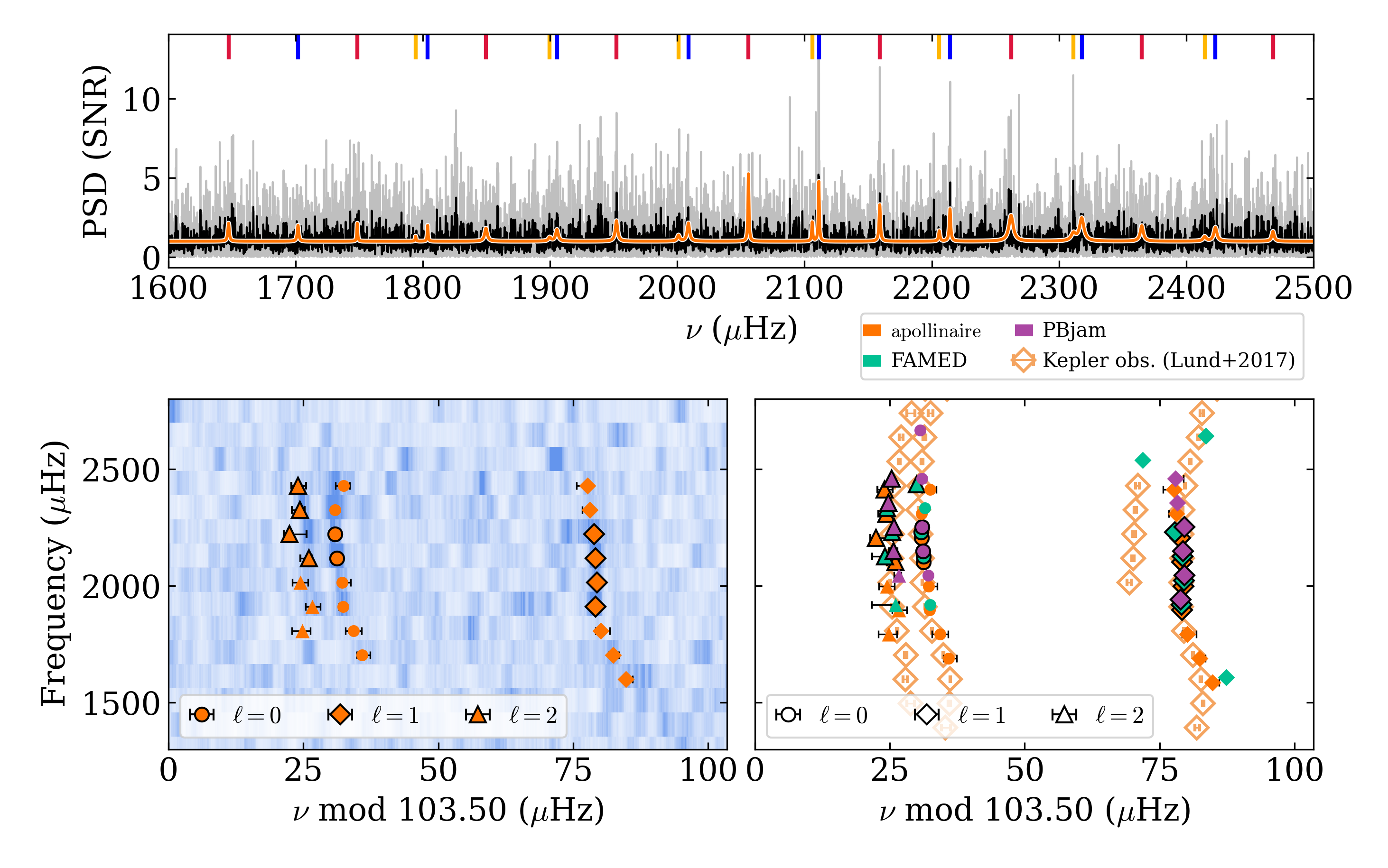}
    \caption{Same as Figure~\ref{fig:354552931_echelle}, but for TIC 27533341 (16 Cyg A). The échelle diagram has been offset by 10$\,\mu$Hz to improve ridges visibility.}
    \label{fig:27533341_echelle}
\end{figure}

\begin{figure}[H]
    \centering
    \includegraphics[width=1\linewidth,trim={0 0 0 0.5cm},clip]{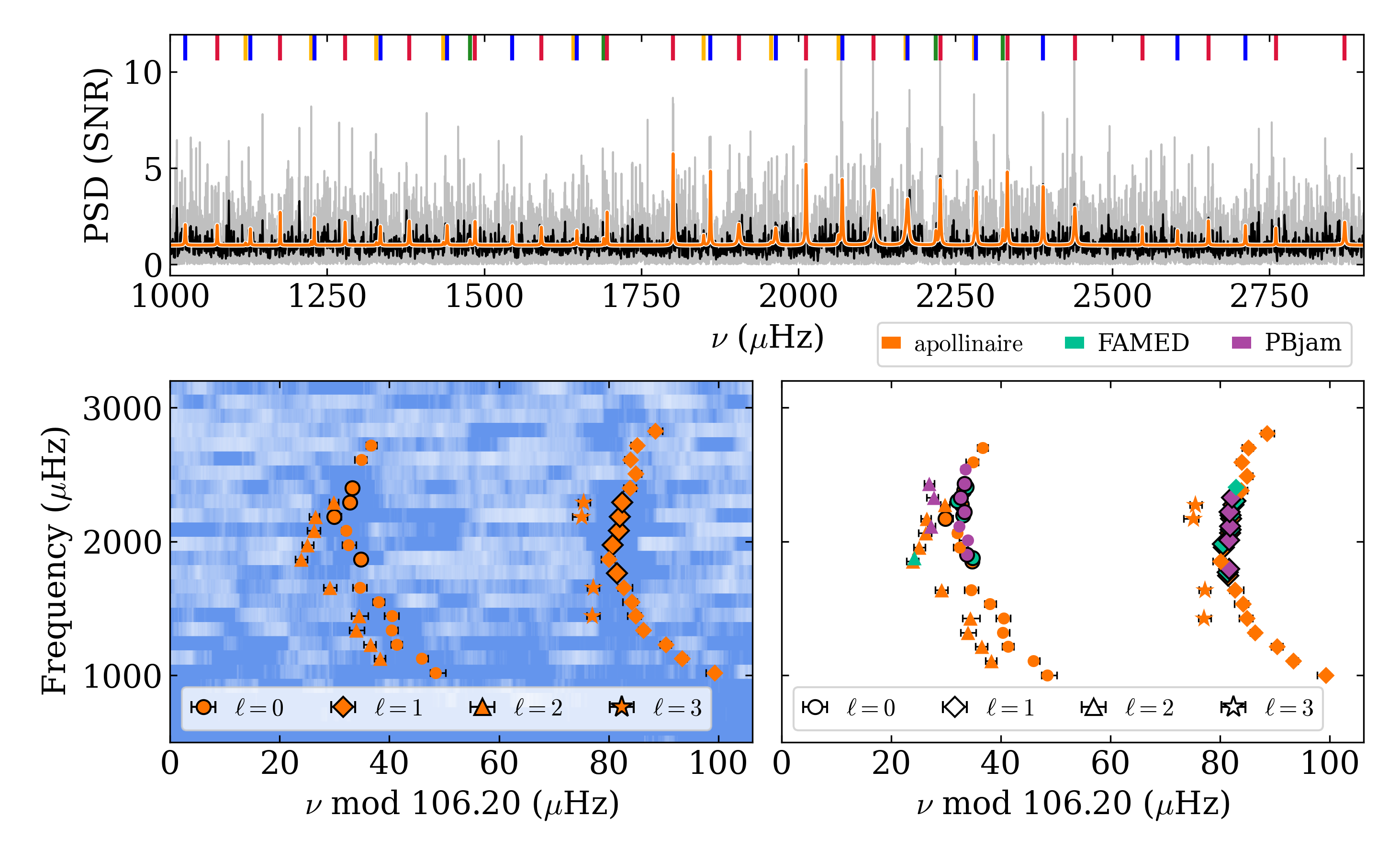}
    \caption{Same as Figure~\ref{fig:354552931_echelle}, but for TIC 9728611 (72 Her). The échelle diagram has been offset by 20$\,\mu$Hz to improve ridges visibility.}
    \label{fig:9728611_echelle}
\end{figure}

\begin{figure}[H]
    \centering
    \includegraphics[width=1\linewidth,trim={0 0 0 0.5cm},clip]{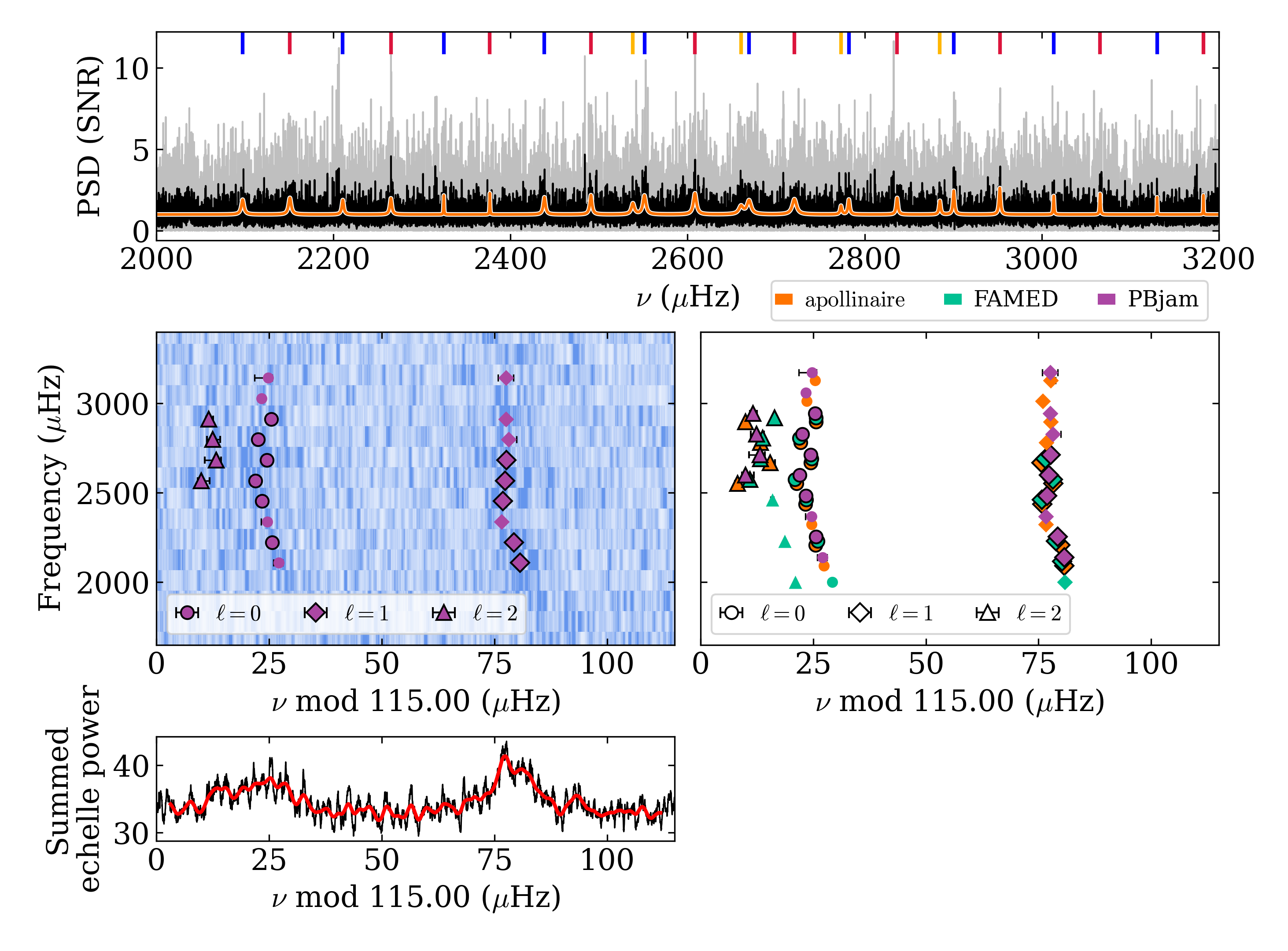}
    \caption{Resulting extracted modes for TIC 289622310 (19 Dra). Top and middle panels are the same as Figure~\ref{fig:354552931_echelle}. The échelle diagram has not been offset. The bottom panel shows the sum of the échelle diagram over frequency as a function of $\Delta\nu$ in black, with its moving mean (window = 2000 points) in red.}
    \label{fig:289622310_echelle}
\end{figure}

\begin{figure}[H]
    \centering
    \includegraphics[width=1\linewidth,trim={0 0 0 0.5cm},clip]{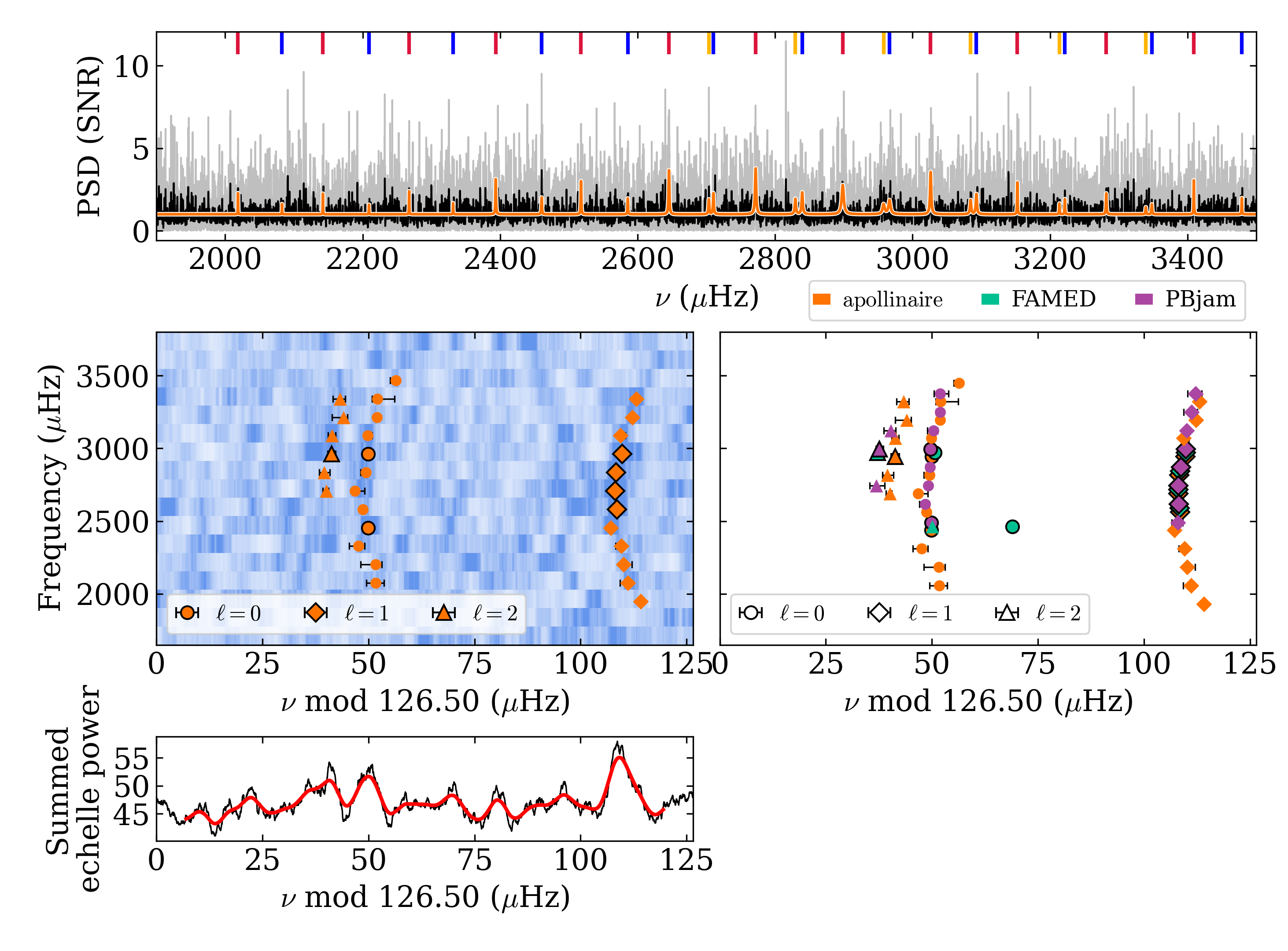}
    \caption{Same as Figure~\ref{fig:289622310_echelle}, but for TIC 20601206 (HD 176051). The échelle diagram has been offset by 7$\,\mu$Hz to improve ridges visibility. The moving mean of the summed échelle power was computed over a window of 1000 points.}
    \label{fig:20601206_echelle}
\end{figure}

\begin{figure}[H]
    \centering
    \includegraphics[width=1\linewidth,trim={0 0 0 0.5cm},clip]{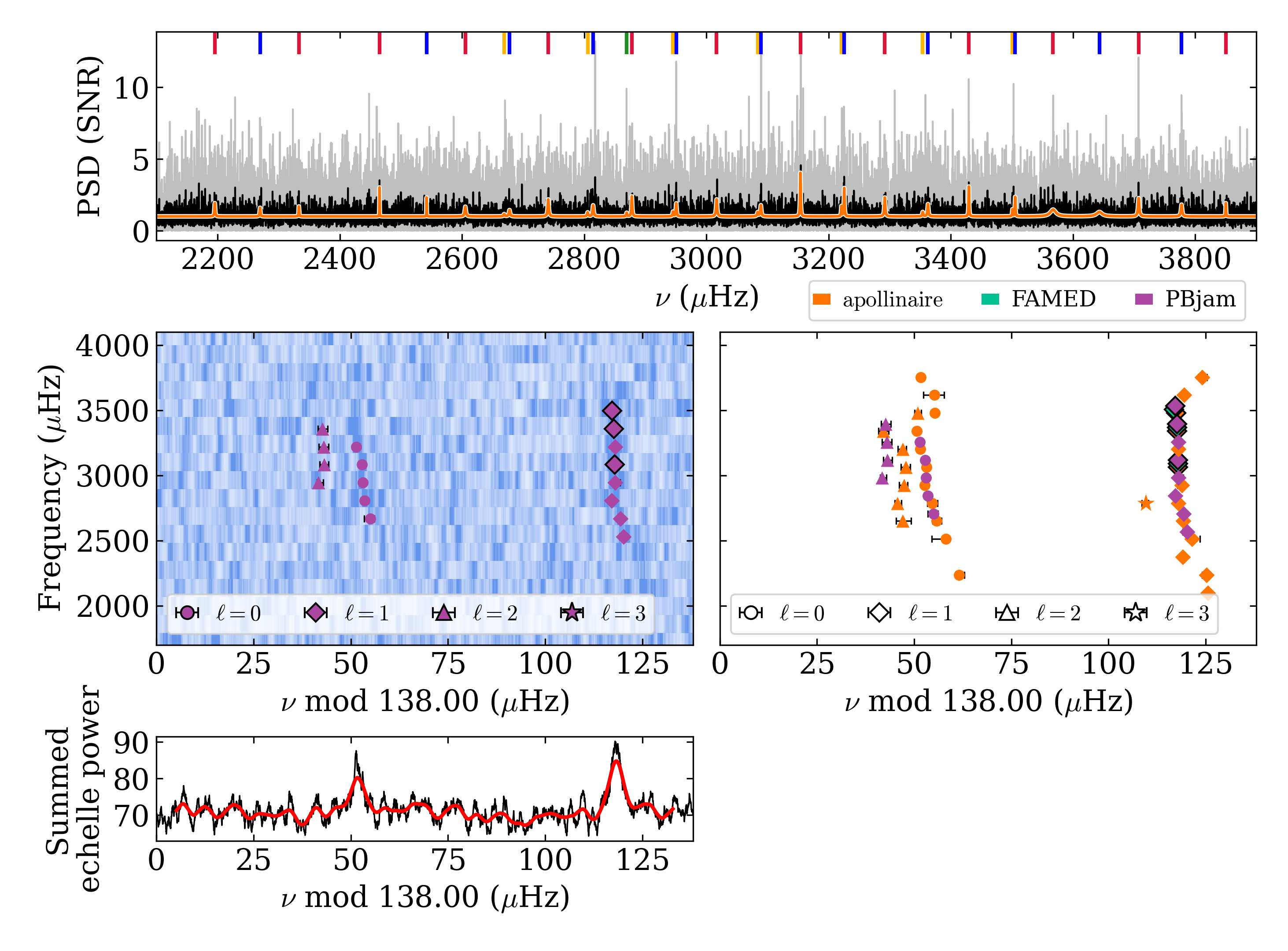}
    \caption{Same as Figure~\ref{fig:289622310_echelle}, but for TIC 403585118 (HD 193664). The échelle diagram has not been offset. The moving mean of the summed échelle power was computed over a window of 2000 points.}
    \label{fig:403585118_echelle}
\end{figure}

\begin{figure}[H]
    \centering
    \includegraphics[width=1\linewidth,trim={0 0 0 0.5cm},clip]{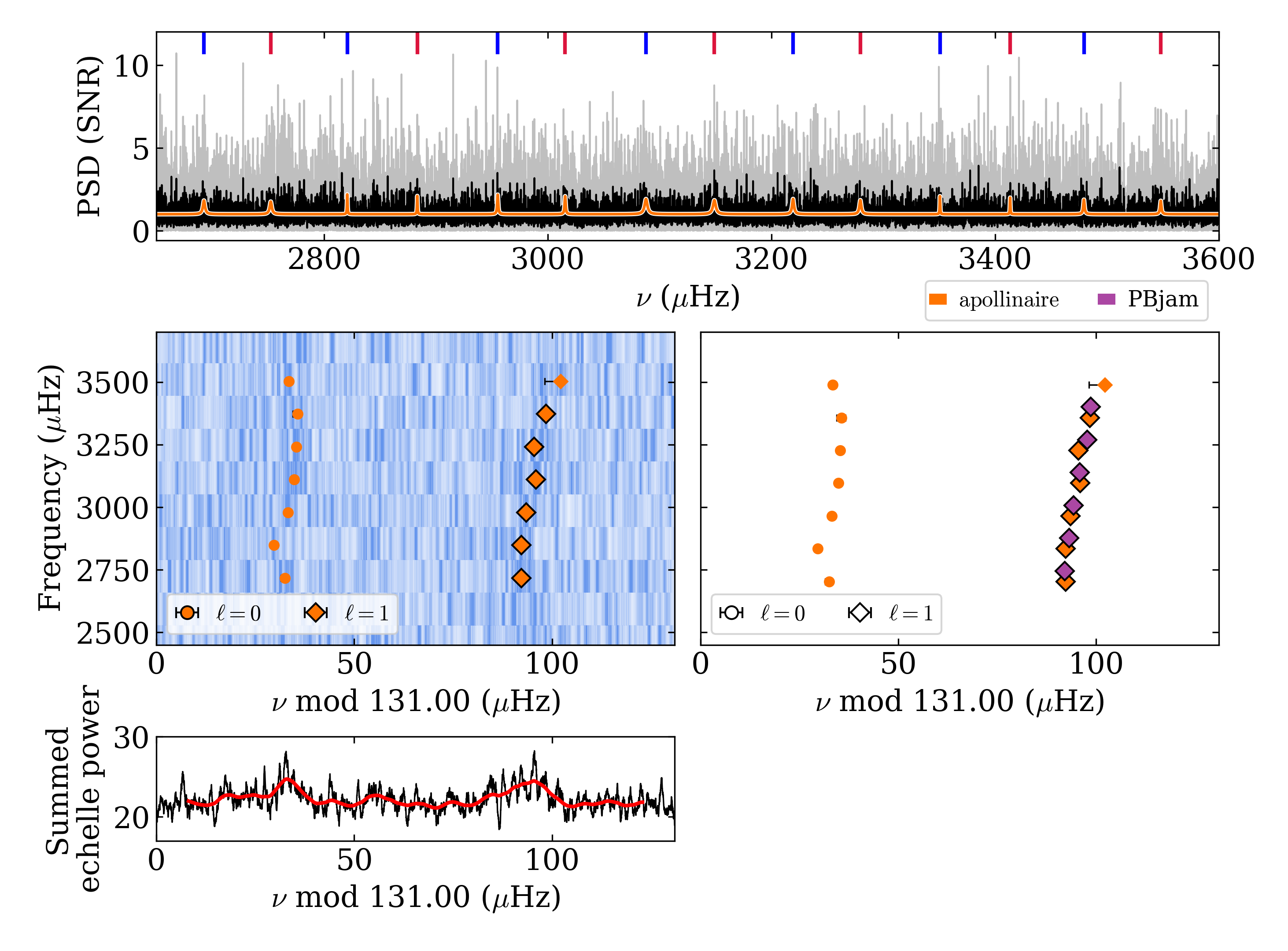}
    \caption{Same as Figure~\ref{fig:289622310_echelle}, but for TIC 219777482 (26 Dra). The échelle diagram has been offset by 40$\,\mu$Hz to improve ridges visibility. The moving mean of the summed échelle power was computed over a window of 5000 points.}
    \label{fig:219777482_echelle}
\end{figure}

\begin{figure}[H]
    \centering
    \includegraphics[width=1\linewidth,trim={0 0 0 0.5cm},clip]{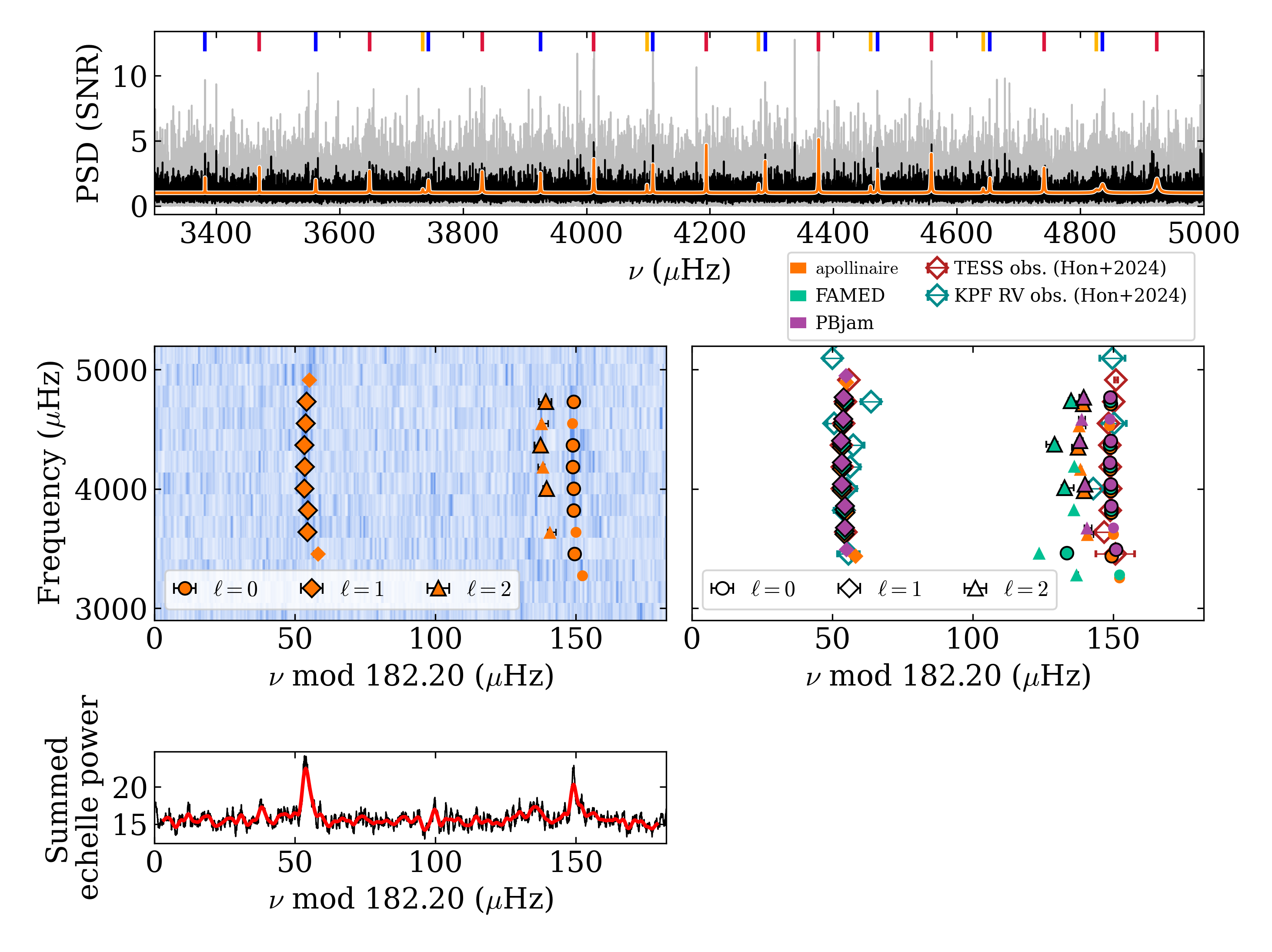}
    \caption{Same as Figure~\ref{fig:289622310_echelle}, but for TIC 259237827 ($\sigma$ Dra). The échelle diagram has been offset by -50$\,\mu$Hz to improve ridges visibility. The moving mean of the summed échelle power was computed over a window of 2000 points.}
    \label{fig:259237827_echelle}
\end{figure}


\FloatBarrier
\newpage
\section{Frequency tables}
\label{app:tables_per_star}

We present here the star-by-star resulting mode frequencies obtained from \texttt{apollinaire}, FAMED and \texttt{PBjam} for the optimal light curve. The other tables are available in machine-readable format.

\begin{table}[!ht]
\caption{List of oscillation mode frequencies obtained from the optimal light curve of TIC 354552931 (HD 36553) using \texttt{apollinaire}, FAMED, and \texttt{PBjam}.}
\label{tab:354552931_freqs}
\centering
\begin{tabular}{cc|ccc|c}
\hline
$n$ & $\ell$ & \texttt{apollinaire} & FAMED & \texttt{PBjam} & flag\\
\hline
10 & 1 & $389.86^{\substack{+1.78 \\ -1.24}}$ & -- & -- & 3\\
11 & 0 & $408.36^{\substack{+1.29 \\ -0.62}}$ & -- & -- & 3\\
11 & 1 & $424.73^{\substack{+0.54 \\ -0.38}}$ & $424.54^{\substack{+0.08 \\ -0.09}}$ & -- & 2\\
11 & 3 & $454.24^{\substack{+1.28 \\ -1.65}}$ & -- & -- & 3\\
12 & 1 & $457.41^{\substack{+0.35 \\ -0.41}}$ & $457.57^{\substack{+0.18 \\ -0.20}}$ & $457.75^{\substack{+0.31 \\ -0.17}}$ & 1\\
13 & 0 & $478.46^{\substack{+0.21 \\ -0.21}}$ & $477.92^{\substack{+0.11 \\ -0.10}}$ & $478.17^{\substack{+0.19 \\ -0.18}}$ & 1\\
13 & 1 & $492.64^{\substack{+0.16 \\ -0.16}}$ & $492.62^{\substack{+0.07 \\ -0.07}}$ & $492.36^{\substack{+0.29 \\ -0.34}}$ & 1\\
13 & 2 & $509.83^{\substack{+0.65 \\ -0.93}}$ & $510.27^{\substack{+0.23 \\ -0.19}}$ & -- & 2\\
14 & 0 & $511.86^{\substack{+0.23 \\ -0.25}}$ & $511.95^{\substack{+0.12 \\ -0.12}}$ & $511.66^{\substack{+0.14 \\ -0.15}}$ & 1\\
13 & 3 & $521.80^{\substack{+0.73 \\ -0.49}}$ & -- & -- & 3\\
14 & 1 & $525.89^{\substack{+0.20 \\ -0.20}}$ & $525.99^{\substack{+0.25 \\ -0.25}}$ & $525.49^{\substack{+0.25 \\ -0.22}}$ & 1\\
14 & 2 & $543.25^{\substack{+0.35 \\ -0.95}}$ & $543.42^{\substack{+0.17 \\ -0.17}}$ & -- & 2\\
15 & 0 & $544.80^{\substack{+0.39 \\ -0.40}}$ & $545.36^{\substack{+0.22 \\ -0.22}}$ & $544.90^{\substack{+0.23 \\ -0.22}}$ & 1\\
14 & 3 & $556.32^{\substack{+0.78 \\ -0.91}}$ & -- & -- & 3\\
15 & 1 & $558.98^{\substack{+0.31 \\ -0.30}}$ & $558.53^{\substack{+0.10 \\ -0.10}}$ & $559.89^{\substack{+0.17 \\ -0.20}}$ & 1\\
15 & 2 & $576.00^{\substack{+0.41 \\ -0.41}}$ & -- & -- & 3\\
16 & 0 & $578.57^{\substack{+0.28 \\ -0.27}}$ & $578.41^{\substack{+0.16 \\ -0.16}}$ & $578.69^{\substack{+0.13 \\ -0.18}}$ & 1\\
15 & 3 & $590.31^{\substack{+0.60 \\ -0.94}}$ & -- & -- & 3\\
16 & 1 & $593.00^{\substack{+0.35 \\ -0.38}}$ & $592.64^{\substack{+0.34 \\ -0.34}}$ & $593.23^{\substack{+0.27 \\ -0.29}}$ & 1\\
16 & 2 & $608.88^{\substack{+0.50 \\ -0.43}}$ & $609.72^{\substack{+0.37 \\ -0.37}}$ & -- & 2\\
17 & 0 & $612.15^{\substack{+0.21 \\ -0.22}}$ & $612.14^{\substack{+0.16 \\ -0.16}}$ & $612.62^{\substack{+0.26 \\ -0.29}}$ & 1\\
17 & 1 & $627.22^{\substack{+0.29 \\ -0.31}}$ & $627.28^{\substack{+0.13 \\ -0.13}}$ & $627.71^{\substack{+0.34 \\ -0.20}}$ & 1\\
17 & 2 & $644.62^{\substack{+0.64 \\ -0.56}}$ & $645.00^{\substack{+0.29 \\ -0.26}}$ & -- & 2\\
18 & 0 & $646.98^{\substack{+0.34 \\ -0.34}}$ & $647.00^{\substack{+0.22 \\ -0.23}}$ & $646.29^{\substack{+0.32 \\ -0.30}}$ & 1\\
18 & 1 & $661.16^{\substack{+0.28 \\ -0.29}}$ & $661.33^{\substack{+0.08 \\ -0.08}}$ & $659.70^{\substack{+0.25 \\ -0.38}}$ & 1\\
18 & 2 & $678.90^{\substack{+0.63 \\ -0.72}}$ & $679.18^{\substack{+0.35 \\ -0.33}}$ & -- & 2\\
19 & 0 & $681.18^{\substack{+0.21 \\ -0.25}}$ & $681.25^{\substack{+0.18 \\ -0.12}}$ & $680.81^{\substack{+0.24 \\ -0.32}}$ & 1\\
19 & 1 & $696.89^{\substack{+0.43 \\ -0.45}}$ & $696.90^{\substack{+0.18 \\ -0.17}}$ & $697.37^{\substack{+0.38 \\ -0.36}}$ & 1\\
19 & 2 & $713.71^{\substack{+0.79 \\ -1.06}}$ & $714.01^{\substack{+0.12 \\ -0.13}}$ & -- & 2\\
20 & 0 & $715.92^{\substack{+0.46 \\ -0.43}}$ & $715.85^{\substack{+0.14 \\ -0.16}}$ & -- & 2\\
20 & 1 & $729.00^{\substack{+0.65 \\ -0.54}}$ & -- & -- & 3\\
21 & 2 & -- & $781.61^{\substack{+0.09 \\ -0.09}}$ & -- & 3\\
22 & 0 & $783.34^{\substack{+1.78 \\ -1.22}}$ & $783.37^{\substack{+0.10 \\ -0.11}}$ & -- & 2\\
\hline
\end{tabular}
\tablefoot{The quoted radial orders ($n$) are indicative. The flag column indicates whether a given mode is included in the minimal (1) or maximal (2) frequency set, or excluded (3).}
\end{table}

\end{appendix}

\end{document}